# Roadmap on Nanoscale Superconductivity for Quantum Technologies

Oleksandr Dobrovolskiy [https://orcid.org/0000-0002-7895-8265][1]
Hermann Suderow [https://orcid.org/0000-0002-5902-1880][2]
Francesco Tafuri [https://orcid.org/0000-0003-0784-1454][3]
Annica M. Black-Schaffer [https://orcid.org/0000-0002-4726-5247][4]
Jose L. Lado [https://orcid.org/0000-0002-9916-1589][5]
Asle Sudbø [https://orcid.org/0000-0001-7550-4017][6]
Daniela Stornaioulo [https://orcid.org/0000-0003-2654-9504][3,7]
Chuan Li [https://orcid.org/0000-0001-5713-2117][8]
Anna E. Böhmer [0000-0001-6836-2954][9]
Lan Maria Tran [0000-0002-8727-0577][10]
Andrzej J. Zaleski [0000-0002-0676-5229][10]
Adrian Crisan [https://orcid.org/0000-0002-0662-0177][11]
Massimiliano Polichetti [https://orcid.org/0000-0002-4534-3301][12]
Armando Galluzzi [https://orcid.org/0000-0002-1372-357X][12]
Ali Gencer [https://orcid.org/0000-0002-9124-3862][13]
Bernd Aichner [https://orcid.org/0000-0002-9976-9877][14]
Neven Barišić [https://orcid.org/0000-0003-4637-0544][15,16]
Wolfgang Lang [https://orcid.org/0000-0001-8722-2674][14]
Tomas Samuely [0000-0001-5618-6965][17]
Martin Gmitra [0000-0003-1118-3028][18]
Tristan Cren [0000-0001-7105-4098][19]
Mateo Calandra [0000-0003-1505-2535][19]
Peter Samuely [0000-0002-0370-4682][18]
Jeroen Custers [https://orcid.org/0000-0001-6974-1489][20]
Rosa Córdoba [https://orcid.org/0000-0002-6180-8113][21]
Vladimir M. Fomin [[https://orcid.org/0000-0002-8959-4831][22,23]
Nicola Poccia [https://orcid.org/0000-0001-7982-0113][3,22]
Pavol Szabó [https://orcid.org/https://orcid.org/0000-0002-1109-5012][18]
Fabrizio Porrati [https://orcid.org/0000-0003-1925-9437][24]
Gleb Kakazei [0000-0001-7081-581X][25]
Jan Aarts [https://orcid.org/0000-0002-4113-0835][26]
Jason Robinson [https://orcid.org/0000-0002-4723-722X][27]
Javier E. Villegas [https://orcid.org/0000-0002-2096-3360][28]
Matthias Althammer [https://orcid.org/0000-0003-1625-6054][29]
Hans Huebl [https://orcid.org/0000-0003-3023-5209][29,30]
Akashdeep Kamra [https://orcid.org/0000-0003-0743-1076][31]
Mathias Weiler [https://orcid.org/0000-0003-0537-9251][31]
J. Hugo Dil [0000-0002-6016-6120][32,33]
Daniil Evtushinsky [0000-0002-5604-6740][32]
Beena Kalisky [https://orcid.org/0000-0002-1270-2670][34]
Yonathan Anahory [https://orcid.org/0000-0002-9368-1129][35]

Simon Bending [https://orcid.org/0000-0002-4474-2554][36]
Peter Liljeroth [0000-0003-1253-8097][37]
Abdou Hassanien [https://orcid.org/0000-0002-6774-7518][38]
Isabel Guillamón [https://orcid.org/0000-0002-2606-3355][2]
Edwin Herrera [https://orcid.org/0000-0002-2606-3355][2]
Alejandro V. Silhanek [https://orcid.org/0000-0001-9551-5717][39]
Joris Van de Vondel [https://orcid.org/0000-0001-6894-7258][40]
Anna Palau [https://orcid.org/0000-0002-2217-164X][41]
Ilya Charaev [https://orcid.org/0000-0002-4036-0778][42]
Maria Sidorova [https://orcid.org/0000-0002-9082-442X][43]
Floriana Lombardi [https://orcid.org/0000-0002-3478-3766][44]
Thilo Bauch [https://orcid.org/0000-0002-8918-4293][44]
Cheryl Feuillet-Palma [0000-0002-8389-5756][45]
Vasily Stolyarov [0000-0002-5317-0818][45]
Dimitri Roditchev [0000-0003-3308-1313][45]
Vladimir M. Krasnov [https://orcid.org/0000-0002-3131-8658][46]
Benedikt Hampel [https://orcid.org/0000-0002-5095-7922][47]
María José Martínez-Pérez [0000-0002-8125-877X][48]
Javier Sesé [0000-0002-7742-9329][48]
Dieter Koelle [0000-0003-3948-2433][49]
Stefano Poletto [0000-0002-3473-3689][50]
Alessandro Bruno [0000-0002-7821-5082][51]
Davide Massarotti [0000-0001-7495-362X][3]

[1]Cryogenic Quantum Electronics, EMG and LENA, Technische Universität Braunschweig, Braunschweig, Germany
[2]Universidad Autónoma de Madrid, Madrid, Spain
[3]Universita Degli Studi Di Napoli Federico II, Napoli, Italy
[4]Department of Physics and Astronomy, Uppsala University, Uppsala, Sweden
[5]Department of Applied Physics, Aalto University, Aalto, Finland
[6]Center for Quantum Spintronics, Department of Physics, Norwegian University of Science and Technology, Trondheim, Norway
[7]CNR-SPIN, Naples, Italy
[8]MESA+ Institute for Nanotechnology, University of Twente, Netherlands
[9]Experimental Physics IV, Ruhr University Bochum, Bochum, Germany
[10]Institute of Low Temperature and Structure Research, Polish Academy of Sciences, Poland
[11]National Institute of Materials Physics, Bucharest, Romania
[12]University of Salerno, Salerno, Italy
[13]Ankara University, Ankara, Türkiye
[14]Faculty of Physics, University of Vienna, Vienna, Austria
[15]Institute of Solid State Physics, TU Wien, Vienna, Austria
[16]Department of Physics, Faculty of Science, University of Zagreb, Zagreb, Croatia

[17]P.J. Šafárik University, Košice, Slovakia
[18]Institute of Experimental Physics, Slovak Academy of Sciences, Košice, Slovakia
[19]INSP, Sorbonne University, CNRS, Paris, France
[20]Department of Condensed Matter Physics, Charles University, Prague, Czech Republic
[21]Institute of Molecular Science (ICMol), University of Valencia, Valencia, Spain
[22]Institute for Emerging Electronic Technologies, Leibniz IFW Dresden, Dresden, Germany
[23]Faculty of Physics and Engineering, Moldova State University, Chișinău, Republic of Moldova
[24]Goethe University Frankfurt am Main, Frankfurt am Main, Germany
[25]Institute of Physics for Advanced Materials, Nanotechnology and Photonics (IFIMUP), Departamento de Fisica e Astronomia, Faculdade de Ciências, Universidade do Porto, Porto, Portugal
[26]Huygens-Kamerlingh Onnes Laboratory, Leiden University, Leiden, The Netherlands
[27]Department of Materials Science & Metallurgy, University of Cambridge, Cambridge, United Kingdom
[28]Laboratoire Albert Fert, CNRS, Thales, Université Paris-Saclay, France
[29]Walther-Meißner-Institut, Bayerische Akademie der Wissenschaften and Technical University of Munich, TUM School of Natural Sciences, Physics Department, Garching, Germany
[30]Munich Center for Quantum Science and Technology (MCQST), Munich, Germany
[31]Fachbereich Physik and Landesforschungszentrum OPTIMAS, RPTU Kaiserslautern-Landau, Kaiserslautern, Germany
[32]Institute of Physics, Ecole Polytechnique Fédérale de Lausanne, Lausanne, Switzerland
[33]Center for Photon Science, Paul Scherrer Institut, Villigen, Switzerland
[34]Bar Ilan University, Ramat Gan, Israel
[35]The Hebrew University of Jerusalem, Jerusalem, Israel
[36]Centre for Nanoscience and Nanotechnology, Department of Physics, University of Bath, Bath, United Kingdom
[37]Aalto University School of Science, Espoo, Finland,
[38]Jozef Stefan Institute, Ljubljana, Slovenia
[39]Université de Liège, Liège, Belgium
[40]Katholieke Universiteit Leuven, Leuven, Belgium
[41]Institute of Materials Science of Barcelona, Barcelona, Spain
[42]University of Zurich, Zurich, Switzerland
[43]Humboldt-Universität zu Berlin and German Aerospace Center, Berlin, Germany
[44]Department of Microtechnology and Nanoscience, Chalmers University of Technology, Gothenburg, Sweden
[45]LPEM, ESPCI Paris, PSL, CNRS, Sorbonne University, Paris, France
[46]Department of Physics, Stockholm University, AlbaNova University Center, Stockholm, Sweden
[47]Institut für Elektrische Messtechnik und Grundlagen der Elektrotechnik, TU Braunschweig, Braunschweig, Germany
[48]Instituto de Nanociencia y Materiales de Aragón (INMA), CSIC – Universidad de Zaragoza, Spain
[49]Physikalisches Institut, Center for Quantum Science (CQ) and LISA+, Universität Tübingen, Germany
[50]Rigetti Computing, 775 Heinz Avenue, Berkeley, CA 94710, USA
[51]QuantWare, 2628 XG Delft, The Netherlands

# Contents

## Abstract

In 2025, the Year of Quantum Science and Technology (https://quantum2025.org/), we celebrate a century of quantum mechanics, witnessing a surge in activities that illuminate its inherent strangeness and drive technological innovation. Superconductivity, discovered 114 years ago, stands as a prime example, offering direct and compelling evidence of macroscopic quantum phenomena. Beyond its ability to conduct immense currents without loss, superconductivity reveals the quantum realm operating on a scale we can directly observe and manipulate. The macroscopic quantum coherence, where an ensemble of particles is described by a single wave function, leads to remarkable consequences: dissipation-less current and flux quantization – the basic properties exploited in superconducting quantum circuit fabrication. This Roadmap has been inspired by intensive discussions and collaborations emerging from the European Cooperation in Science & Technology COST-Action CA21144 (SuperQuMap – Superconducting Nanodevices and Quantum Materials for Coherent Manipulation). The aim of the COST Action SuperQuMap is to establish a strong European network centered on macroscopic quantum behavior in superconductors, bringing together groups of different backgrounds and more than 30 countries. The roadmap outlines the network's concrete activities, driving advancements in superconductor-based quantum technologies and charting future directions. Spanning fundamental research to practical applications, the roadmap incorporates insights from industry partners developing quantum computation. It begins by exploring quantum materials, highlighting how topology and electronic correlations could catalyze a quantum leap in technology. We then delve into manipulating the superconducting phase, leveraging advancements in magnetism, 3D fabrication, and tunable correlations. Further, we showcase the advanced microscopy techniques—such as angle-resolved photoemission spectroscopy and scanning probes—used to visualize quantum behavior. Finally, and crucially, we detail the quantum devices developed within the network, and their transformative impact on modern quantum computing approaches.

## Introduction

*Oleksandr Dobrovolskiy*[1], *Hermann Suderow*[2], *and Francesco Tafuri*[3]

[1]Technische Universität Braunschweig, Germany

[2]Universidad Autónoma de Madrid, Spain

[3]Universita Degli Studi Di Napoli Federico II, Italy

### Introduction

The field of superconducting quantum electronics has experienced explosive growth in recent years, marked by a rapid proliferation of nanodevices and hybrid systems for quantum coherent manipulation and sensing. A promising avenue for advancing superconducting electronics, including enhanced device stability and coherence, lies in exploring novel quantum materials and phenomena. This ambitious endeavor necessitates a collaborative effort spanning condensed matter physics, materials science, and engineering. To address this challenge, the COST Action CA 21144 [1] has united numerous research groups in more than 30 European countries to pursue three key objectives: (i) the synthesis and characterization of quantum materials with novel topological properties, (ii) the fabrication of sensors and devices leveraging innovative superconducting functionalities, and (iii) the generation and coherent manipulation of superconducting states to unlock new opportunities in quantum electronics. This synergistic approach, integrating theoretical modeling, design, fabrication, characterization, and validation, paves the way for the exploitation of nanoscale superconductors in emerging quantum technologies.

Driven by the dynamic convergence of quantum materials, superconductivity, and nanotechnology, this Roadmap offers a concise guide for researchers from academia, research institutes, industry, and other stakeholders, as funding agencies. This document provides guidance on foundational knowledge, cutting-edge techniques, and emerging trends, empowering the development of a skilled workforce to propel future innovations. By delving into the complexities of topology, dimensionality, and order parameter hybridization, researchers and engineers are poised to unlock unprecedented advances in science and technology. This roadmap envisions a future where nanoscale superconducting systems become pivotal in shaping the next generation of electronic and information processing systems.

The Roadmap comprises contributions that address the objectives of the SuperQuMap COST Action, namely quantum materials, nanoscale structures, and quantum devices. These include advances in superconducting quantum materials, descriptions of three-dimensional and hybrid structures, properties of nanoscale superconductors, and of superconducting quantum devices.

The Roadmap begins with the introduction of superconducting quantum materials, covering recent advances in fabrication methods and outlining emerging phenomena occurring in low-dimensional systems, at interfaces and in systems with competing interactions.

Section 1, authored by Black-Schaffer, Lado, and Sudbø, explores topological matter, focusing on its non-trivial electronic structure and the resulting topological protection of states for quantum information processing. They also discuss the crucial roles of the superconducting symmetry, material quality, and advanced probing techniques in characterizing topological superconducting states.

In section 2, Stornaiuolo and Li explore 2D materials and interfaces as promising solid-state platforms for spintronics and the realization of qubits. They detail novel oxide interfaces and surface superconductivity in Weyl semimetals, and discuss their potential for downscaling future electronic components.

Section 3 by Böhmer, Tran and Zaleski reviews latest advances in iron-based superconductors. Their unique superconducting order parameters, coupled with relatively high transition temperatures offer significant potential for spintronics and high-field magnet applications. They describe efforts to achieve a comprehensive understanding of unconventional superconductivity, a key challenge for the development of future superconducting quantum technology.

Section 4 by Crisan, Polichetti, Galluzzi and Gencer discusses critical currents in superconductors, addressing both macroscopic and nanometer scales. They emphasize that the development of new superconducting materials for both large-scale and nanoscale applications necessitates thorough studies of vortex matter, dynamics, and pinning across a wide range of time scales and material parameters.

Section 5, authored by Aichner, Barišić, and Lang, focuses on copper-oxide superconductors, a subject of ongoing debate since their 1986 discovery. Despite their complexity and diverse conductivity regimes, the authors identify universal behaviors within the cuprate phase diagram. They also present focused He+ beam irradiation as a method for selectively suppressing superconductivity.

In section 6, T Samuely, Gmitra, Cren, Calandra and P Samuely focus on misfit layer superconductors as versatile platforms for engineering quantum effects through vertical stacking. They describe Ising superconductivity in atomically thin transition metal dichalcogenides and in bulk misfit layer compounds, Fermi level shifts via tuneable charge doping, and proximity-induced spin-orbit coupling and exchange interactions in 2D material heterostructures as sources of novel quantum phenomena, engineered through elaborate stacking.

Section 7, by Custers, focuses on heavy fermion compounds with strong electronic correlations, addressing the universality of quantum phase transitions, dimensionality, and the development of a microscopic theory for unconventional superconductivity. The emphasis is on synthesizing ultra-clean crystals and advancing experimental techniques.

Then, the roadmap describes nanofabrication, hybrid structures and advanced techniques for the visualization of superconducting behavior.

In section 8, Dobrovolskiy, Córdoba and Fomin introduce superconductor 3D nanoarchitectures, where complex 3D geometry dictates the nontrivial topology of screening currents. They highlight strain-driven self-rolling and focused ion/electron beam nanoprinting as established 3D nanofabrication techniques, and discuss potential applications in fluxonic superconducting electronics.

In section 9, Poccia, Szabó and Porrati explore emergent states of matter in twisted and mesoscopic superconductors. They highlight how artificial design enables control over these states for specific device functions, focusing on cuprate twistronics, artificially fabricated layered systems, and superconductor-metal-insulator quantum phase transitions.

In section 10, Kakazei provides a review of magnetic nanoelements and their arrays, highlighting their role as essential building blocks for hybrid systems, specifically superconductors engineered with pinning site arrays. He discusses the characteristics of magnetic vortex states, their microwave response in the context of magnonics, a field leveraging collective spin precessions for data processing, and underscores an increasing interest in merging magnetic phenomena with superconductivity.

In section 11, Aarts, Robinson, and Villegas review superconductor-ferromagnet hybrids, focusing on how a magnetic layer can create unconventional Josephson junctions or carry spin-polarized supercurrents for dissipationless spintronics. They emphasize the need for interdisciplinary collaboration to design and understand these hybrid systems.

In section 12, Althammer, Huebl, Kamra, and Weiler explore superconducting spin currents in heterostructures, focusing on spin-orbit torque detection in superconductor-ferromagnet hybrids. They discuss potential applications as magnetic memories, spin-torque oscillators, diode effects, and Josephson π-junctions for flux qubits utilizing spin- triplet supercurrents in conventional superconductors.

In section 13 Dil and Evtushinsky review advancements in angular resolved photoemission (ARPES) as a high-resolution tool for mapping electronic structures. They outline that ARPES experiments would benefit from reducing the spot size and incorporating an external magnetic field or mechanical strain for unravelling the entangled nature of the superconducting state.

In section 14, Kalisky and Anahory detail scanning microscopy techniques based on using a superconducting quantum interference device (SQUID), including SQUID-on-chip, SQUID-on-tip and SQUID-lever designs. They highlight applications in vortex physics and for exploring exotic order parameters, and anticipate improvements in spatial resolution, bandwidth, and combination with other probes.

In section 15, Bending reviews scanning Hall microscopy, a non-invasive mapping technique which can be used over wide temperature and magnetic field ranges. Anticipated developments include wafer-scale graphene fabrication methods targeting the preservation of high carrier mobilities in devices with nanometer-scale leads.

In section 16, Suderow, Liljeroth, Hassanien, Guillamon and Herrera review scanning tunnelling microscopy highlighting its ability to visualize atoms and electronic property changes at the atomic scale. They emphasize STM’s role in creating quantum dot states for studying 2D electronic systems.

Finally, the focus is directed towards superconducting devices.

In section 17, Silhanek, Van de Vondel and Palau explore tuning of superconducting devices with an electric field. They discuss gating and electromigration, which enables controllable charge carrier concentration changes without increased disorder. They anticipate advancements in CMOS compatibility and nanofabrication optimization.

In section 18, Charaev and Sidorova review superconducting single-photon detectors, crucial for quantum optics and quantum information science, including secure communication via quantum key distribution. Their integration of detectors in telecommunications, environmental monitoring and medical diagnostics, underlines their pivotal role in modern photonics.

In section 19, Lombardi and Bauch discuss recent advances and perspectives in high-Tc Josephson junctions. New strategies include Dayem bridges, high-energy ion irradiation of wide bridges, and direct He focused ion beam irradiation. Further research will focus on minimizing 1/f noise and critical current density variability.

In section 20, Feuillet-Palma, Stolyarov, Cren and Roditchev delve into Josephson vortices in proximity junctions, focusing on the emergence of novel quantum functionalities from the interplay of non-superconducting material properties and superconducting correlations. They examine the fast, low-dissipation dynamics of these vortices in the context of challenges in developing scalable cryogenic memories compatible with superconducting microwave technologies.

In section 21, Krasnov and Hampel details the application of superconducting Josephson junctions in terahertz (THz) frequency domain. They highlight the potential of THz electronics for next-generation telecommunications and ultrafast computing emphasizing the broad frequency tunability of superconducting THz sources. However, they also acknowledge the challenge of achieving high far-field emission power.

In section 22, Martínez-Pérez, Sesé and Koelle review the progress on nanoscale Josephson junctions for quantum technology applications. They highlight progress in fabrication techniques for improved Josephson junctions and nanoSQUIDs, the shift towards constriction-like designs, and the utilization of high-Tc superconductors, which collectively enable new applications in quantum sensing, magnonics, and spin-based quantum computing.

Poletto, Bruno, Massarotti, and Tafuri, in Section 23, review the current state and future prospects of superconducting qubits. They highlight that superconducting qubit-based quantum processors and technologies have reached a critical juncture, where operational quality and device scaling enable valuable applications. However, they also emphasize that achieving a universal quantum computer necessitates further advancements in both physics and engineering.

# I. Superconducting quantum materials

## 1. Superconductivity and topology

*Annica M. Black-Schaffer*[1]*, Jose L. Lado*[2]*, and Asle Sudbø*[3]

[1]Uppsala University, Sweden

[2]Aalto University, Finland

[3]Norwegian University of Science and Technology, Norway

### 1.1 Status

The discovery of various topological states of matter has revolutionized physics over the past two decades. Topological matter is defined by a *global non-trivial topology* in its electronic structure, giving rise to unconventional properties not observed in conventional matter. Topology provides protection against perturbations, exemplified by protected conductivity in the quantum Hall state. For superconductors, topological protection enables quantum information protection, potentially overcoming major challenges in quantum computing. This global approach differs from traditional classifications, where *local order parameters* are instead the foundation within the Ginzburg-Landau paradigm, including for superconductivity. In topological superconductors (TSCs), these global and local views merge, creating exceptional possibilities for new quantum states.

Even a conventional spin-singlet *s*-wave superconductor, when combined with a complex normal state band structure, can form a TSC. An example is an effective spinless *p*-wave TSC when spin-orbit coupling and magnetism are present, see Fig. 1.1(a) [1]. Non-trivial topology, or winding, in the superconducting order parameter itself can also create TSCs. Examples include chiral *p*- and *d*-wave superconductors, see Fig. 1.1(b), which can be dictated by the crystal structure for pairing beyond phonon-based mechanisms [2]. Superconducting heterostructures provide additional possibilities, combining distinct materials into effective TSCs (note these do not necessarily have to be TSCs but they can become if, for example, spin-orbit coupling (SOC) is involved), while twisting 2D materials (c.f. Section 9) can alter both the superconducting order and the normal-state band structure to produce TSCs [3].

A main consequence of topology in matter is protected zero-energy boundary states, ensured by the bulk-boundary correspondence. TSCs are unique because these zero-energy states can manifest as Majorana excitations, particles that are their own antiparticles. Depending on material dimensionality, they can be localized Majorana bound states in 1D TSCs, chiral Majorana states in 2D TSCs, or flat zero-energy Majorana bands in nodal TSCs. Superconducting vortices also create effective boundaries hosting topological zero-energy states. These Majorana excitations are behaving effectively a "half" an electron and, is thus an entirely new emerging particle.

Current research aims to discover new TSCs in bulk materials, heterostructures, and twisted matter, understand their properties, and explore applications, including topological quantum computing via the non-Abelian statistics of Majorana states [1]. Even more ambitiously, superconductivity within fractional Chern insulators would realize parafermions and Fibonacci anyons, see Fig. 1.1(c), enabling the creation of universal topological qubits [4].

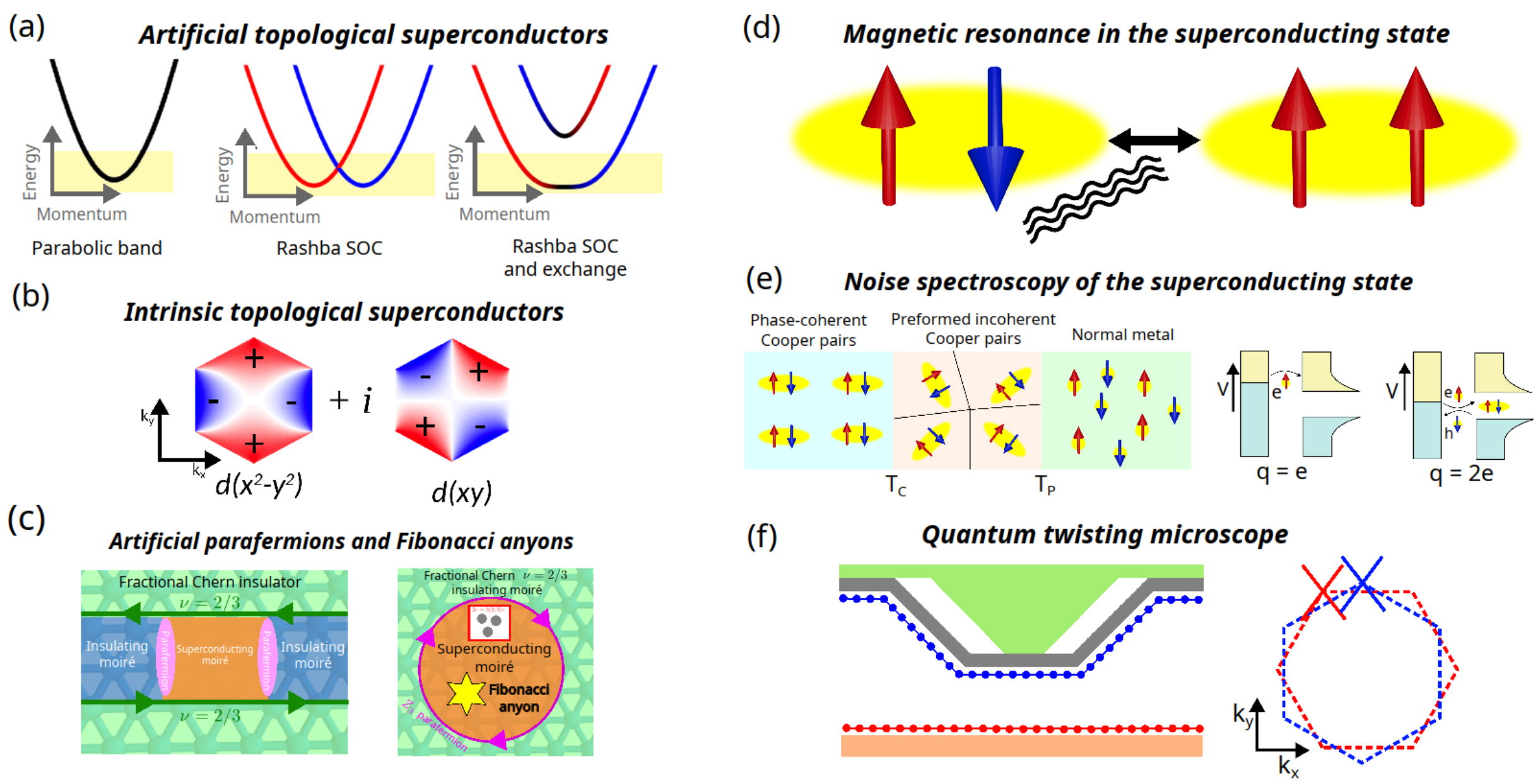

**Figure 1.1.** (a,b,c) TSC states and (d,e,f) probing methods for TSCs. Panel (a) depicts the strategy to create artificial TSCs, (b) an example of an intrinsic TSC order in a honeycomb lattice, and (c) the strategy to create parafermions and Fibonacci anyons. Panel (d) depicts spin transition that would allow gaining spin resolution into Cooper pairs, (e) noise spectroscopy that allows probing Cooper pairs in the absence of phase coherence, and (f) shows the quantum twisting microscope technique that allows gaining momenta resolution in spectroscopy.

## 1.2 Current and Future Challenges

TSCs offer immense promise for both fundamental advances and technical applications, but multiple challenges are tied to working with each material or system. As with all superconductors, a key factor for TSCs is determining the superconducting order parameter, particularly its symmetry, as this dictates the material's physical behavior. For any superconductor beyond the conventional s-wave superconductors, this involves identifying both the spin and orbital dependencies of the Cooper pairs. Progress in this area requires joint experimental and theoretical efforts.

Another critical aspect for all superconductors is material quality. Not only is disorder present in all materials, but TSCs are often sensitive to all forms of disorder. Many correlated bulk superconductors also require chemical doping, which introduces additional disorder. The problem is compounded in heterostructures, where functionality often relies on interfaces (c.f. Section 2), making interface quality control crucial. Thus, developing clean materials and high-quality interfaces is essential for discovering, characterizing, and utilizing TSCs.

An additional challenge for TSCs is determining their topology. Theoretically, this involves identifying the topological invariant inside the bulk. While mathematically well-defined, this is challenging to determine experimentally. Instead, only the consequences of topology are usually measured. The simplest of these is the existence of zero-energy boundary states. However, superconductors can host zero or near-zero energy states for various reasons, including magnetic impurities, crystal inhomogeneities, vortices, interface scattering, or a nodal order parameter. Therefore, the presence of zero-energy states at the boundary does not guarantee non-trivial topology, making it necessary to go beyond such indicators.

A standard solution in topological systems is to measure quantized Hall conductance, as it directly links to the topological invariant. However, because superconductors do not preserve charge, such quantization is not possible in TSCs. Instead, spin or thermal Hall signatures are required, which can be more challenging to measure [5]. Ultimately, directly measuring the topological invariant or at least unique consequences thereof is necessary. One intriguing example is the non-Abelian statistics of Majorana excitations, which are accessible through particle braiding. Although particle braiding is very challenging to achieve, it would confirm the topological nature of the system and also directly open up applications in quantum computing.

### 1.3 Advances in Science and Technology to Meet Challenges

Advances needed to address the research challenges focus on both generating and detecting TSCs. A key issue is energy scales: superconducting states in correlated materials often appear at very low temperatures, even challenging for current cryogenic techniques, making experimental work with many TSCs difficult. Paradigmatic examples include unconventional superconducting states in heavy-fermion systems (Section 7) and twisted van der Waals materials, which typically feature much smaller energy scales than bulk cuprate (Section 5), pnictide (Section 3), or nickelate superconductors. TSCs at higher temperatures and/or improved energy resolution in experimental probes, such as scanning tunneling microscopy (STM) (Section 16) and magnetic scanning probes (Sections 14 and 15), are critical areas for advancement. Enhanced energy resolution in angle-resolved photoemission (ARPES) (Section 13), which allows for imaging of the superconducting Fermi surface, is also necessary.

Further progress in material quality is also essential. Both advancements in material synthesis to improve quality and better theoretical methods for modelling impurity effects at the atomic scale are vital for future developments. Twisted van der Waals materials may offer benefits here, as they allow exploration of doping dependence through gating without inducing disorder, although they are instead sensitive to twist disorder [6]. Twist disorder occurs at larger length scales (tens of nanometers), which demands new strategies for accurate theoretical modelling at the moiré and even super-moiré scales.

Finally, development of new innovative probing methods to examine the superconducting state more effectively is desirable. Promising recent developments include electrically-driven parametric resonance with scanning probe microscopy, which has enabled probing of spin-flip excitations with sub-thermal resolution and may facilitate probing of spin transitions in spin-triplet superconductors, see Fig. 1.1(d) [7]. Furthermore, noise spectroscopy can identify regimes where Cooper pairs form, see Fig. 1.1(e) [8] without superconducting phase coherence, a ubiquitous regime in unconventional superconductors. Additionally, measuring the momentum dependence of the superconducting gap, particularly in van der Waals materials, may be achievable with the scanning twisting microscope technique, see Fig. 1.1(f) [9], potentially leading to breakthroughs in understanding unconventional superconductivity.

### 1.4 Concluding Remarks

Combining superconductivity with topology offers unprecedented possibilities for uncovering new fundamental phenomena and enabling ground-breaking technological breakthroughs, ultimately including universal topological quantum computers. The field, however, faces multiple challenges, not only those well-known to exist when identifying unconventional superconductivity, but also challenges tied to the need to determine the true microscopic and topological nature of potential TSCs. These challenges range from material synthesis and characterization to theory and modelling.

Nevertheless, the field is rapidly advancing through a flurry of activities, resulting in ever cleaner materials, improved heterostructures, and entirely new measurement techniques that allow for probing the internal structure of TSCs. On the theoretical side, new algorithms to understand superconducting states are essential and is being developed, especially for superconducting states that emerge at large length scales, as in moiré materials with unit cells of hundreds of thousands of atoms. Moreover, new strategies that leverage techniques traditionally not used in superconducting research are expected to enable significant advances and likely lead to unforeseen discoveries, as the potential for combining topology and superconductivity is vast.

**Acknowledgements**

AMBS acknowledges financial support from the Knut and Alice Wallenberg Foundation through the Wallenberg Academy Fellows program, KAW 2019.0309. JLL acknowledges financial support from the Research Council of Finland and the Finnish Quantum Flagship.

## 2. 2D materials and interfaces

*Daniela Stornaiuolo*[1] *and Chuan Li*[2]

[1]University of Naples and CNR-SPIN, Italy
[2]University of Twente, The Netherlands

### 2.1 Status

2D materials and interfaces in complex heterostructures exhibit a variety of properties interesting for spintronics and future quantum technologies [1]. Indeed, they are promising solid-state platforms for the realization of qubits and for topological quantum computing elements. This contribution focuses on selected recently developed material platforms, highlighting their potentiality in these fields.

**Novel oxide interfaces.** Since the striking discovery of the 2-dimensional electron gas (2DEG) a the $LaAlO_3/SrTiO_3$ (LAO/STO) interface, the development of atomically precise growth techniques enabled the creation of thin-film heteroepitaxial interfaces based on a wide class of oxide materials. Not only do these materials exhibit interesting physics, but they have also opened up new perspectives in oxide electronics and spintronics. Recently, the attention of the oxide research community has focused on $KTaO_3$ (KTO)-based heterostructures. KTO has a crystal structure similar to STO (Fig. 2.1(a)), with the additional advantage of a spin-orbit coupling (SOC) more than one order of magnitude larger (Fig. 2.1(b)), thanks to the presence of 5d tantalum ions. This property, locking together charge and spin degrees of freedom thereby allowing an electrical control of the spin, is the basic ingredient for spintronics (see also Sections 11 and 12). In the last decade, KTO-based 2DEG were realized at a variety of interfaces. Besides confirming a Rashba coefficient comparable to semiconductor structures, other interesting properties emerged [2] (Fig. 2.1(c)). For instance, 2D superconductivity was revealed in (111) and (110) oriented LAO/KTO and (111) EuO/KTO interfaces, and planar Hall effect and anisotropic magnetoresistance was measured at LVO/KTO interface, strongly resembling Weyl materials.

**Surface superconductivity in Weyl semimetals.** Weyl semimetals are materials where Weyl fermions appear as low-energy excitations due to the topological nature of their band structure related to the symmetry breaking (see also Section 7). A key feature of these materials is the presence of Fermi arcs, open surface states that connect Weyl nodes of opposite chirality [4]. It has been shown that these surface states can become superconducting when subjected to proximity-induced superconductivity [5]. More recently, experiments have provided direct evidence of superconducting Fermi arcs in trigonal $PtBi_2$ [6]. This discovery has validated some of these predictions and opened new avenues for studying surface superconductivity in topological systems. This has spurred great interest in understanding the relationship between topology and superconductivity in these materials, especially with the possibility of realizing Majorana bound states (see also Sections 3, 5, 6, 9, and 16) – quasiparticles that are promising for quantum computing.

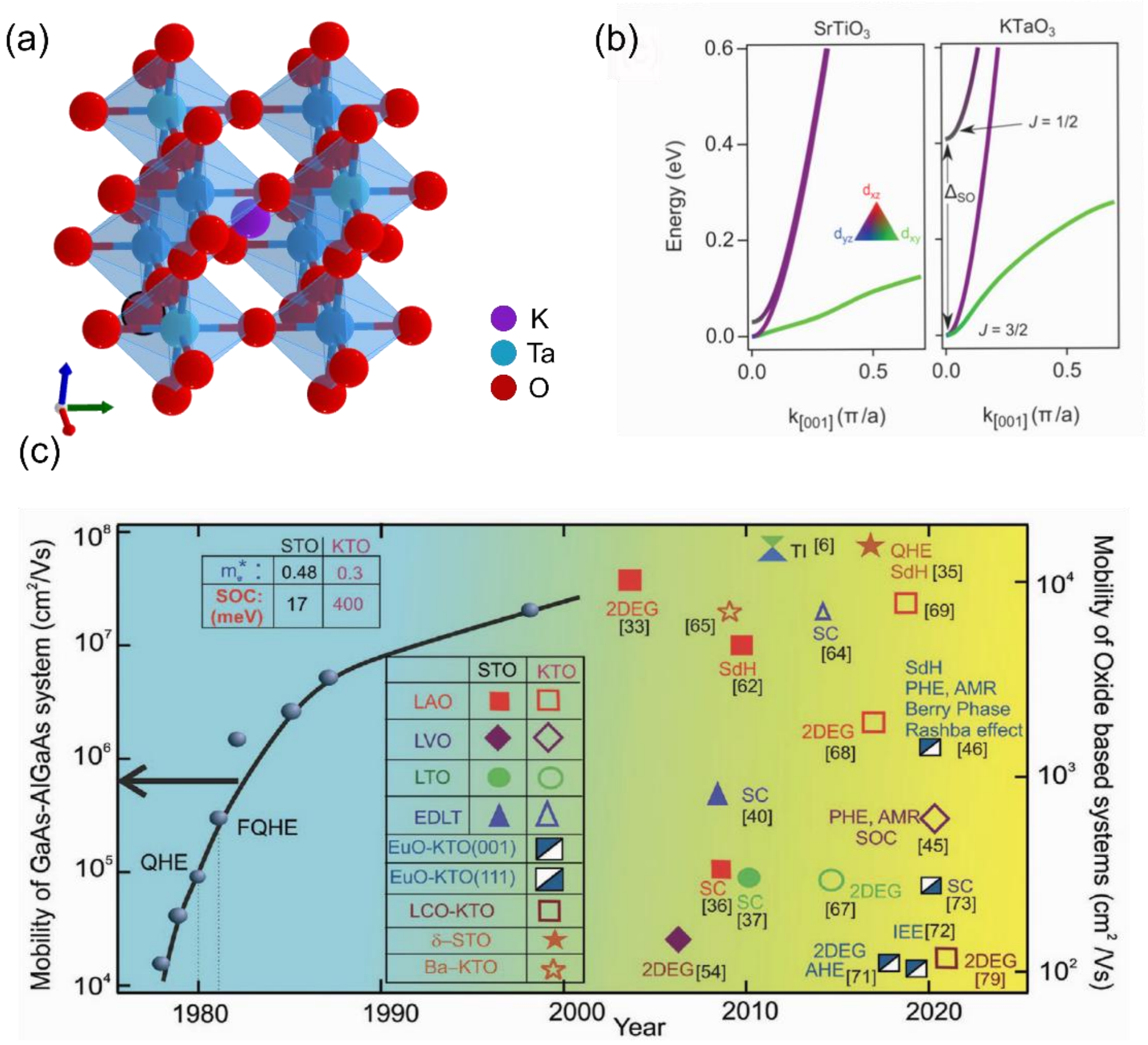

**Figure 2.1.** (a) Sketch of KTO crystal structure, (b) The bulk band structure of STO and KTO (adapted from Ref. [3]) (c) Evolution of charge carriers mobility in GAs-AlGaAs based semiconductor systems (left axis) and oxide-based systems (right axis). The oxide systems include STO and KTO, and their composites. The observed properties are abbreviated as: QHE = Quantum Hall Effect, FQHE=Fractional QHE, PHE=Planar Hall Effect, AHE=Anomalous Hall Effect, 2DEG=2D Electron Gas, SdH=Shubnikov–de Haas effect. Panel (c) is reproduced from [2], John Wiley & Sons. [Copyright © 2022 WILEY-VCH Verlag GmbH & Co. KGaA, Weinheim].

### 2.2 Current and Future Challenges

***Understanding superconductivity in KTO 2DEGs.*** A superconducting critical temperature as high as 2K has been measured for (111) EuO/KTO interfaces [7]. This value is one of order of magnitude larger than that of STO based 2DEGs, making fundamental studies and future devices application of interfacial oxide 2DEGs more easily attainable. Additionally, the values of the critical temperature and magnetic field are different when current is applied along the two orthogonal in-plane directions. This anisotropy has been linked to the emergence of 1D structures due to strong (Ta)5d–(Eu)4f coupling in low-carrier-density samples, implying the emergence of superconducting stripes from a band-filling-dependent ferromagnetic proximity [7]. Further studies of this heterostructure will allow a deeper understanding of the exotic superconducting phases arising from the interplay with magnetism.

***Exploiting SOC in KTO 2DEGs.*** A second property with great potential for future applications is linked to spin-orbit coupling. Spin–charge interconversion efficiency in KTO 2DEGs was evaluated using direct and inverse Edelstein effect and thermal spin injection. For the 2DEG realized depositing a thin Al film on (001)-oriented KTO substrate, a spin–charge conversion efficiency, measured through inverse Edelstein effect, of -3.5nm was revealed [8]. This value is among the highest reported in the literature, and an order of magnitude larger than those observed with transition metals such as Pt.

***Distinguishing the surface states from the bulk in Weyl semimetals.*** Weyl semimetals are gapless systems, and in most of the materials where they have been realized, numerous bulk states exist at the Fermi level. Given the potential use of surface states in future devices, detecting these states through transport measurements is crucial. Consequently, distinguishing between surface and bulk states presents a significant challenge. One possible approach leverages the long coherence length of the surface states in proximity effect—due to their topological properties, specifically spin-momentum locking and parity protection. By varying the length of the junctions, it becomes possible to tune the ratio between the bulk and surface channels, which have different superconducting coherence lengths.

***Understanding unconventional superconductivity in Weyl semimetals.*** The presence of superconducting Fermi arcs implies the unconventional superconductivity, such as p-wave pairing, which is considered as one of the basics for non-Abelian anyons. So far, it has not been demonstrated experimentally. Therefore, clarifying the pairing nature is strongly desired in the field.

**2.3 Advances in Science and Technology to Meet Challenges**

***Engineering oxide interfaces***

• Oxide interfaces exhibiting large SOC could be used in a novel class of low power consumption devices, such as mgnetoelectric spin–orbit logic devices for information read-out. For this application, improving the quality and reproducibility of the heterostructures is of primary importance. Further efforts are required to improve the fabrication techniques needed to realize oxide heterostructures nanodevices.

• Another major challenge is the development of manufacturing solutions making these materials compatible with silicon semiconductor technologies. A promising route to obtain this goal is to realize freestanding oxide heterostructures [9]. Recent works have exploited a water-soluble sacrificial layer or strain engineering. In the latter case, a strained LAO film is deposited under specific conditions, on a STO substrate, in order to subsequently induce a strain relaxation process that fragments the surface in regularly shaped LAO/STO flakes up to several microns in size. This promising approach could be applied also to other oxide heterointerfaces.

***Engineering Weyl semimetals***

• Interface engineering: For the proximity effect induced superconductivity, improving the quality of interfaces between Weyl semimetals and superconductors is key to inducing stable superconductivity in surface states. Advances in thin-film growth techniques, such as molecular beam epitaxy (MBE), could enable higher-quality heterostructures that facilitate the desired proximity effect. Controlled doping and strain engineering might also help optimize the conditions for superconducting Fermi arcs. For the intrinsic surface superconductivity, the identification of the order parameter phase symmetry is one of the key steps towards the p-wave superconductivity detection. In order to realize such kind of devices [10], a high-quality tunnel barrier interface is crucial.

• Material engineering: In order to reduce the bulk contribution, for the type I Weyl semimetal, it is possible to tune the Fermi level by substituting the Pt atoms.

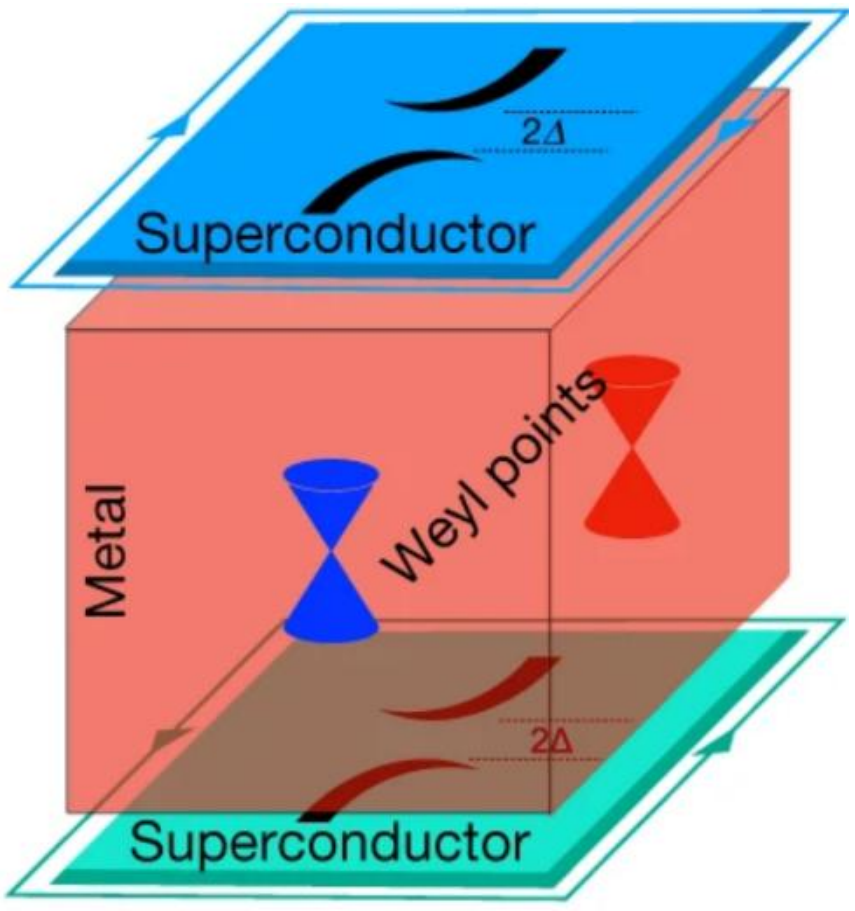

**Figure 2.2.** Schematic drawing of the Weyl semimetal and the superconducting Fermi arcs. Adapted from [6]. CC BY 4.0.

### 2.4 Concluding Remarks

The continuous reduction in size of electronic systems required in the recent years the scaling of electronic components, which are now approaching atomic-scale dimensions, where properties are dominated by quantum physics. The connection between materials quantum properties and quantum devices is especially apparent in the field of 2D materials. These systems offer a broad range of properties and the ability to form artificial quantum states of the matter. Furthermore, they enable the fabrication of novel heterostructures with functionalities unavailable in bulk materials. In this contribution we outlined some of the recent discoveries on two classes of 2D materials of great interest for future quantum electronics applications, namely KTO based 2DEGs and surface states in Weyl semimetals. The rich physic of these 2D systems likely hides many intriguing phenomena still to be uncovered.

### Acknowledgements

D.S. acknowledges support from the European Union's Horizon Europe research and innovation programme under grant agreement n. 101115190 (IQARO) and from the Italian Ministry of University and Research (project PRIN 2022 STIMO, n. 2022TWZ9NR). C.L acknowledges Dutch Research Council (NWO) for the financial support of the project SuperHOTS with file number VI.Vidi.203.047.

## 3. Iron-based superconductors

*Anna E. Böhmer*[1], *Lan Maria Tran*[2], *Andrzej J. Zaleski*[2]

[1]Ruhr University Bochum, Germany

[2]Polish Academy of Sciences, Poland

### 3.1 Status

Magnetic ions, such as Fe, have been known to suppress superconductivity since many decades. The discovery of widespread superconductivity in the iron-pnictide and iron-chalcogenide materials in 2008 came therefore as big surprise, and quickly triggered an intense worldwide research effort. The Fe-based superconductors (FeSCs) rapidly turned out to be a rich testbed for unconventional superconductivity [1]. As the second family of high-temperature superconductors, following the cuprates, they reach critical temperatures up to 50 K and critical fields of more than 50 T. Unlike the cuprates, many FeSCs are ductile intermetallics and exhibit relatively low anisotropy, making them promising candidates for applications in superconducting wires.

FeSCs indeed display a remarkable materials variety, with parent compounds of different stoichiometries and a multitude of dopants/chemical substitutions that can induce superconductivity. Such chemical tuning, as well as tuning by physical pressure, leads to intricate phase diagrams (see Fig. 3.1). Itinerant antiferromagnetism is prominent in the FeSCs, typically in the form of stripe-type antiferromagnetic order. A rotational-symmetry breaking lattice distortion, driven by an electronic "nematic" order parameter is intimately coupled to the magnetism. Most notably, superconductivity tends to arise when these orders are suppressed to lower temperatures, but superconductivity may also coexist and interact strongly with magnetic order, showing that they are not mutually exclusive. The superconducting order parameter has been found to be mainly of the sign-changing $s\pm$ type, with exceptions in several materials. The multi-orbital nature of the FeSCs, where all 5 Fe-3*d* orbitals are relevant for their properties, is certainly a key factor contributing to this rich phenomenology.

The investigation of FeSCs has driven diverse methodological and intellectual advances. This includes the use of strain as a tuning parameter, and the study of the strain derivatives of many experimental quantities (from electrical resistance to ARPES spectra), which are now applied for a much broader range of materials and phenomena. The identification of unconventional superconducting pairing mechanisms and gap symmetries, the orbital differentiation, and the understanding of electronic correlations driven by Hund's interaction are some of the highlights in theoretical physics.

Overall, the FeSCs represent an enormous material basis harbouring intriguing opportunity that range from studying exotic gap symmetries to the possible applications as high-performance superconducting wires. Many of the themes that were known from the study of heavy-fermion or cuprate superconductors (see Sections 5 and 7) are also found in the FeSCs, where they may be better investigated because of more accessible temperature ranges, a wide availability of single-crystalline samples, and the many possibilities to modify their properties via chemical substitution (see Fig. 3.1). Delineating the bridges between these material classes may lead to a broader understanding of correlated superconductors and superconducting quantum materials in general.

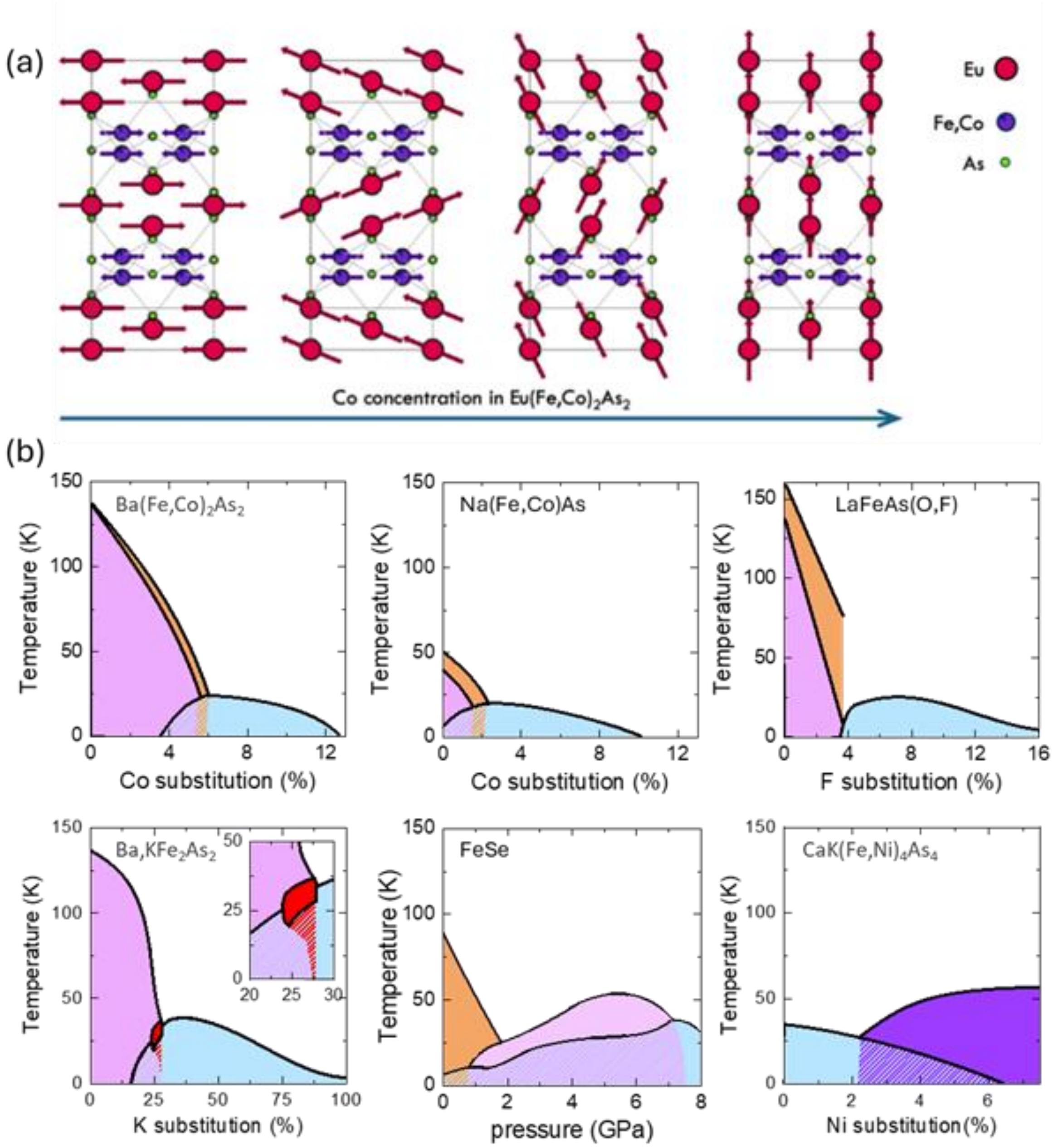

**Figure 3.1**. (a) Crystal structure of $Eu(Fe,Co)_2As_2$, with Co substitution ranging from 0 to 18% [2]. The arrows represent the magnetic moments on Eu (red) and Fe (violet), exemplifying the richness of magnetic orderings in the FeSCs. (b) Examples of phase diagrams of FeSCs. A blue shade indicates the superconducting phase, the pink shade is stripe-type Fe antiferromagnetism (light pink for FeSe indicates a degree of uncertainty), the orange shade indicates the nematic phase. Red and purple shades indicate different types of magnetic order: the spin-charge-density wave and the spin-vortex crystal, respectively. Coexistence of phases is also indicated.

### 3.2 Current and Future Challenges

The arrival of superconducting wire technology based on FeSCs has been much slower than initially hoped for. With the high critical temperatures and critical fields, superconducting wires based on FeSCs are, however, much desired. With critical temperatures above 20 K, they are ideal for a future energy infrastructure based on liquid hydrogen. Notably, the next generation of particle colliders will require high magnetic fields that exceed the capabilities of current magnet technology. For example, the current technical design report of the Circular Electron Positron Collider, hosted in China, calls for wires based on iron-based superconductors to create the high magnetic fields required to guide protons around the ring – projected to be 20 T. Here, $Sr_{1-x}K_xFe_2As_2$ is a promising material, of which superconducting tapes have already been produced in feasibility studies [3]. Moderate anisotropy and relatively high ductility are key advantages of the FeSCs in this respect. Nevertheless, there are still many technical and engineering challenges to be overcome before FeSCs can be used widely in superconducting wires. For example, even if the superconducting core would have perfect properties, there remains the challenge of finding the right barrier material, and a suitable way to incorporate the core in the barrier.

Moreover, FeSC are unique among magnetic superconductors with some intriguing phenomena. Some FeSCs show signs of exotic superconducting order parameters. This includes a suggested *s*+i*d* superconducting order parameter in $Ba_{1-x}K_xFe_2As_2$ [4], an "ultranodal" state in Fe(Se,S) [5] and a smectic pair density wave in $EuRbFe_4As_4$ [6]. Such exotic superconducting order parameters might be useful for spintronic applications, an avenue that has not been explored extensively to date. Other iron-based materials, such as Fe(Se,Te) have topological properties and are suggested to host non-Abelian Majorana quasiparticles. These might provide a platform for high-temperature topological quantum information processing [7].

Overall, the enormous research effort of the past 17 years has led to a vast body of experimental and theoretical results. Following this "gold-rush" era, the field is now transitioning to a new stage of maturity. One of the main challenges to date appears to be finding universal trends and underlying themes within a large treasure trove of data. With over 10 000 papers published in the field, this may be a worthy challenge for big data and machine learning methods.

### 3.3 Advances in Science and Technology to Meet Challenges

Tackling the challenges of design of development of viable superconducting wires for high-field applications, the challenges of identification and understanding of exotic superconducting states, and the challenges related to finding essential themes within a large amount of experimental and theoretical results, will necessitate advances on multiple fronts. With progress from materials engineering to data science, many of the current challenges appear surmountable, however.

The 122-type iron-pnictides, of which $SrFe_2As_2$ is an exemplary parent compound, have received significant attention due to their high critical current density $J_c$, and upper critical field $H_{c2}$, making them promising for high-field magnet applications. Thin films of these materials have achieved $J_c$ values exceeding 1 MA/cm², and recent advancements in 122 wires and tapes have improved $J_c$ to 105 A/cm² at 4.2 K and 10 T. However, these high $J_c$ values have typically been limited to short samples (2-3 cm), with performance deteriorating in longer wires. Despite this, the $J_c$ of long pnictide wires remains acceptable for practical engineering uses (see also Section 4). Lately, there was a study which reports the fabrication and superconducting properties of 100-meter-long $Sr_{0.6}K_{0.4}Fe_2As_2$ tapes and the successful creation and testing of superconducting coils [3]. The results highlight the correlation between tape and coil performance, with an observed anisotropy of less than 2, consistent with single crystal behaviour. Significant progress has been made in developing high-performance iron-pnictide conductors for large-scale applications.

Specialized and very sensitive experimental and theoretical tools have been developed to study the unique features of FeSCs. These include the development of strain-based probes, such as elastoresistance [8], or a formalism to identify sign-changing superconducting order parameters from quasiparticle interference data obtained in a scanning tunnelling microscope [9]. One of the most sensitive and most recent techniques is a high-resolution measurement of the elastocaloric effect (the entropy change induced by an infinitesimal strain) which can be obtained at various large offset strains [10], see Fig. 3.2. Many of these techniques are now applied in different contexts, where they have enlarged the toolbox of solid-state physics. Pushing these new techniques to their boundaries – be it in resolution, temperature or magnetic field range, or simply in their accessibility to many researchers, will, among other things, help to identify exotic superconducting states and their mechanisms.

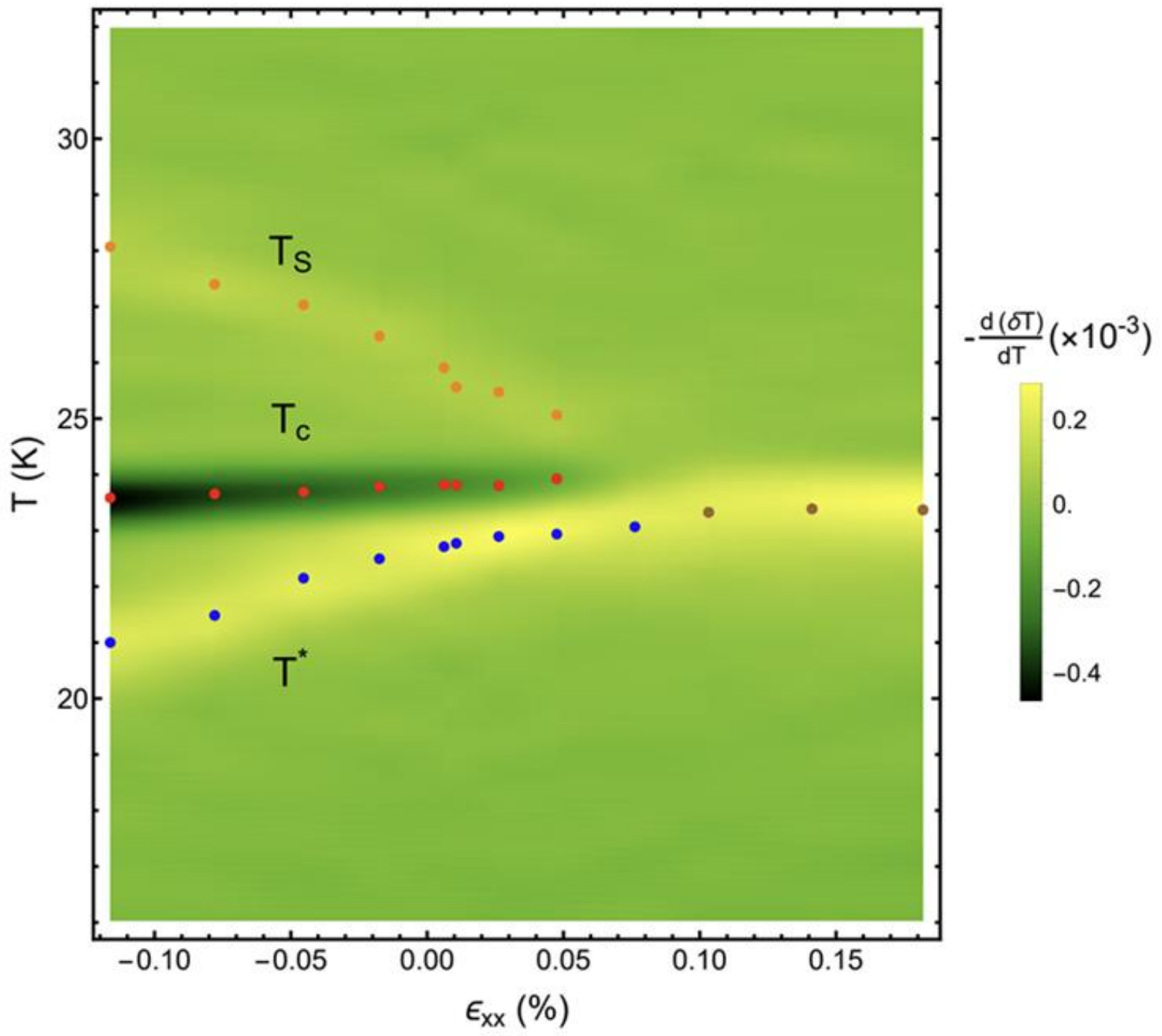

**Figure 3.2.** The temperature derivative of the elastocaloric effect of the FeSC $Ba(Fe_{0.938}Co_{0.062})_2As_2$ as a function of uniaxial strain $\epsilon_{xx}$, revealing hitherto unimagined detail. Uniaxial strain is used as a tuning parameter creating a phase diagram with the superconducting transition $T_c$ and the nematic transition at $T_s$. $T^*$ is a new characteristic temperature within the superconducting phase, suggested to be a transition between different superconducting states. Reproduced from with permission [10].

The FeSCs realize intriguing superconducting phenomena, and there are certainly some that have not yet been discovered. With the vast amount of data on the FeSCs, the overall progress in data science and deep learning for materials physics and materials science will certainly have an impact on the field, even though that may still be some years ahead.

### 3.4 Concluding Remarks

A universal understanding of unconventional superconductivity remains elusive, and the prediction of new unconventional superconductors is still impossible to date. Notably, there are many common themes among different classes of unconventional superconductors, such as heavy-Fermion systems (Section 7), cuprates (Section 5) and FeSCs. These common themes include complex phase diagrams with multiple interacting phases, moderate to strong electronic correlations, signatures of quantum criticality, and sign-changing superconducting order parameters. Making progress within the field of FeSCs and finding the bridges to other material classes is therefore an important goal. This is naturally a challenge based on a large amount of data. Significant progress on the front of materials data initiatives may lead to a new era in the research of FeSCs, and unconventional superconductors in general. New materials discoveries, better understanding of physical phenomena and further progress with technical challenges will certainly reveal new exotic superconducting phenomena and expand the use of (iron-based) superconductivity for diverse applications. Superconducting wire technology based on FeSCs appears within reach, promising decisive advantages.

**Acknowledgements**
Work in Bochum is partially supported by the ERC grant Distort-to-Grasp (No. 101040811), and by the Deutsche Forschungsgemeinschaft (DFG) under CRC/TRR 288 (Project A02). We have benefitted from collaborations through the EU program COST CA21144 (superqumap.eu).

## 4. Critical Current in Superconductors

*Adrian Crisan[1], Massimiliano Polichetti[2], Armando Galluzzi[2], Ali Gencer[3]*

[1]National Institute of Materials Physics, Bucharest, Romania

[2]University of Salerno, Salerno, Italy

[3]Ankara University, Ankara, Türkiye

### 4.1 Status

Advancements in the research of superconducting materials are commonly linked with their progress in the critical current density of the superconductor, $J_c$, and its critical temperature, $T_c$, both demarking the transition between nearly zero electrical resistivity seen in superconductors and their dissipative behaviour. Regarding the critical temperature, today's cryogenic industry is sufficiently mature to support superconductivity applications at temperatures as low as the liquefaction temperature of nitrogen (<77.35 K), hydrogen (<20.15 K), or helium (<4.15 K), with compact designs that are efficient and cost-effective. Furthermore, the push towards the energy transition, with its component hydrogen economy, will mean that liquid hydrogen could become a cheap and abundant product, so many of the superconducting materials having critical temperatures higher than 20 K could be of practical interest. The advancement of real-world applications and market penetration is likely to be influenced by the advancement of a single key parameter, i.e., its $\boldsymbol{J}_c$, because this crucial parameter delineates the threshold for a superconductor's current-carrying capacity and, consequently, its potential to carry high currents and to generate or sustain powerful magnetic fields. The concept of critical current density is present in various research areas, from fundamental microscopic condensed matter physics to applied electrical engineering, with material science and chemistry in between. In fact, for studying the physical mechanisms leading to acceptable engineering parameters, one necessarily ends up with the mesoscopic vortex physics (see also Sections 5, 9, 14) and, eventually, with the microscopic superconducting coupling mechanisms.

At the *macroscopic* level, for applied electrical engineering, electromagnetic fields are evaluated in terms of the Maxwell equations and material laws that are well-defined as averages in macroscopic volume elements. Meaningful material laws and their connection to specific superconducting samples can be understood only by considering the physics at the *nanometer* scale because this is the range that dominates the *interactions of vortices*, which are the basic elements in practical superconductors at high magnetic fields. In "classical" (low temperature) superconductors, where the thermal energy can usually be neglected, critical currents were comprehensively treated by Campbell and Evetts [1], while in the case of high temperature superconductors where the influence of thermal energy on vortex dynamics and pinning is important one can consult the seminal theoretical work of Blatter et al [2]. More recently, vortex dynamics and pinning in high-performance high-temperature superconductors and the impact on critical current, with emphasis on coated conductors, including very suggestive numerical simulations, were treated by Kwok et al. [3].

### 4.2 Current and Future Challenges

With the continuous discovery of new superconducting materials, the field is still very important from both theoretical and applications points of view. Returning to the critical current density, in most superconducting materials $J_c$ is decreasing monotonically with both the increasing temperature and increasing field, the surface $J_c(T,H)$ in a 3-D plot having a bell shape. This was the common feature in classical superconductors. With the discovery of cuprate high-temperature superconductors and, later on, of iron-based superconductors, $J_c\ (T,H)$ can display very unusual features, due to much different vortex matter, dynamics and pinning. Two such examples are shown in Fig. 4.1, for $CaKFe_4As_4$ [4] (left-hand side), and, respectively, for $BaFe_2\ (As_{0.68}P_{0.32})_2$ [5] (right-hand side) superconducting single crystals.

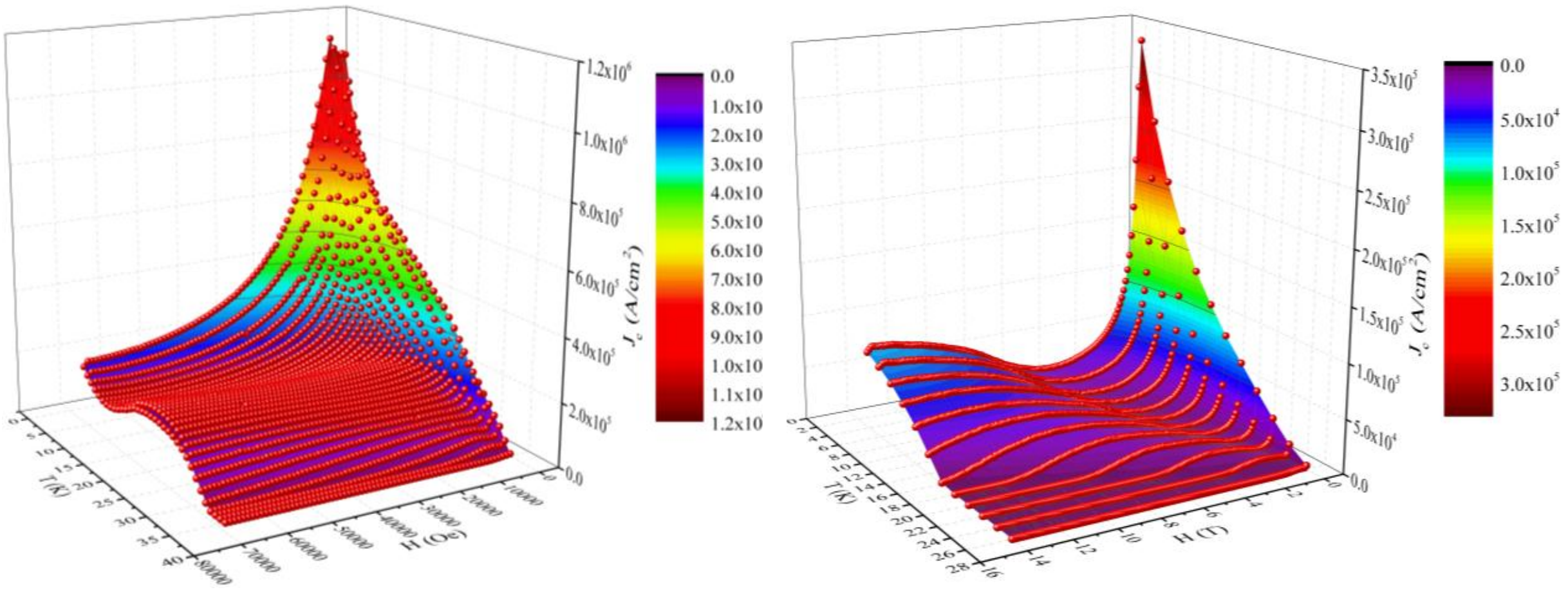

**Figure 4.1.** Temperature and DC field dependence of the critical current density in $CaKFe_4As_4$ (left-hand side), and, respectively, $BaFe_2\ (As_{0.68}P_{0.32})_2$ (right-hand side). Reproduced through authors' copyright from [4] and [5].

The feature on the left-hand side of Fig. 4.1, namely a second maximum in the $J_c(H)$ for various $T$ is due to the so-called Second Magnetization Peak (SMP), which is a very well-known phenomenon discovered initially in high-temperature superconductors. Polichetti et al. [6] showed that, regardless of many specific proposed scenarious, the SMP phenomenon can be framed in a general scenario of a vortex lattice crossover from a configuration with less effective pinning to another one with higher pinning efficiency. Although there are many examples of more spectacular SMP, we have chosen to present $CaKFe_4As_4$ because of the feature at around 20 K, a temperature relevant to the future hydrogen economy, with a quite high $J_c$ for a large scale of fields. $J_c(T,H)$ of $BaFe_2\ (As_{0.68}P_{0.32})_2$, shown in the right-hand side of Fig. 1, shows not only the usual SMP, but also an unusual anomalous peak in $J_c(T)$ at several constant DC fields, which was shown to be related to a rhombic-to-square transition (RST) of the Bragg vortex glass [7]. The RST was shown to be also the cause of a magnetic memory effect in the temperature dependence to AC susceptibility [5]. Due to the anisotropy of most of the superconducting materials, in most cases is important also the dependence of the critical current density on the angle of the sample with the magnetic field, $J_c(T,H,\vartheta)$.

For most applications of superconducting materials in the energy transition employing high magnetic fields, the undesired dissipation due to the Lorentz force and/or thermal energy can be limited by the introduction of artificial pinning centres (APCs). The study of the influence of APCs on $J_c(T,H,\vartheta)$, although being more than two decades old, is still a field of very important interest.

**4.3 Advances in Science and Technology to Meet Challenges**

Any advancement in the discovery of novel superconducting materials with potential applications in either larger-scale power/energy applications and/or superconducting and hybrid cryo-electronics should be accompanied by comprehensive investigations of vortex matter, dynamics and pinning, in a large domain of time-scales (DC and AC fields, microwaves, etc.) since new phenomena are likely to be discovered. Regarding the improvement of $J_c(T,H,\vartheta)$ in REBCO films ("building block" for 2-nd generation coated conductors), through various APCs, of different materials, architectures, etc., there are hundreds of published results, a large amount being reviewed in [8], while more recent advances in improving $J_c(T,H,\vartheta)$ are presented in the work by MacManus-Driscoll and Wimbush [9].

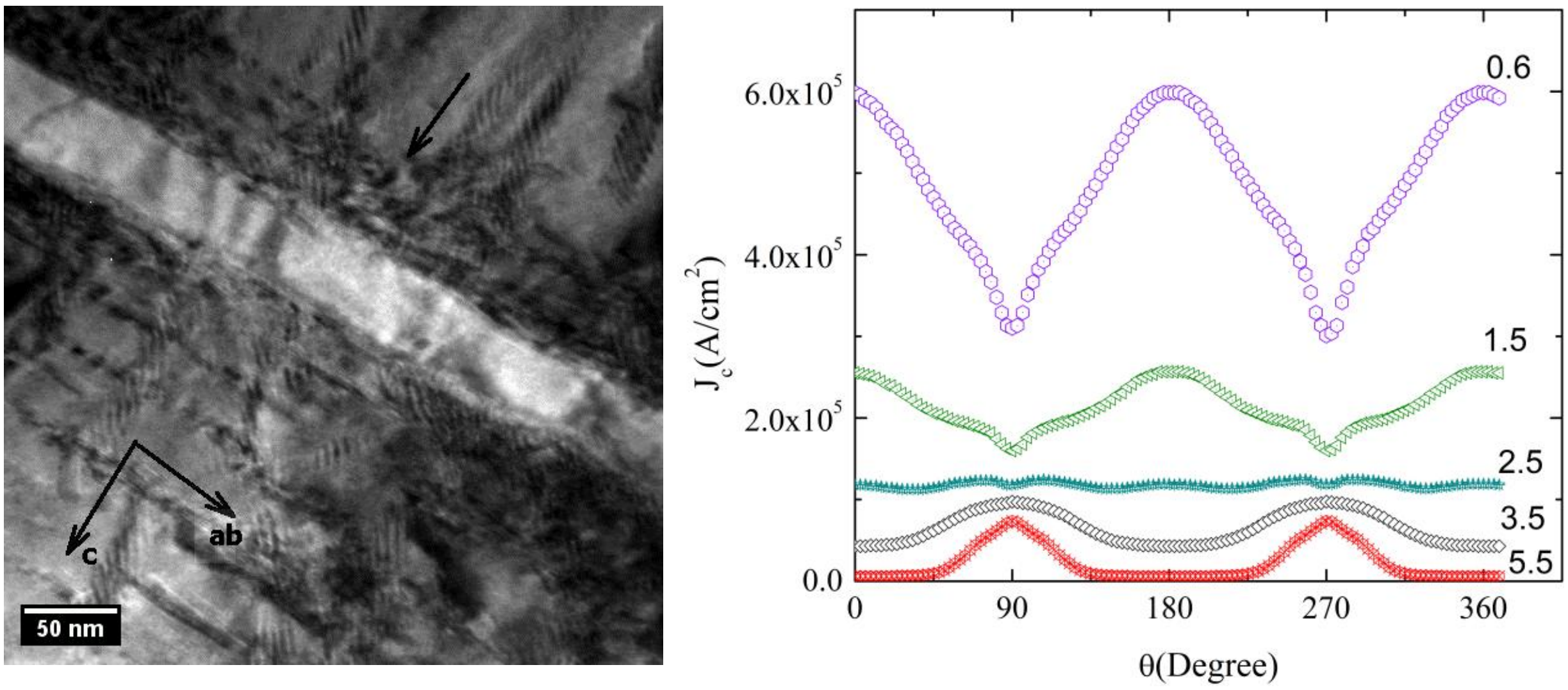

**Figure 4.2.** Transmission electron microscopy image of the interfaces between two $BaZrO_3$-doped YBCO layers grown on substrates decorated with Ag nano-islands and a 30 nm thick $SrTiO_3$ nanolayer (left-hand side); and dependence of the critical current density, at 77 K, on the angle between the applied magnetic fields (in T as indicated for each curve) and the c-axis direction (right-hand side). In the left-hand panel, arrows indicate the direction of the ab-planes (parallel to the substrate), and, respectively, of the c-axis (perpendicular to the substrate). The arrow in the upper part of the figure indicates a $BaZrO_3$ nanoparticle.

It became evident that the best results are obtained by synergetic pinning centres in which several techniques are employed (substrate decoration, secondary phase(s) in the target, quasi-superlattice approach). Figure 4.2 (left) shows a TEM image of the interface between $BaZrO_3$ (BZO)-doped YBCO layers grown on substrates decorated with Ag nano-islands and a 30 nm thick SrTiO 3 nanolayer [10]. It can be seen that, apart from the c-axis oriented $BaZrO_3$ nanorods (and some isotropic nanoparticles), a high density of APCs oriented in the a-b plane are formed due to the interfacial stress, leading to very interesting $J_c(T,H,\vartheta)$.

From the right-hand side of Fig. 4.2 it can be seen that for fields smaller than 2.5 T, the maximum in $J_c(T,H,\vartheta)$ occurs for field parallel to the c-axis due to the c-axis correlated BZO nanorods, while for higher fields the dependence is the "normal" one, i.e., critical current density larger for fields oriented parallel to the a-b planes. A very important feature is at 2.5 T, with a $J_c(T,H,\vartheta)$ almost constant, which is a very desirable property of coated conductors for coils and solenoids applications, in which parts of the superconducting tape/wire is subjected to various orientation of the magnetic field. Fine-tuning of the synergetic pinning centres approach will lead to a "pinning by design" scenario, in which, for certain applications in certain fields, a "receipt" for optimum pinning architecture could be deviced.
APCs can also be produced by irradiation with various particles. In this case, pinning by design can be achieved more easily since one can choose different particles with different energy while fluence (dosage) and angle can be varied at will.

### 4.4 Concluding Remarks

For the future advancement of the field, understanding $J_c(T,H,\vartheta)$ for both old and newly-discovered superconducting materials and bridging the gaps between the advances in materials science, theoretical developments, and engineering applications is essential. This has to be done in correlation with comprehensive studies of vortex matter, dynamics and pinning since new phenomena are to be expected in the case of new families of superconducting materials. For power applications, nanoengineering of APCs will ensure a better market penetration since through various landscapes of synergetic pinning centres one can achieve the most desirable pinning by design, which means optimizing $J_c(T,H,\vartheta)$ for particular applications, in certain magnetic fields. Manipulating defects using irradiation techniques is very important for the energy sector, because, on the one hand, the pinning properties of advanced superconductors can often be improved by suitably choosing the type and energy of the striking particles while controlling the density of defects by tuning the irradiation fluence; and on the other hand, because the radiation hardness of Jc in HTS materials is critical for the successful development of the fusion energy sector.

**Acknowledgements**

We acknowledge the support from the Core Program of the National Institute of Materials Physics, granted by the Romanian Ministry of Research, Innovation and Digitalization under the Project PC2-PN23080202, the PRIN 2022 PNRR Project QUESTIONs Grant No. P2022KWFBH (Italy) and COST Action 21144 SUPERQUMAP (EU).

## 5. Superconductor cuprates

*Bernd Aichner*[1], *Neven Barišić*[2,3], *Wolfgang Lang*[1]

[1] University of Vienna, Austria

[2] Technische Universität Wien, Austria

[3] University of Zagreb, Croatia

**5.1 Status**

Copper-oxide superconductors, or cuprates, have captivated condensed matter physics since the 1986 discovery of high-temperature superconductivity by Bednorz and Müller. Despite decades of research, the mechanism behind their superconductivity remains debated. Understanding these materials is a fundamental challenge with the potential to revolutionize technology by enabling superconducting devices at accessible temperatures.

All lines of inquiry to understand cuprates fall broadly into two limits. The first involves quasi-localized or heavily renormalized objects and models, such as polarons, Zhang-Rice singlets, the resonating valence bond theory, and the Hubbard model. The second explores the unusual evolution of scattering rates, discussed in terms such as the Ioffe-Regel limit, bad metals, strange metals, quantum critical behaviour, anisotropic scattering rates, and the Planckian limit. Both limits challenge the validity of the quasiparticle (Fermi liquid) concept [1].

Both limits above reflect, to some extent, the extremes of the cuprate phase diagram depicted in Fig. 5.1: the parent compounds, which are antiferromagnetic insulators at half-filling with exactly one localized hole per planar $CuO_2$ unit, and the overdoped side, which exhibits Fermi-liquid behaviour with $1 + p$ charges, where $p$ corresponds to doping.

Despite the complexity, a series of surprisingly simple and universal behaviours have been identified [2]. Based on these, it was shown that the phenomenology of cuprates across the phase diagram is fully captured by the simple charge conservation relation $1+p = n_{loc}+n_{eff}$. Here, $n_{eff}$ is the carrier density and $n_{loc}$ is the density of localized charge within a $CuO_2$ unit. The corresponding superfluid density is related to both components $\rho_S = n_{eff} \cdot (O_S n_{loc})$, where all terms can be experimentally determined directly. The charge $n_{loc}$ is responsible for all the strangeness of these compounds, which includes the pseudogap phenomenon and the superconducting glue. The compound-dependent constant, $O_S$, is fine-tuned by the local crystal structure. It arises from the $p$–$d$–$p$ fluctuation by the Cu-localized holes visiting the neighbouring planar–oxygen atoms and can be determined from NMR spectrometry.

More recently, focusing exclusively on the overdoped regime, two electronic subsystems have also been identified [3]. However, the $n_{loc}$ has been interpreted as incoherent ($n_{\mathrm{incoh}}$) and is exclusively associated with superconductivity. Such an approach involving anisotropic scattering rates entirely neglects the presence of itinerant charges ($n_{eff}$) in the superconducting phenomenon. Nevertheless, in both cases, the simplicity of the observed phenomenology, after four decades of extensive research, finally seems to unveil the secrets of high-$T_c$ superconductivity.

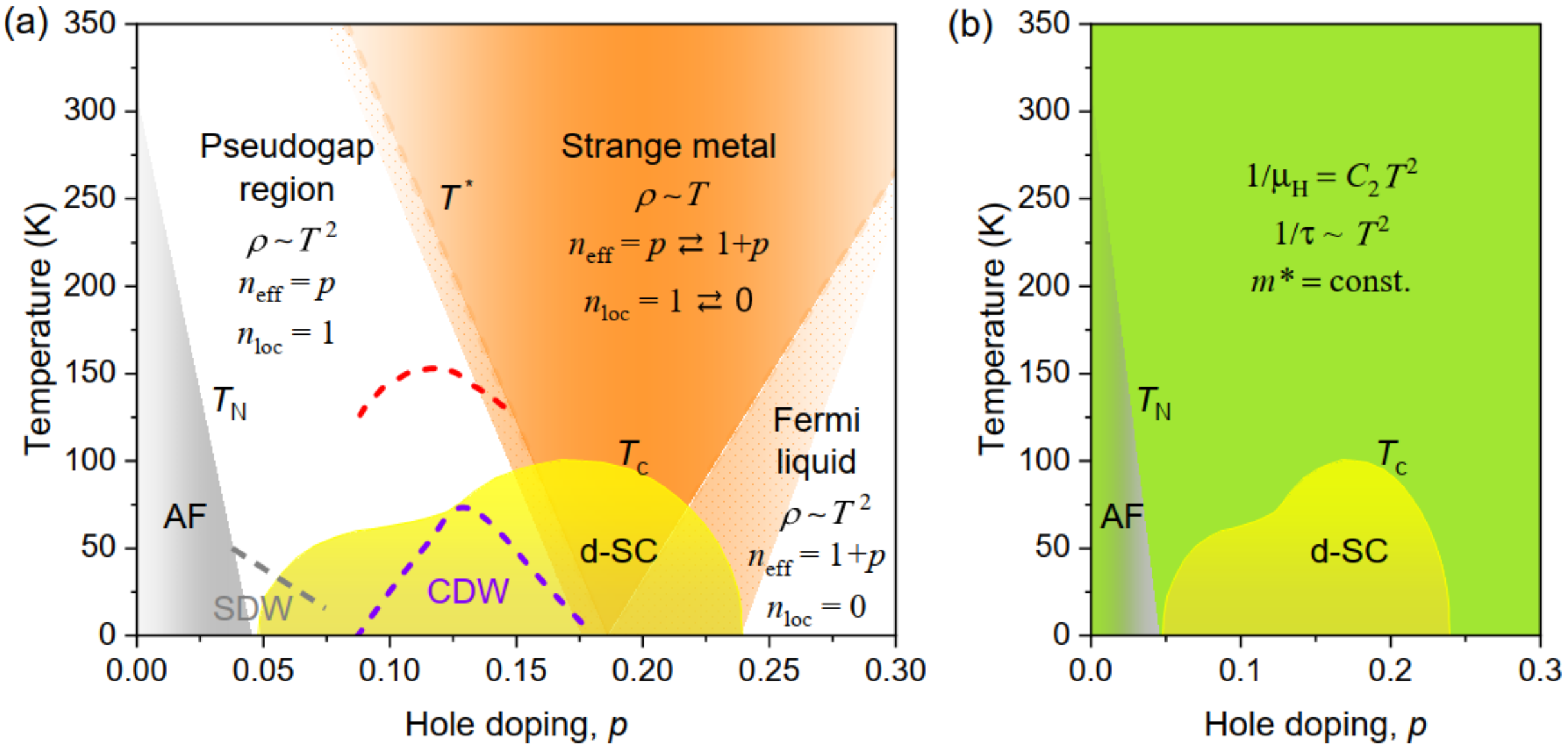

**Figure 5.1.** (a) The schematic temperature versus hole-doping ($p$) phase diagram of cuprates, adapted from Keimer et al. [1], highlights the complexity of these compounds, showcasing the emergence of selected phases and ordering tendencies. These include antiferromagnetic order (AF), *d*-wave superconductivity (d-SC), the pseudogap region, the strange metal regime, and the overdoped Fermi liquid region with carrier density of $n_{eff} = 1 + p$. Characteristic temperatures such as the Néel temperature ($T_N$), the superconducting critical temperature ($T_c$), and the pseudogap temperature ($T^*$), are also marked. The onset of spin-density waves is represented by a dashed grey line, onset of the Nernst effect and/or fluctuating short-range charge-density-wave order by a dashed red line, while a charge-density wave order stabilized by high-magnetic fields is shown by a dashed lilac line. In the pseudogap regime, the Fermi-liquid nature of itinerant charges is evidenced by the Fermi-liquid scaling of the optical scattering rate and adherence to the Kohler rule. Correspondingly, resistivity [$\rho = m^*/(n_{eff}e^2\tau)$] is quadratic in temperature, with $n_{eff} = p$. This implies that, in the strange metal regime, carrier density must evolve from $p$ to $1 + p$. In this regime, resistivity exhibits linear-like temperature dependence. (b) The inverse Hall mobility [$1/\mu_H = m^*/(e\tau)$] in cuprates exhibits essentially a universal quadratic dependence on temperature, characterized by a universal prefactor $C_2$. Consequently, the phase diagram presented is experimentally justified and the simplest possible. It implies the universality of the effective mass and the Fermi-liquid scattering rate, which is quadratic in temperature. Such a straightforward and surprising relationship between these observables allows for carrier density ($n_{eff}$) determination from resistivity and the Hall coefficient [$R_H = 1/(n_{eff}e)$], offering detailed insights into the evolution of carrier density and localized charge ($n_{loc} = 1 + p - n_{eff}$) [2].

### 5.2 Current and Future Challenges

Besides understanding the mechanisms of superconductivity in cuprates, translating their unique properties into practical quantum devices presents numerous challenges. Their anisotropic electrical properties complicate integration into devices requiring low-resistive electrical contacts or interfacing with silicon-based technology. Moreover, grain boundaries and other intrinsic defects significantly limit the maximum current a device can carry. They also reduce the reproducibility in the fabrication of devices, which is particularly problematic for circuits based on many Josephson junctions requiring identical properties.

Most superconducting quantum devices rely on the growth of thin films on suitable substrates. Besides chemical methods, such as the sol-gel route and metal-organic chemical vapor deposition, high-quality films are typically fabricated using pulsed-laser deposition (PLD), which requires precise control of temperature and atmospheric conditions. However, scaling up PLD to deposit homogeneous thin films over areas larger than a few square centimetres is challenging.

A major bottleneck is patterning cuprate thin films to the often required sub-micrometer scale. Standard lithography and etching processes may damage cuprate materials due to their chemical sensitivity and are practically incapable of producing structures smaller than about 100 nanometers. While gallium focused-ion beam (Ga-FIB) techniques offer higher resolution, contamination of the cuprates by Ga ions poses a severe problem. Moreover, creating open side faces by removing cuprate material leads to oxygen out-diffusion, which reduces the critical temperature $T_c$ and deteriorates other essential superconducting properties.

Cuprate superconductors often exhibit shorter coherence times than conventional superconductors, limiting their usefulness in quantum computing applications. Also, the *d*-wave symmetry of the superconducting order parameter in cuprates introduces additional noise channels, complicating application in quantum technologies.

### 5.3 Advances in Science and Technology to Meet Challenges

To achieve advances in science and technology, the quest for a better understanding of the origin of superconductivity in cuprates will be supported by progress in using neural computing networks to approximate many-body wavefunctions and reduce computational demands. In addition, quantum computers could, in principle, simulate quantum many-body systems more efficiently than classical computers. However, further developments in quantum hardware are demanded to handle the complexity of cuprate superconductors. Utilizing cloud-connected community computing resources might provide flexible performance scaling options.

Due to their significantly higher $T_c$ and upper critical field $B_{c2}$, cuprate superconductors offer a much broader operational range than their metallic counterparts. To address the remaining challenges, several strategies have been developed. The first involves creating high-performance cuprate conductors by fabricating biaxially textured films on metallic substrates (coated conductors), which circumvent the detrimental effects of grain boundaries. The second strategy enhances performance in strong magnetic fields by introducing artificial pinning centers (APCs), such as non-superconducting nanocomposites that self-assemble into nanorods, serving as vortex pinning defects. Combining various defect architectures can further improve the volume pinning force. Recently, enhancing the charge carrier concentration into the overdoped region has been proposed to improve the intrinsic parameters governing the self-field critical current and to create synergistic effects with the APCs [4].
A novel method for engineering well-defined patterns in cuprate superconductors is emerging, using a focused helium-ion beam (He-FIB) [5]. This technique primarily displaces oxygen atoms in $YBa_2Cu_3O_{7-\delta}$ (YBCO) thin films, creating point defects while maintaining the crystallographic framework and preserving the film's surface integrity. These point defects break the superconducting pairs due to the *d*-wave nature of the superconducting gap, thereby suppressing $T_c$.

By employing a helium-ion microscope [6], schematically shown in Fig. 5.2(a), 30 keV $He^+$ ions create profiles of point defects that locally suppress superconductivity in YBCO. This method has enabled the realization of Josephson junctions [7] (see also Section 19), vortex pinning lattices with matching fields up to 6 T [8] as shown in Fig. 5.2(b), and complex structures like quasi-kagomé lattices [9]. The technique can be further advanced to fabricate Josephson junction arrays (see also Section 9), superconducting quantum interferometers (see also Section 22), single-flux-quantum flip-flop devices, and superconducting diodes, potentially leading to breakthroughs in disordered neural networks for neuromorphic computing.

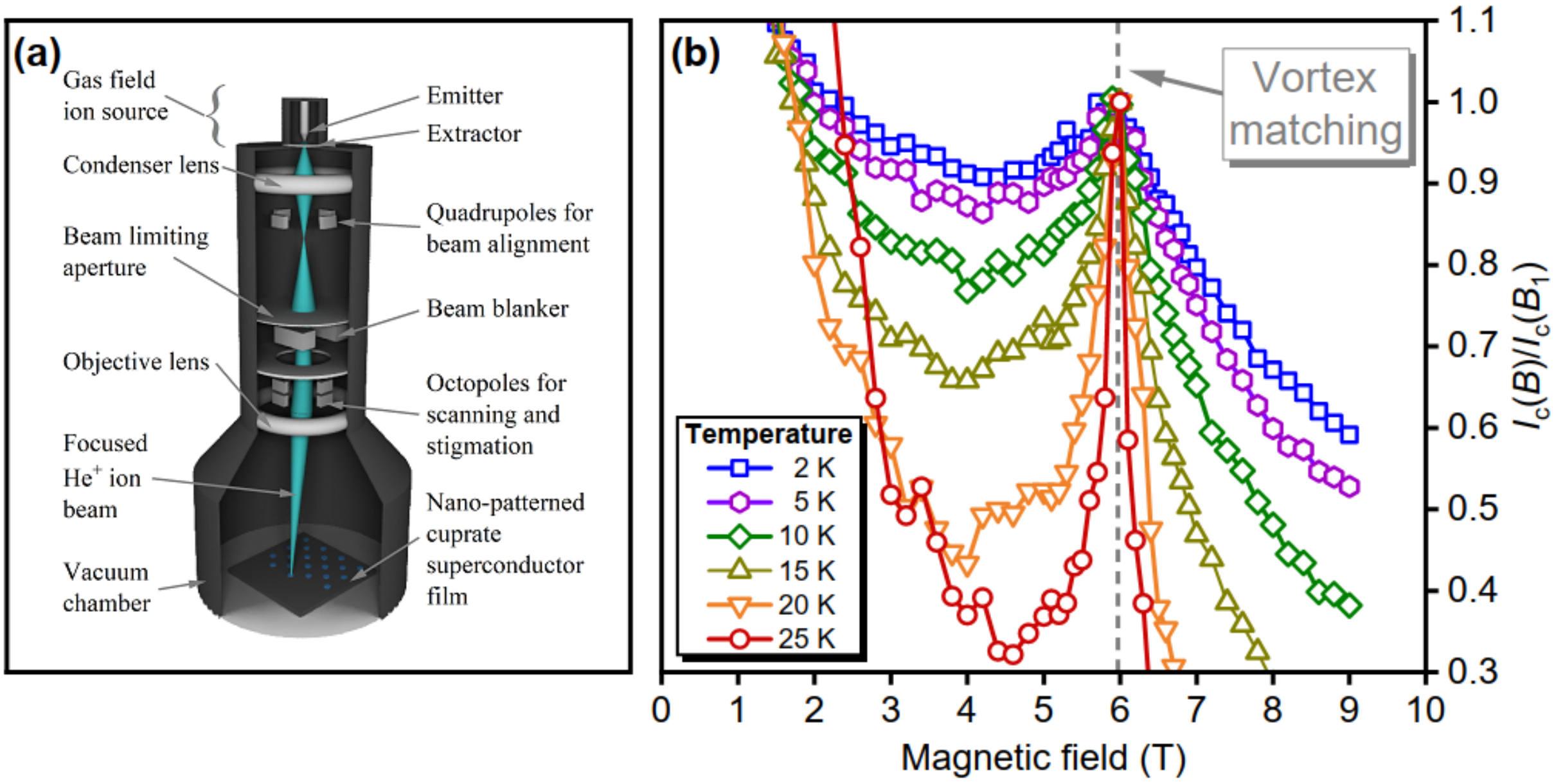

**Figure 5.2.** (a) Schematic illustration of a helium-ion microscope used to create columns of point defects via focused $He^+$ beam irradiation, suppressing superconductivity in targeted regions. (b) Critical current as a function of applied magnetic field at various temperatures, normalized to the value at the matching field $B_1$, corresponding to one vortex per defect column. The sharp peaks indicate enhanced vortex pinning due to the commensurate arrangement of vortices with the artificial defect lattice. Data adapted from [8].

While commercially available superconducting nanowire single-photon detectors (SNSPDs) are predominantly based on metallic superconductors like NbN, recent research has explored implementing them with cuprate superconductors [10]. The higher $T_c$ values of cuprate superconductors offer the potential to extend the operational temperature range of SNSPDs above liquid helium temperatures. This advancement could simplify cooling requirements and broaden the practical applications of SNSPDs (see also Section 18).

### 5.4 Concluding Remarks

The study of copper-oxide superconductors is a rich and challenging field at the intersection of material science, condensed matter physics, and quantum mechanics. The complexity of these materials, from their unconventional superconductivity and unusual normal state properties, which includes pseudogap formation and interplays of competing orders, presents significant hurdles. Overcoming these challenges requires collaborative efforts combining advanced experimental techniques, novel theoretical approaches, and innovative material synthesis methods.

Progress in understanding cuprates not only promises to solve one of the most significant puzzles in physics but also paves the way toward developing superconductors that function at higher temperatures, potentially transforming energy transmission, superconducting magnet technologies, and electronic quantum devices.

### Acknowledgements

This research was funded by the Austrian Science Fund (FWF) grant number I4865-N and is based upon work from COST Actions SuperQuMap CA21144 and Polytopo CA23134 (European Cooperation in Science and Technology). The work at TU Wien was supported by FWF Project P 35945-N, while the work at the University of Zagreb was supported by Croatian Science Foundation under Project No. IP-2022-10-3382.

# 6. Misfit layer superconductors

*T Samuely[1], M. Gmitra[2], T Cren[3], M Calandra[3], P Samuely[2]*

[1]P. J. Šafárik University, Slovakia
[2]Slovak Academy of Sciences, Slovakia
[3]INSP, Sorbonne University, CNRS, Paris, France

### 6.1 Status

Atomically thin layered materials are systems with zero limit bulk-to-surface ratio. Their physical properties are determined by two-dimensionality (2D) and strongly affected by interfacing with other systems. Therefore, they represent an accessible platform for the abundance of quantum effects that can be engineered by combining them into vertical stacks. Two types of layered systems are considered here – artificially prepared (exfoliated) van der Waals nanostructures [1], and naturally layered systems showing quasi 2D behaviour already in a bulk form. A special class of naturally layered materials is misfit structures combining atomic layers of hexagonal (H) transition metal dichalcogenides (TMD) and slabs of tetragonal (T) ionic rare-earth monochalcogenides in the same superlattice [2]. Both types of layered systems feature a new state of quantum matter, the Ising superconductivity extremely resilient to external magnetic field. A giant electron doping, natural to the misfit structures, can lead to topological superconductivity. Both systems can also be assembled into heterostructures combining different constituents. Layered 2D heterostructures have a large number of implications for many potential applications in solid-state devices, solar cells, photodetectors, semiconductor lasers, light-emitting diodes, and biosensors [3]. One of the most challenging applications of topological superconductors is quantum computation with putative Majorana zero-energy modes which are quasiparticles with the properties of non-Abelian anyons. Braiding of such quasiparticles constitutes the basis of topologically protected qubits, a robust solution to the problem of decoherence and unitary errors of the state-of-the-art quantum computers. In spite of extensive theoretical elaboration, the physical realisation of the topological quantum bits is in its infancy, and fundamental research in the field of topological superconductivity is needed. This includes also understanding of physical properties of the materials used as building blocks of artificial heterostructures.

### 6.2 Current and Future Challenges

One way to topological superconductivity is Ising superconductivity (IS). IS has been discovered in atomically thin TMD films, namely in ion-gated $MoS_2$ and in monolayer $NbSe_2$ [4] showing extremely large in-plane upper critical magnetic fields much above the theoretical Pauli limit, where the Zeeman energy equals to condensation energy. Within a single TMD layer, finite in-plane electric fields are allowed by broken inversion symmetry leading through strong spin-orbit interaction to effective *k*-dependent magnetic fields polarized perpendicularly to the layer locking the orientation of the electron spin to the out-of-plane direction (Ising) and making the action of experimentally achievable in-plane external magnetic field highly irrelevant. A careful transport study of a sequence of a monolayer, double layer, etc., of $NbSe_2$ shows a suppression of the upper critical field with increasing number of layers so the effect is restricted to 2 dimensions.

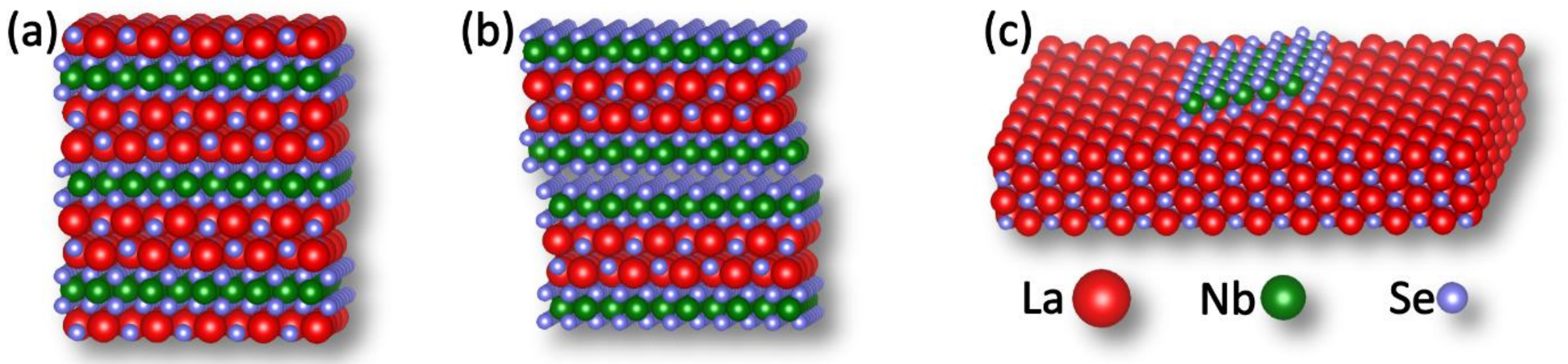

**Figure 6.1.** Crystal structure of misfit layer superconductors. (a) 1T1H or $(LaSe)_{1.14}(NbSe_2)$ ionocovalent crystal with $T_C$ = 1.3 K and in-plane upper critical field of 20 T. (b) 1T2H or $(LaSe)_{1.14}(NbSe_2)_2$ where HTH trilayers are van der Waals bound, with $T_C$ = 5.7 K and in-plane upper critical field of 50 T. (c) Artificial heterostructure where a monolayer of hexagonal $NbSe_2$ is placed on top of tetragonal LaSe crystal.

Our team studied the family of bulk misfit layer compounds where mutually misfitting hexagonal $NbSe_2$ and tetragonal LaSe layers are alternating which exhibit large superconducting anisotropy and quasi 2D character. The most surprisingly even more extreme in-plane upper critical magnetic fields $B_{c2//ab}$ strongly violating the Pauli paramagnetic limit have been observed compared to monolayers. Namely, the misfit layer $(LaSe)_{1.14}(NbSe_2)$/1T1H single crystal (Fig. 6.1(a)) with $T_C$ = 1.23 K shows $B_{c2//ab}$ = 20 T and the crystal $(LaSe)_{1.14}(NbSe_2)_2$/1T2H (Fig. 6.1(b)) with $T_C$ = 5.7 K exhibits $B_{c2//ab}$ = 50 T, which is 10 and 5 times more than the respective Pauli limit [5]. Complementary experimental methods of transport measurements, ARPES (Section 13), STM (Section 16) and QPI in combination with first-principles calculations bring surprising evidence that it is due to IS surviving in bulk $NbSe_2$-based materials despite the fact that the in-plane inversion symmetry should be preserved [6]. It was shown that the band structure of 1T2H and 1T1H crystal is very similar to that of the monolayer $NbSe_2$ with a rigid band shift (Fig. 6.2). Namely, in 1T2H the LaSe layer donates about 0.6 electron per $NbSe_2$ chemical unit corresponding to the Fermi level shift of 0.3 eV in comparison with monolayer $NbSe_2$. 1T1H is even more heavily doped and the Fermi level is shifted by 0.5 eV still not completely filling the hole Nb band. Moreover, ARPES proved that the misfits are quasi-2D systems with a very little dispersion of the electronic band along the *z*-direction. Being the electronic equivalent with $NbSe_2$ monolayer with spin-split bands around the K(K') points in the Brillouin zone makes the misfits Ising superconductors. In [7] a theoretical model shows that in 1T1H the coupling between the superconducting $NbSe_2$ layers through the insulating LaSe is so small that 1T1H creates an infinite stack of monolayers, each without inversion symmetry. In 1T2H, the structure can be viewed as a stack of seemingly centrosymmetric 2H-$NbSe_2$ bilayers separated by LaSe misfit layers. But LaSe layers below and above the $NbSe_2$ bilayer are oriented differently with respect to $NbSe_2$, thus breaking the total inversion symmetry in the misfit. Hence, the hierarchy of energy scales of interlayer coupling, namely, breaking of inversion symmetry by the LaSe layer, spin-orbit coupling and the magnitude of the superconducting gap in the 1T1H and 1T2H compounds, provides a consistent picture for the strong Ising protection observed in these bulk compounds. This is different from the IS in the monolayer where only the last two energy scales are present.

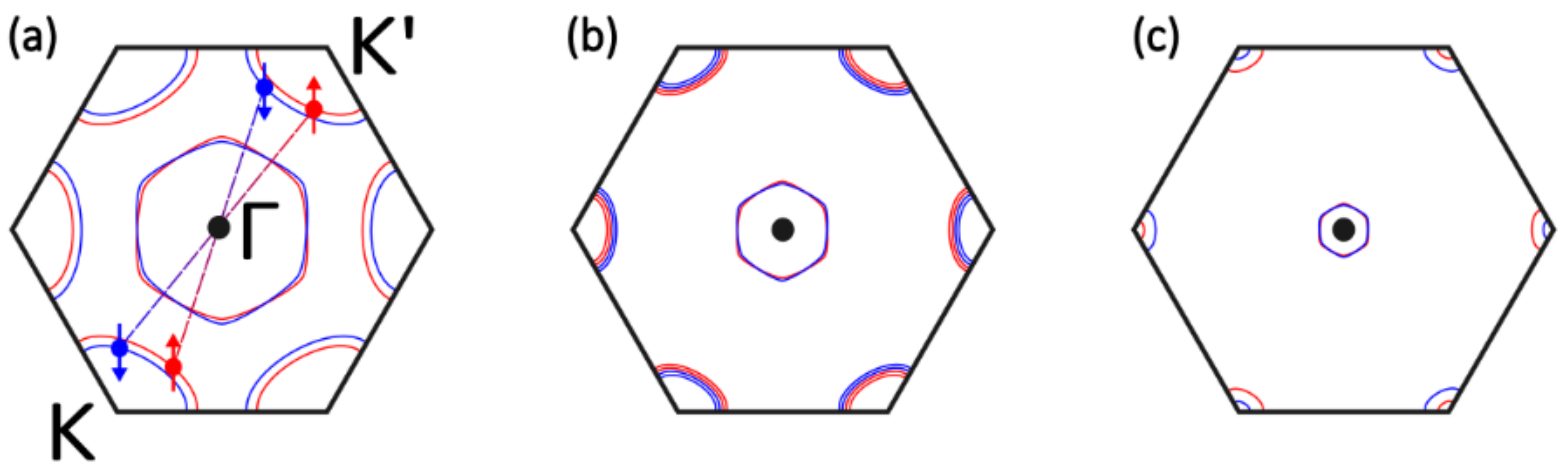

**Figure 6.2.** Brillouin zone and Fermi surfaces of (a) monolayer $NbSe_2$, (b) 1T2H or $(LaSe)_{1.14}(NbSe_2)_2$ and (c) 1T1H or $(LaSe)_{1.14}$(NbSe2) showing gradual shrinking of the hole Fermi surfaces due to doping of $NbSe_2$ layers from LaSe. Spin splitting around K and K' points is responsible for Ising superconductivity.

### 6.3 Advances in Science and Technology to Meet Challenges

Tunability of properties is of primary importance in quantum design. The success of 2D materials in different fields of science is in their better tunability in comparison with the bulk materials. One of the knobs in doping. By a double-layer FET geometry $10^{14}$ electrons per $cm^2$ can be introduced to the monolayer of $NbSe_2$. Somewhat larger doping can be achieved via deposition of alkali atoms (such as potassium). The large electron charge transfer in the misfits allows to obtain doping fractions that largely encompass both those obtained in the case of ionic-liquid-based FETs and K adatom deposition. This large transfer from LaSe to the TMD is due to the chemical properties of the highest occupied states of bulk LaSe that are formed only by atomic La states. As a consequence, LaSe essentially behaves as a donor, an effect not depending on the kind of TMD used to build the misfit structure. This suggests that similarly large doping can be achieved by sandwiching LaSe with other metallic TMDs. It is also possible to vary the rocksalt layer, opening a tremendous space of possibilities and a fertile ground for the discovery of highly innovative materials. For instance, assuming La3+, Pb2+, and Se2− in the compounds $(La_{1-y}Pb_ySe)_{1+x}(NbSe_2)_2$, a simple charge balance calculation shows that the charge transfer from the rock salt layer $(La_{1-y}Pb_ySe)$ to the $(NbSe_2)_2$ double layer could vary continuously from 0 to $1 + x$ by reducing the lead content. Therefore, the charge transfer could be tuned by appropriate substitution in misfit layer compounds, pretty much in the same way as doping can be varied in FET, but in a much broader range [6].

Tuneable charge doping in the misfits introduces the possibility of shifting the Fermi level to be in between the two spin-split bands, which is predicted to give rise to topological superconductivity [8]. It is plausible that these insights can be extended to other misfit compounds and bulk structures comprising TMD layers, where large in-plane critical magnetic fields that exceed the Pauli limit have been reported but not identified as evidence of Ising superconductivity. This includes works on $(SnSe)_{1.16}(NbSe_2)_2$, $Ba_6Nb_{11}S_{28}$, cation-intercalated $NbSe_2$, and organometallic intercalated compounds of 2H-TaS2 (references in [7]). Future experiments and theory work to determine the electronic and structural properties for this broad range of potential bulk Ising superconductors will be of great interest.

Few layer transition-metal dichalcogenides and their misfit structures can be easily incorporated in 2D heterostructures (Fig. 6.1(c)). Proximity-induced spin-orbit coupling and exchange interactions in heterostructures made of 2D materials are sources of novel quantum effects engineered through elaborate stacking. Using this approach, there have been created artificial systems where proximity effects are combined to yield entirely new physical qualities (*p*-wave superconductivity in ferromagnetic/*s*-wave superconductor systems, Yu-Shiba-Rusinov states, Majorana bound states, etc.) [9].

### 6.4 Concluding Remarks

Abundance of highly innovative materials of untapped properties can be created in the form of misfit layer compounds. Their unit cells are naturally growing heterostructures of different stacking from tetragonal rocksalt layers and hexagonal TMD with various electronic and magnetic properties. Unprecedent tuning of their properties is controlled by the charge transfer/doping from T layers to H layers. Concerted effect of charge-transfer, defects, reduction of interlayer hopping, and stacking enables Ising superconductivity. It provides a possible pathway to design of bulk superconductors that are resilient to magnetic fields. Ising spin-orbit coupling can also be tuned to topological superconductivity (see also Section 1). Like traditional layered materials, the misfits are often exfoliatable and incorporatable as units of artificially stacked heterostructures.

**Acknowledgements**

This work was supported by Projects COST Action No. CA21144 (SUPERQUMAP), APVV-23-0624, APVV SK-FR-22-0006, VEGA 1/0472/25, VEGA 2/0073/24, Slovak Academy of Sciences IMPULZ IM-2021–42.

# 7. Strongly correlated materials: Heavy Fermions

*Jeroen Custers*

Charles University, Prague, Czech Republic

### 7.1 Status

The defining characteristic of strongly correlated materials is that their electrons cannot be treated as non-interacting particles. Due to the strong coupling among electrons, analyzing the behavior of individual particles offers limited insight into the macroscopic properties of these materials. Strongly correlated materials often emerge when electrons face two competing tendencies: kinetic energy from hopping between atomic orbitals, which favors band behavior, and Coulomb energy from electron-electron repulsion, which facilitates atomic behavior. When these two energy scales are comparable, neither the itinerant nor atomic perspective alone can adequately describe the system's physics. The canonical pathway to strong correlations is through materials with highly localized orbitals such as heavy fermion (HF) f-electron systems. Typically, these compounds constitute of a periodic lattice of localized spins, i.e., f-electron elements like Ce, Yb, U, in a metallic environment and are characterized by a competition between a local Kondo singlet formation and the development of a cooperative magnetic state via the Ruderman–Kittel–Kasuya–Yosida (RKKY) interaction [1]. The first report of a heavy fermion state was on $CeAl_3$ in 1975. However, the intense interest in heavy fermion systems started with the discovery of superconductivity in $CeCu_2Si_2$ in 1979 by Steglich et al. [2]. It was revealed that the Cooper-pairs are composed from heavy fermions, quasiparticles with up to 1000 times the bare electron mass. Additionally, the hierarchy of energy scales significantly deviates from the premises in BCS theory, indicating that the pairing cannot be phonon-mediated. Soon after, HF superconductivity was demonstrated in $UBe_{13}$, $UPt_3$ and $URu_2Si_2$ [3]. In these initial HF superconductors, the superconductivity develops from a non-magnetic heavy Fermi liquid state. While a mutual relationship between magnetism and HF superconductivity has been theorized the missing proof was provided by the discovery of a superconducting dome in pressurized $CePd_2Si_2$ and $CeIn_3$ that appeared in the phase diagram where the respective Néel temperatures vanished to zero [4]. Furthermore, in the vicinity of this magnetic to non-magnetic quantum phase transition strange metal, coined non-Fermi liquid behavior, was reported manifested by a linear in temperature dependence of the resistivity. It is widely believed that the magnetic quantum fluctuations associated with the quantum critical point both act as a pairing mechanism for Cooper-pair formation and are at the origin of the observed non-Fermi liquid behavior. This picture is believed to apply more broadly to any kind of strongly correlated materials; that is, unconventional superconductivity, nematic phases and any other exotic quantum phase of matter emerge near a quantum phase transition [5], see Fig. 7.1.

Despite extensive research over many years, the fundamental physics of many of these systems remains elusive, and we lack a comprehensive understanding of strong electron correlations. Additionally, our ability to predict behaviors in such systems is still limited. The modern interest in heavy fermion, or Kondo physics has two reasons, (1) certain heavy fermion compounds are believed to possess non-trivial topological characteristics. For instance, reports have emerged indicating the possibility of Weyl topological superconductivity (see also Section 1), a state characterized by the breaking of time-reversal symmetry within the superconducting phase of $UTe_2$ [6]. A fundamental question in this context is whether Majorana fermions manifest in these materials which would open a gateway to fault-tolerant quantum computing. (2) Heavy fermion physics is also observed in frustrated lattices [7], van der Waals (vdW) heterostructures, for example CeSiI [8], and Moiré superlattices [9]. The common denominator among these diverse material classes, including heavy fermion compounds, is the presence of flat bands. This naively suggests that heavy fermion physics may have a more universal application.

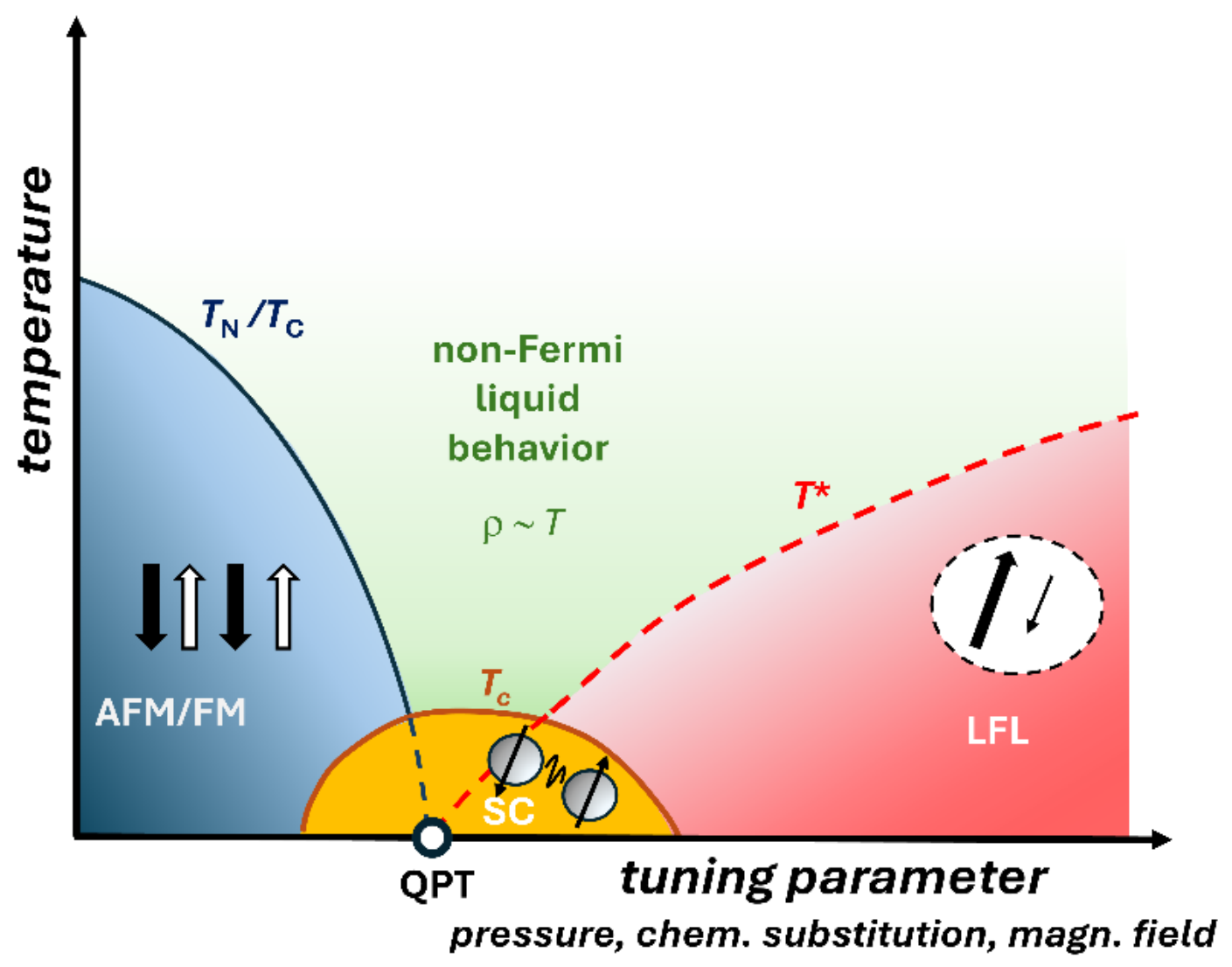

**Figure 7.1.** Typical phase diagram of heavy Fermion materials.

### 7.2 Current and Future Challenges

Unequivocally, the field of strongly correlated materials harbors a plethora of challenges. While many questions have been answered, there are numerous intriguing and unresolved issues to address. Some are as old as the topic itself, others surfaced in the wake of new materials. Outlined below are a few key challenges currently debated [10]:

- There is a long-standing question about universality of quantum phase transitions. This idea is rooted in the observation of similar physical properties across a variety of different material classes, for instance, the iron-based superconductors exhibit transport phase diagrams that strongly resemble the cuprate strange metal region and the heavy Fermion low temperature quantum critical state. What are the scaling relations and the values of the critical exponents and more intriguing, do topological quantum phase transitions fall into the same universality class?
- Developing a microscopic theory for unconventional superconductivity, akin to the BCS theory for conventional superconductors, remains an ambitious goal. This pursuit suggests a single ubiquitous Cooper pairing interaction for unconventional superconductors. It is often suggested that spin-fluctuation mediated pairing is the common thread linking. However, the presence of multiple phases in the proximity of the superconducting transition as seen in examples like Fe(Se,Te), exhibiting nematic and antiferromagnetic states, opens the discussions if not various interactions contribute to Cooper-pair formation so that the pairing mechanism cannot be ascribed to a single process.

In the last decade there has been a surge of work on interplay between strong correlations and topology. The identification of $SmB_6$ as a heavy fermion Kondo insulator provided a new interpretation of the persistent low-temperature residual conductivity through robust metallic surface states. Generally, examples of materials exhibiting both strong correlations and topological properties are rare. $Ce_3Bi_4Pd_3$ and $CeRu_4Sn_6$, classified as Weyl-Kondo semiconductors, and the Weyl-semimetal $(TaSe_4)_2I$ stand out as notable instances (see also Section 3). The quest for bulk topological superconductors remains a focal point. Yet, a conclusive example of a strongly correlated topological superconductor is absent. However, several unconventional superconductors, such as $UPt_3$ (if it is even parity), $URu_2Si_2$, $U_{1-x}Th_xBe_{13}$, and $PrOs_4Sb_{12}$, are projected to possess topological features giving rise to a Bogoliubov Fermi surface - a topologically protected region of zero-energy excitations. The physical properties of a BFS are hitherto elusive.

- The topic of dimensionality has been sparingly explored in the past. Reducing dimensionality enables access to 2D electronic states, where quantum confinement and anisotropic interactions enhance correlations, leading to new quantum phases. This includes high-temperature heavy Fermion states and superconductivity, as demonstrated in the vdW material CeSiI and the superlattice $NbS_2$ - $Ba_3NbS_5$, respectively. It is an interesting path to follow.

### 7.3 Advances in Science and Technology to Meet Challenges

From a materials perspective, one emphasis is on synthesizing ultra-clean crystals. The histories of $UTe_2$ and $SrRuO_4$ illustrate how disorder can obscure the superconducting state. Furthermore, well-known techniques such as flux growth, floating zone, Bridgman, molecular beam epitaxy, pulsed laser deposition, and chemical vapor deposition must be advanced by extending them to new frontiers, such as synthesizing under high pressure or higher temperatures, to create novel materials. The fabrication of materials using superlattices is particularly intriguing. By varying the stacking of $CeIn_3$/$LaIn_3$ layers, the impact of dimensionality has been demonstrated, but the technique also allows for the creation of novel materials. Freestanding films produced using wet-etch techniques, plasma growth, or exfoliation can act as innovative substrates, expanding the range of lattice parameters and crystal symmetries available for the (epitaxial) growth of strongly correlated electron materials.

Equally import is pushing the boundaries of existing experimental techniques. Measurements of electrical resistivity below 1 mK enabled the mapping of the superconducting phase in $YbRh_2Si_2$ and allowed the determination of crucial superconducting parameters. The observation of superconducting pocket in $UTe_2$ in magnetic fields as high as 65T emphasizes the importance of expanding the current limitations. To detect the de Haas van Alphen signal of the strongly renormalized (heavy) bands of the Fermi surface high fields in combination with low temperatures are required. Measuring the thermal Hall effect is crucial for probing topological surface states. However, this technique is often inaccessible for many materials due to poor signal-to-noise ratios related to the sample dimension and/or because the method hasn't been developed for the specific temperature (pressure) regions. Essentially, addressing the challenges necessitates lower temperatures, higher fields, and increased uniaxial/hydrostatic pressures, ideally with enhanced sensitivity of the experimental methods employed.

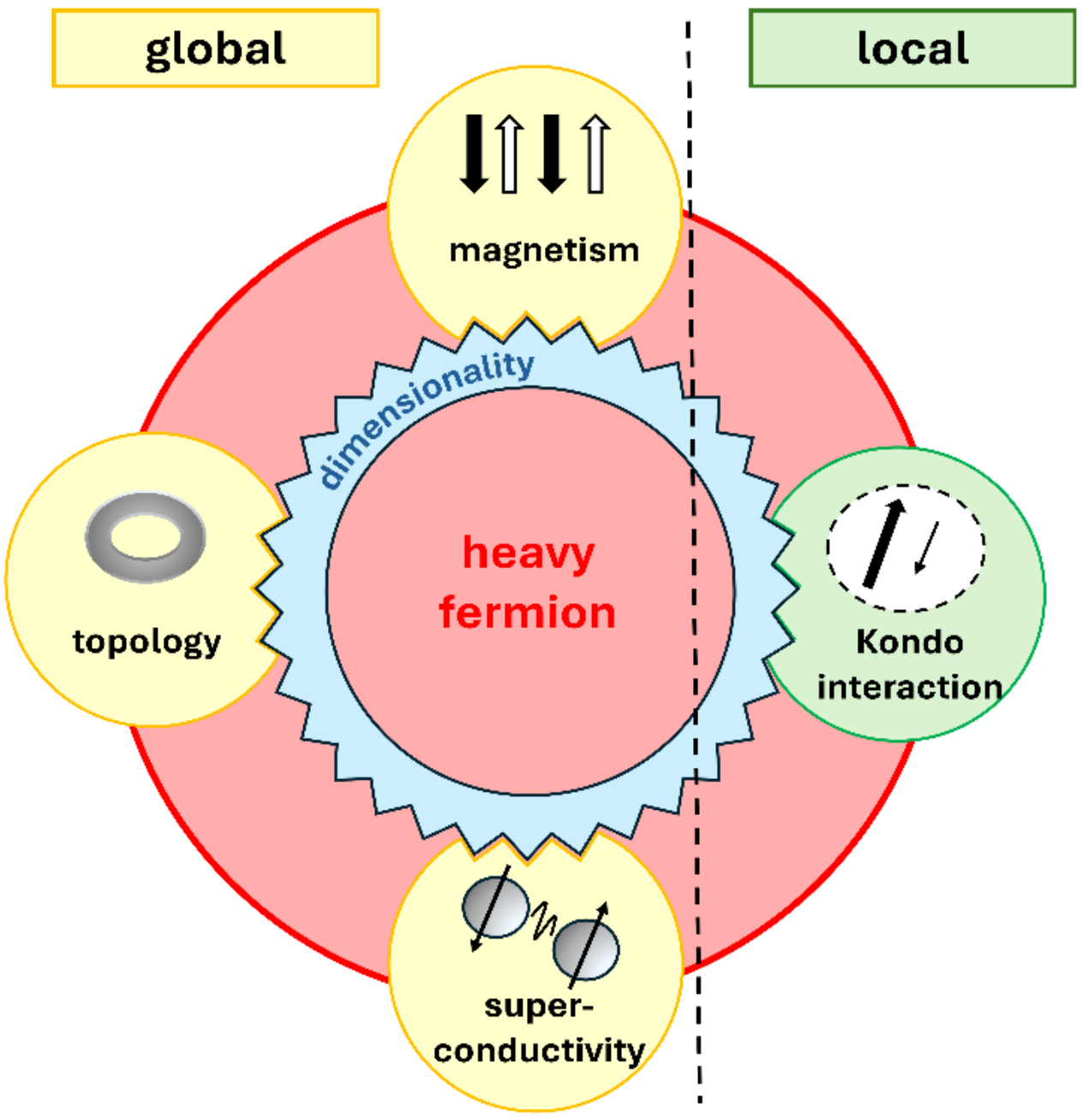

**Figure 7.2.** In heavy fermion materials, properties like the Kondo effect, magnetism, superconductivity and topology coexist/compete. Global (local) refers to whether the property is of long-range (atomic) type. Dimensionality can tune each of these properties.

### 7.4 Concluding Remarks

The field of strongly correlated materials is vast and recently rapidly advancing, with discoveries of new materials and phenomena announced almost daily, see Fig. 7.2. It is impossible to do justice to this field in a single section. The field of strongly correlated materials started some 70 years ago with the Kondo problem which is still actual in modern problems of topological matter. From a theoretical standpoint, the profound question lingers whether a general framework to understand strong electronic correlations is possible or if we have to settle with the undeniable reality that each strongly correlated is strongly correlated on its own. Experimentalists on the other hand should pioneer new frontiers in the experimental techniques. Pushing the boundaries of their setups and merging multiple extreme environments may unveil pivotal information of underlying mechanism at work. Synthesizing and exploring new materials are important to discover novel and exotic phenomena, but it is equally crucial to focus on growing clean crystals of known compounds, as disorder may have masked their true ground state properties. The fabrication of artificial superlattices deserves more attention. It offers an intriguing pathway to explore dimensionality effects and allows creating new compounds with tailored properties.

### Acknowledgements

J.C. acknowledges the support of the Czech Ministry for Education, Youth and Sports program INTER-COST (Grant No. LUC24139) as well as many valuable discussions with L. Havela and P. Brydon.

# II. Probing nanoscale superconductors

## 8. Superconductor 3D nanoarchitectures

*Oleksandr Dobrovolskiy*[1], *Rosa Córdoba*[2], *and Vladimir M. Fomin*[3,4]

[1]TU Braunschweig, Germany
[2]Institute of Molecular Science (ICMol), University of Valencia, Spain
[3]Leibniz IFW Dresden, Germany
[4]Moldova State University, Republic of Moldova

### 8.1 Status

The application of superconductors in magnetic and electromagnetic field sensing, quantum optics and quantum information technology requires their miniaturization to the nanoscale, operation in a broad (dc to THz, see also Section 21) frequency range, and the integrability with other technologies. In this regard, extension of planar superconductors into the third dimension (see also Section 9) allows for the device's footprint reduction, its thermal decoupling from the substrate, and the development of multi-terminal devices and circuits with complex interconnectivity [1].

The major effects originating from the extension of a planar superconductor into the third dimension are [2] (i) the inhomogeneity of the magnetic field induced by the 3D geometry, see the left panel in Fig. 8.1(a). This makes possible (ii) the coexistence of vortex-rich and vortex-free states in different parts of the 3D nanostructure and (iii) non-trivial topology (multiple connectedness) of the superconducting screening currents, giving rise to (iv) emerging phenomena in the dynamics of magnetic flux quanta and slips of the phase of the superconducting order parameter [3].

The major techniques for the fabrication of superconductor 3D nanoarchitectures are (i) *self-organization* driven by relaxation of mechanical strain, (ii) *3D nanoprinting* using focused beams of particles (electrons, ions, photons), and (iii) deposition of superconducting materials onto various *3D scaffolds*. The principles of the two former techniques are the following.

The strain-relaxation-driven roll-up technology [4] implies that a bilayer of two materials with different lattice constants is deposited onto an etchant-sensitive layer. After selective etching of this sacrificial layer, each material tends to acquire its inherent lattice constant. If the material of the bottom layer has a larger lattice constant than that of the material of the top layer, the bilayer bends upward, forming an open nanotube after one complete revolution, see the right panel in Fig. 8.1(a). An SEM snapshot of a 3D Nb nanohelix bolometer [4] fabricated by this technique is presented in Fig. 8.1(c).

In the process of 3D nanoprinting using focused ion/electron beams [5], precursor molecules are supplied via a gas injection system. Under the impact of the focused particle beam, the molecules are decomposed into volatile compounds and the non-volatiles, the latter forming a nanostructure, see the left panel in Fig. 8.1(b). SEM snapshots of a 3D WC nanobridge [5], a 3D WC nanohelix [6] and 3D WC hollow WC nanocylinder [7, 8] fabricated by this technique are presented in the left panel in Fig. 8.1(b), the right panel in Fig. 8.1(c) and in the left panel in Fig. 8.2(a), respectively.

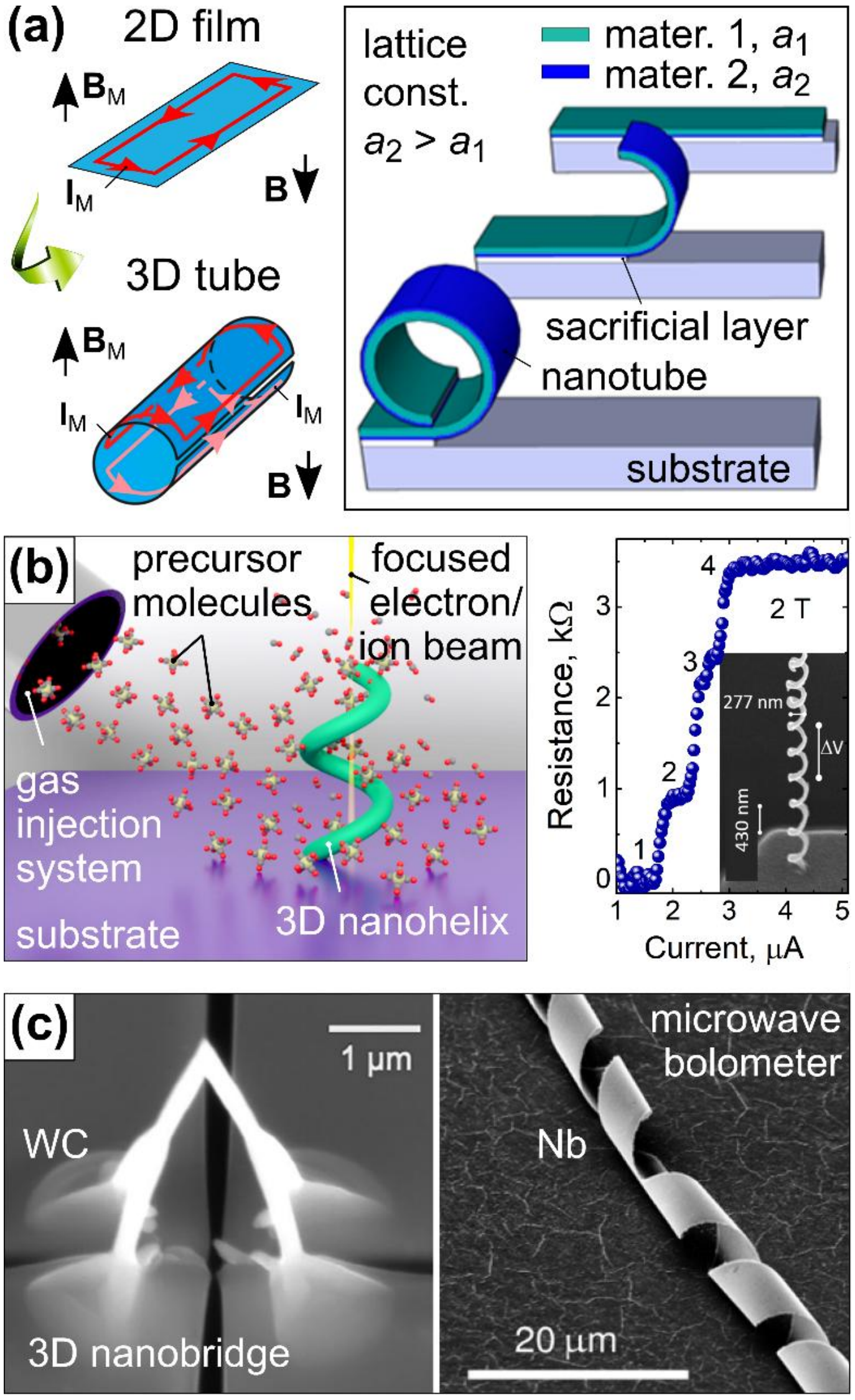

**Figure 8.1.** (a) Conceptually, rolling up of a 2D superconductor film into a 3D tube leads to inhomogeneity of the external magnetic field **B** and nontrivial topology of the superconducting screening currents $I_{M.}$ Adapted with permission from [2]. (b) Cartoon of 3D nanoprinting by focused ion/electron beam induced deposition. The transitions of different parts of the 3D WC nanohelix to the normally conducting state are reflected in resistive steps with increasing transport current. Inset shows an SEM snapshot of the nanohelix. Adapted with permission from [6]. (c) SEM snapshots of a 3D WC nanobridge (left) and a 3D Nb nanohelix bolometer (right) fabricated, correspondingly, by 3D nanoprinting and the roll-up technology driven by relaxation of mechanical strain. Adapted with permission from [5] and [4], respectively.

For numerical modeling of 3D superconductor micro- and nanostructures, the time-dependent Ginzburg-Landau (TDGL) equation has been proven the most versatile tool [1-3]. For instance, a TDGL equation-based analysis of the spatio-temporal evolution of the superconducting order parameter in 3D WC nanohelices [6] has allowed for the explanation of resistance steps observed with increase of the transport current because of sequential transitions of different helix half-windings to the normal state, see the right panel in Fig. 8.1(b).

## 8.2 Current and Future Challenges

For understanding of unconventional topologies of superconducting states and vortex dynamics in 3D nanoarchitectures, several challenges should be met regarding their fabrication, characterization and theoretical modelling.

The superconductor materials demonstrated with 3D nanoprinting are currently limited to NbC and WC. The direct writing by focused particle beams is a perfect technique for fabricating high-resolution, complex nanoarchitectures, either by connecting nanowires or through layer-by-layer growth [6-8]. However, it is characterized by a relatively low throughput. By contrast, the rolled-up technology is a parallel process which can yield multiple structures in one fabrication step. However, the geometries achievable with this technology are limited to open and multi-wall tubes, and helical structures (see, e.g., the right panel of Fig. 8.1(c) and Ref. [1]).

The superconducting and resistive properties of 3D micro- and nanostructures can be experimentally investigated using both integrated-response techniques – such as magnetometry, electrical magnetotransport, and microwave power absorption measurements – and local-probe methods. However, the adaptation of local-probe techniques presents greater challenges compared to integrated-response techniques. For instance, scanning Hall probe microscopy, magnetic force microscopy and scanning SQUID-on-tip microscopy require the probe to raster across the sample surface at a distance finely adjusted to accommodate the superconductor's topography which is challenging for 3D nanoarchitectures.

Imaging 3D superconductor nanoarchitectures necessitates techniques with nanoscale spatial resolution, sensitivity to magnetic flux, and the ability to operate at cryogenic temperatures [1]. Currently, the primary candidate techniques for such studies can include transmission electron microscopies, electron holography, magneto-optical imaging (e.g., magneto-optical Kerr effect, MOKE), x-ray microscopy, and scanning tunnelling microscopy and spectroscopy (STM/S). However, the application of these techniques for 3D nanoarchitectures faces significant challenges. TEM encounters difficulties due to shadowing effect, while MOKE and x-ray microscopy are more complicated by the need to use a magnetic indicator film. STM/S, on the other hand, is barely feasible due to the stringent requirements imposed by tunneling conditions.

The challenges for the modelling of 3D superconductor nanoarchitectures include an account for (i) structural imperfections giving rise to vortex pinning and nonreciprocity in vortex motion, (ii) proximity effects caused by non-superconducting leads and scaffolds, and (iii) heating effects which become critical at strong transport currents given the very few points of thermal contact of free-standing 3D structures to the substrate. (iv) Extension of studies towards mK temperatures and exploration of superconducting fluctuations in 3D nanoarchitectures are essential for applications in quantum information processing, and (v) nonequilibrium phenomena appearing as a consequence of a strong dc current (e.g., flux-flow instability), microwave stimuli (e.g., stimulation of superconductivity) and resulting from the absorption of infrared-to-optical-range photons (see also Section 18) appeal for theoretical studies.

## 8.3 Advances in Science and Technology to Meet Challenges

In the studies so far, global effects of curvature in 3D nanoarchitectures were mediated by the topology of the superconducting screening current. Local effects of curvature can be anticipated when downscaling the systems to reach the curvature radii on the scale of the coherence length (~5 nm for WC), which is a severe experimental challenge. 3D nanoprinting using $He^+$ focused ion beam induced deposition (FIBID) allows one to realize hollow nanowire wall thicknesses down to ~20 nm and helix curvature radii down to ~50 nm, see Fig. 8.2(a) [6-8].

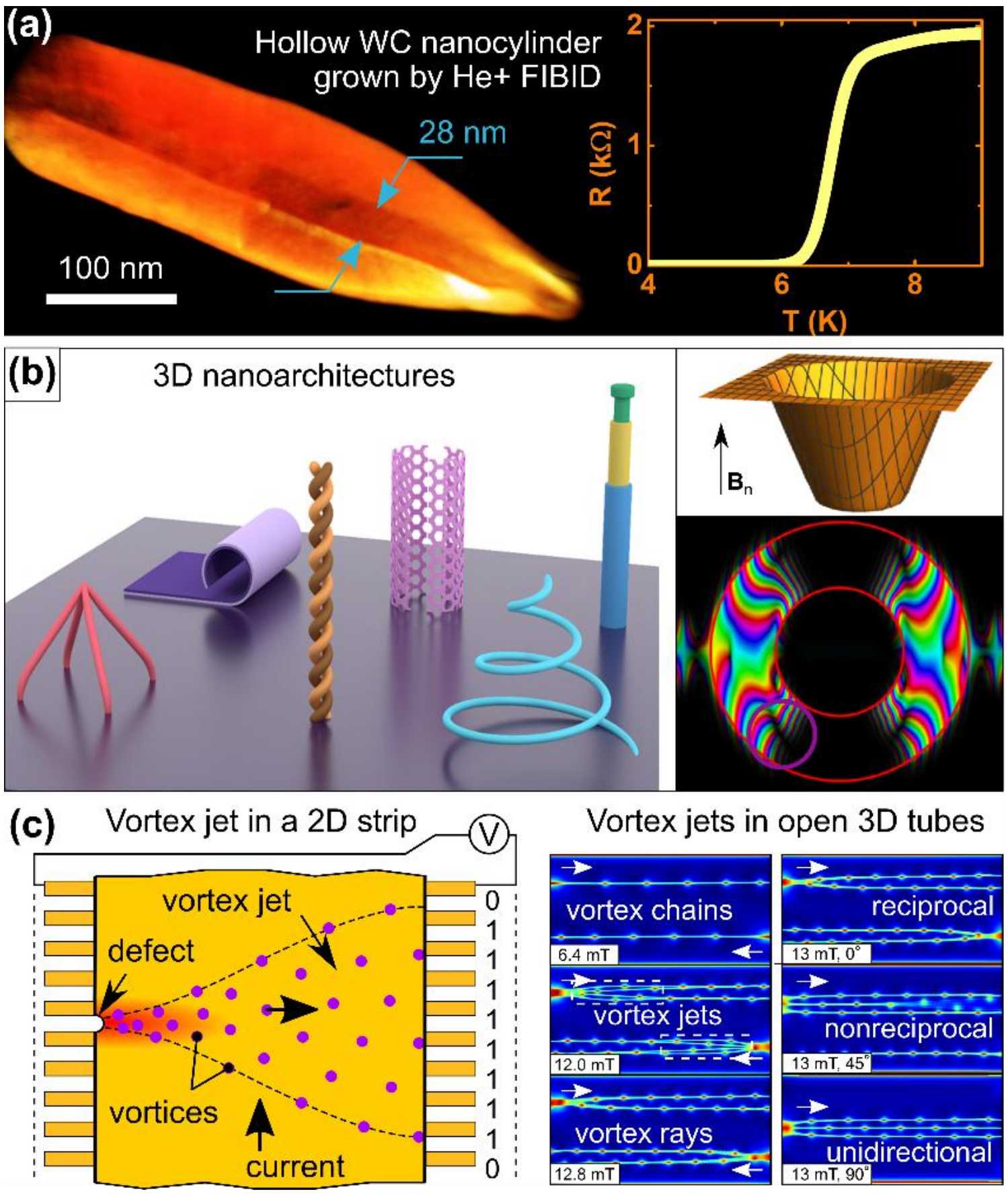

**Figure 8.2.** (a) Snapshot of the 3D tomographic reconstruction and the superconducting transition of a WC hollow nanocylinder grown using $He^+$ FIBID, with an outer diameter of 142 nm and an inner diameter of 28 nm. Adapted with permission from [8]. (b) Prototype nanoarchitectures for 3D nanoprinting (left) and the order-parameter distribution predicted theoretically for a superconducting membrane shaped as a cup (right) in the magnetic field under the transport current that flows from left to right. Black color: states of suppressed superconductivity. (c) Concept of a fluxonic multiterminal device based on a diverging jet of vortices which enter a 2D strip at an edge defect and give rise to transverse voltages (the state "1" corresponds to a vortex crossing the line connecting a pair of transverse voltage leads) which is extended to 3D open tubes. In the latter, the vortex jets are not diverging so that flux quanta can be steered as carriers of information to the given point of the structure and the regimes of reciprocal, nonreciprocal, and unidirectional flux transport can be realized via tilting the direction of the applied magnetic field with respect to the nanotube slit. Adapted with permission from [9, 10].

One of the challenges in 3D nanoprinting techniques, Fig. 8.2(a) left, is expanding the currently narrow range of superconducting materials (currently limited to two) that can be used. Addressing this challenge open up the opportunity to develop novel precursor materials, preferably those based on carbonyl groups, as they decompose more readily, and on metallic elements that exhibit superconducting properties at low temperatures. It is prospective to fabricate novel hybrid superconductor-magnetic 3D nanoarchitectures, like cup-shaped structures in Fig. 8.2(a) right, for engineering of both the superconducting and magnetic properties through the proximity effects. For the local characterization of micrometer-scale 3D structures, we anticipate that micro-Hall magnetometry can be applied. Though the diffraction-limited size of the laser probe in low-temperature scanning microscopy restricts its use to microarchitectures, the "far-field" nature of the laser probe should, in principle, allow for mapping of surfaces of 3D nanostructures via *xy*-scanning in conjunction with focus adjustment.

A few directions can be formulated for enhancing the theoretical analysis of superconductor 3D nanoarchitectures. These include (i) the establishment of a link with the quantum-fields and gravity theories, (ii) expanding diversity of nontrivial geometries, and (iii) the use of differential-geometry-based algorithms in conjunction with GPU-based numerical calculations. Since TDGL equation-based simulations are rather time-consuming, alternative approaches are needed such as AI/Deep Learning with their capability to empirically recognize patterns of the order parameter for complex-shaped systems via supervised training and deep neural networks. One can anticipate that the use of AI/Deep Learning would allow for optimization and improvement of reproducibility in 3D nanopriting in near-real time.

**8.4 Concluding Remarks**

Extension of 2D manifolds into the third dimension has become a trendline across various disciplines ranging from solid-state physics over chemistry and biology to mathematics [1]. In superconductivity, the geometry-induced nonuniformity of the external magnetic field and nontrivial topology of the screening currents give rise to emerging states of vortex matter unseen in 2D counterparts. The 3D geometry offers additional mechanisms for steering of vortex dynamics and the associated voltage responses. For example, distinct vortex and phase-slip regimes, attributed to topologically non-trivial screening currents and confinement potentials, have been evidenced in nanohelices [6]. While operations with individual vortices and their ensembles forming a "jet" have recently been proposed for 2D superconductors strips, see Fig. 8.2(c) [9], vortex jets become confined to the 3D tube areas where the normal component of the magnetic field is maximal [10]. Furthermore, by tilting the magnetic field relative to the slit location, one can achieve reciprocal, nonreciprocal and even unidirectional magnetic flux transport. The nonreciprocity in superconducting systems represents a topic of increasing interest nowadays [9]. Its enhancement and tailoring for superconductor 3D chiral structures is expected to lead to novel applications in the years to come. Realization of the above-described ambitious goals for experiment, supported by theory and numerical modelling in the fascinating realm of superconductor 3D nanoarchitectures, will introduce structural freedom in complex geometries generating nontrivial topologies superior over existing superconducting electronics, provide devices with tailored on-demand characteristics and easier fabrication procedures down to nanoscale, and thus advance the strategic application potential for nanoscience and nanotechnology [1].

**Acknowledgements**

O.D. acknowledges financial support from the Austrian Science Fund (FWF) under Grant No. I6079-N (FluMag) and the Deutsche Forschungsgemeinschaft (DFG, German Research Foundation) under Germany's Excellence Strategy – EXC-2123 QuantumFrontiers – 390837967, project FF-145 (3DSuper).

R.C. acknowledges financial support from the Generalitat Valenciana (SEJIGENT/2021/012T), MCIN/AEI/10.13039/501100011033 and “ESF Investing in your future” (RYC2020-029075-I). All authors acknowledge support from the European Cooperation in Science and Technology COST Action CA21144 (SuperQuMap), and R.C. is also an Action Member of FIT4NANO (CA19140) and grateful for the collaborations that COST Actions enable her to have.

## 9. Emergent states of matter in twisted and mesoscopic superconductors

*Nicola Poccia*[1,2], *Pavol Szabó*[3], *Fabrizio Porrati*[4]

[1]University of Naples, Italy
[2]Leibniz IFW, Germany
[3]SAS Košice, Slovakia
[4]Goethe University Frankfurt am Main, Germany

### 9.1 Status

Semiconductivity and superconductivity are two cornerstones of modern electronics. However, the focus today is not on the study of these states of matter because of their intrinsic interest, but primarily on the artificial design of either superconducting or semiconducting materials for controlling new emergent states, in line with the desired function of the device. Popular media often highlights the discovery of materials with higher superconducting critical temperatures, but the significant advances in controlling superconductivity at the nanoscale receive less attention. These developments have revolutionized material science, enabling the creation of functional heterostructures with new emergent states. One major breakthrough was the introduction of molecular beam epitaxy, allowing precise control of copper-oxide superconductors, which hold the record for critical temperature under ambient pressure.

Another significant technological advance in the fabrication of heterostructures was the use of exfoliated flakes of layered systems to fabricate stacked two-dimensional (2D) nanostructures. Even though most of the processes start from a simple scotch tape, the level of technology that now is required to realize a stack of twisted graphene is quite remarkable. Direct writing methods are increasingly employed since regular shape for the most advanced quantum devices based on exfoliated 2D layers are demanded. In twisted graphene for example, it is shown that a fibre-laser adapted to a transfer stage can be used to shape and cut graphene layers [1]. These experimental innovations have led to a plethora of theoretical predictions on the quantum geometry of the van der Waals heterostructures [2]. However, the ideal picture often presented in theory needs, as discussed in the Section 1 to progress together with the knowledge and the development of advanced technologies by experimentalists because of the reproducibility problem that is affecting in general 2D materials and twisted architectures [3]. This problem is particularly evident if the constituting elements of the 2D layers are chemically complex and the number of twisted layers is increasing. Using some of these experimental approaches, on the basis of superconducting dichalcogenides there have been created artificial systems where tunnelling [4] and proximity effects are combined to yield entirely new physical qualities, such as p-wave superconductivity in ferromagnetic/s-wave superconductor systems [5].

More recently, a novel cryogenic stacking technology has emerged [6], enabling the preparation of atomically sharp interfaces between copper-oxide crystals with specific twist angles [7], see Fig. 9.1. This has led to the observation of high-temperature, spontaneous time-reversal symmetry-broken states without external fields, marking a new era in controlling emergent states of matter. The introduction of this twist angle adds a new dimension to the phase diagram of cuprates, opening up exciting possibilities in the field of superconductivity.

Closely related to the advancements in fabrication techniques, mesoscopic superconductivity is nowadays the heart of quantum technologies, with devices such as single-photon detectors, Josephson junctions and SQUIDs which are largely employed in astronomy, metrology and quantum communications. Since more than four decades, mesoscopic superconductors are prepared by techniques such as photon-/electron-beam lithography, focused ion beam milling, or by self-assembly. Currently, these fabrication techniques have reached a high degree of precision in engineering planar structures for the investigation of, e.g., Majorana bound states in topological superconductors (Section 1), fast vortex dynamics in hybrid superconducting/ferromagnetic structures (Section 11) and the superconducting diode effect. Emerging states in mesoscopic superconductivity relate to the control of matter at the nanoscale, where size-effects start to play a role. Furthermore, the extension of the geometry in the third dimension, see Fig. 9.2, promises novel phenomena and functionalities absent in the corresponding planar case.

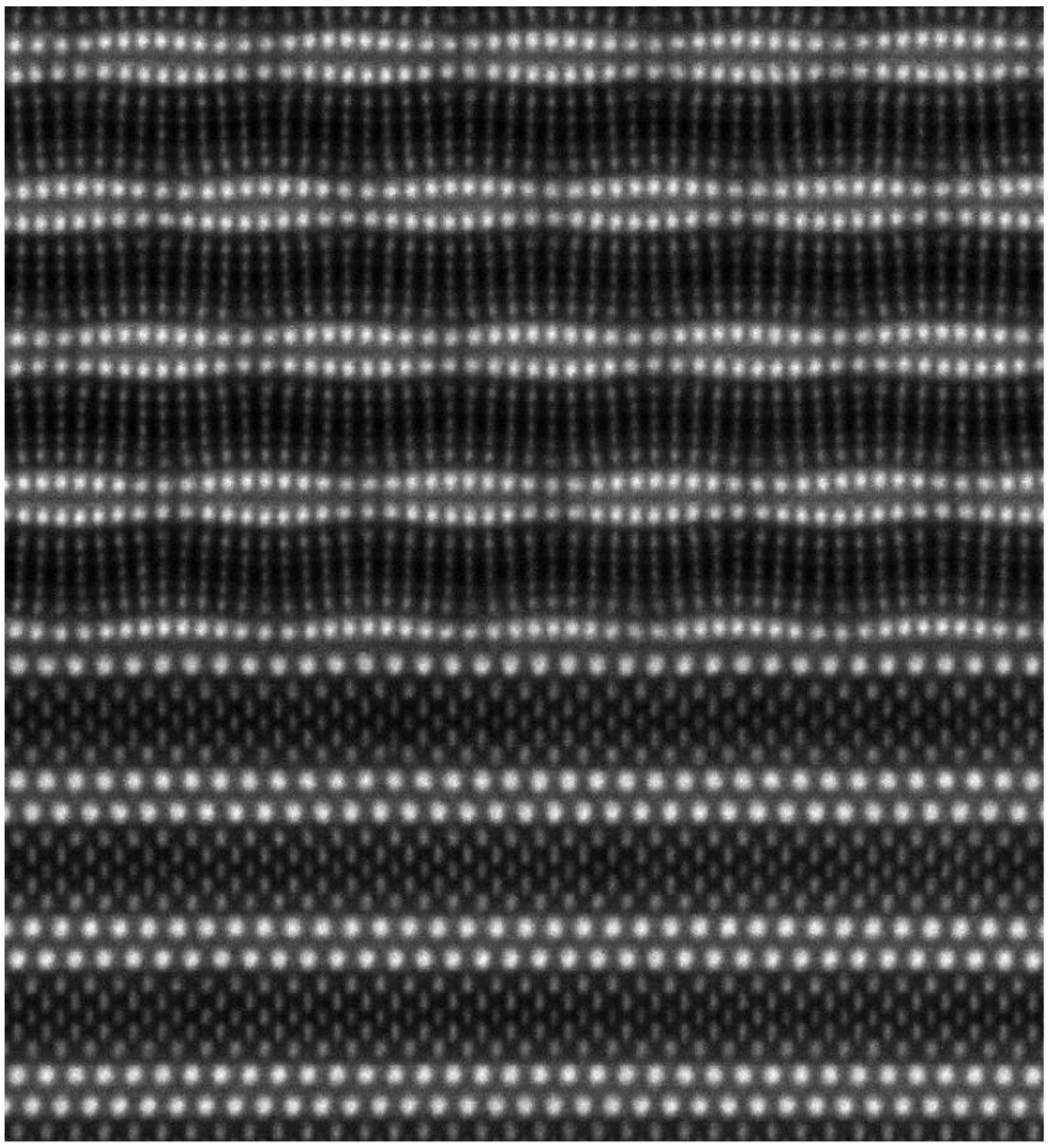

**Figure 9.1.** Cross-sectional annular dark field scanning TEM image of ϑ = 46° cuprate superconducting Josephson junction, realized by the cryogenic stacking technology and showing bulk-like crystalline order at the interface. Bright spots are columns of atoms. Reprinted with permission from [6].

### 9.2 Current and Future Challenges

A current challenge in cuprate twistronics is identifying the distinctive topological features of emergent quantum states arising from the d+id superconducting order parameter at the interface. Following the breakthrough in creating ultra-clean interfaces via cryogenic transfer, focus is also shifting towards molecular beam epitaxy. Researchers are beginning to synthesize oxide nanomembranes [8] to explore twist dependencies, but a major challenge remains: achieving interface qualities comparable to cryogenic transfer without lattice reconstructions. Another key challenge in cuprate twistronics is integrating complex circuits to study these emergent states, moving beyond just a few electrical contacts [9] as needed for devices discussed in Section 23. Looking ahead, a significant goal is synthesizing multilayered structures and heterointerfaces between chemically and artificially different van der Waals layers. Finally, in merging photonic, for example at the THz frequency as discussed in the Section 21 and quantum technologies with artificial van der Waals heterostructures made from cuprate layers, it is foreseen as one of the future challenges.

Artificially fabricated layered systems of different van der Waals materials have enormous variability and allow the study of various phenomena that can be manifested only in quantum systems, such as Ising superconductivity, time-reversal symmetry-breaking effects on superconductor/ferromagnet interfaces, topological superconductivity, Majorana states, etc. The challenge, besides the preparation of heterostructures, is also to master the methodology of low temperature tunneling measurements on these systems, both using scanning tunneling microscope and planar tunneling layers [4].

One main research topic in mesoscopic superconductivity is the study of superconductor-to-metal/insulator quantum phase transitions [10], which is often achieved via proximity-coupled Josephson junctions arrays (JJA), i.e., a 2D array of superconducting islands/dots on a metallic film, see Fig. 9.2(a). During the years studies focused on the Kosterlitz-Thouless transition, the nature of an ubiquitous anomalous/quantum metallic state and of competing vortex states. A current challenge is the investigation of JJA with dot size below 10 nm, where size-effects enter in competition with the proximity effect of neighboring dots. A further trend in mesoscopic superconductivity is the investigation of 3D curved geometries, such as nanowires, nanohelices and core-shell nanobridges [11], with predicted novel phenomena in the dynamics of vortices and phase slips. Current challenges mainly relate to the development of new materials and to the improvement of characterization techniques.

### 9.3 Advances in Science and Technology to Meet Challenges

To fully harness the fragile quantum states created by the reduced dimensionality and the twist, it is crucial to improve how we design electrical circuits, ensuring they are as competitive as those in the more established semiconductor industry. Already in 2D dichalcogenides, there are efforts in integrating them in complex electrical circuits that can rival with the competitive silicon based semiconducting circuits. However, cuprates are the most fragile material when the size and shapes are reduced and heterostructures with a twist are created. Here the goal of creating complex circuits is particularly challenging. Indeed, detrimental disorder in the twisted cuprate interfaces can be easily created as they are highly sensitive to heat, solvents, and polymers. Furthermore, reducing heterostructures to atomically thin layers requires environments even cleaner than the best gloveboxes can provide. Advances in ultra-high vacuum technology, combined with progress in nanomembrane circuit design, and ultra-high vacuum spectroscopic tools, are expected to provide a powerful solution to the challenges faced by the emerging field of 2D dichalcogenides and cuprate twistronics.

In mesoscopic superconductivity size effects and superconductivity suppression are relevant for island sizes in the sub-10 nm range (Anderson criterion). Electron-beam lithography and helium ion-beam lithography, with beam sizes <1 nm, can target the sub-10 nm regime. However, the development of suitable resists is required for the fabrication of large scale high resolution periodic nanostructures. Alternatives are direct-writing techniques like focused electron/ion beam-induced deposition (FEBID/FIBID) with record resolution of 3 nm for dots fabricated in a SEM. These techniques own as well high flexibility in growing 3D complex geometries, see Fig. 9.2. In this field current research focus on the optimization of the growth strategy to perform layer-by-layer 3D nanoprinting (see also Section 8) and on the synthesis of single-source precursors suitable to fabricate superconductors at nanoscale. Finally, the characterization for 3D structures is up to now mainly limited to electrical measurements [12], [13]. Micro-Hall magnetometry and point-contact spectroscopy appear to be feasible methods for additional characterizations.

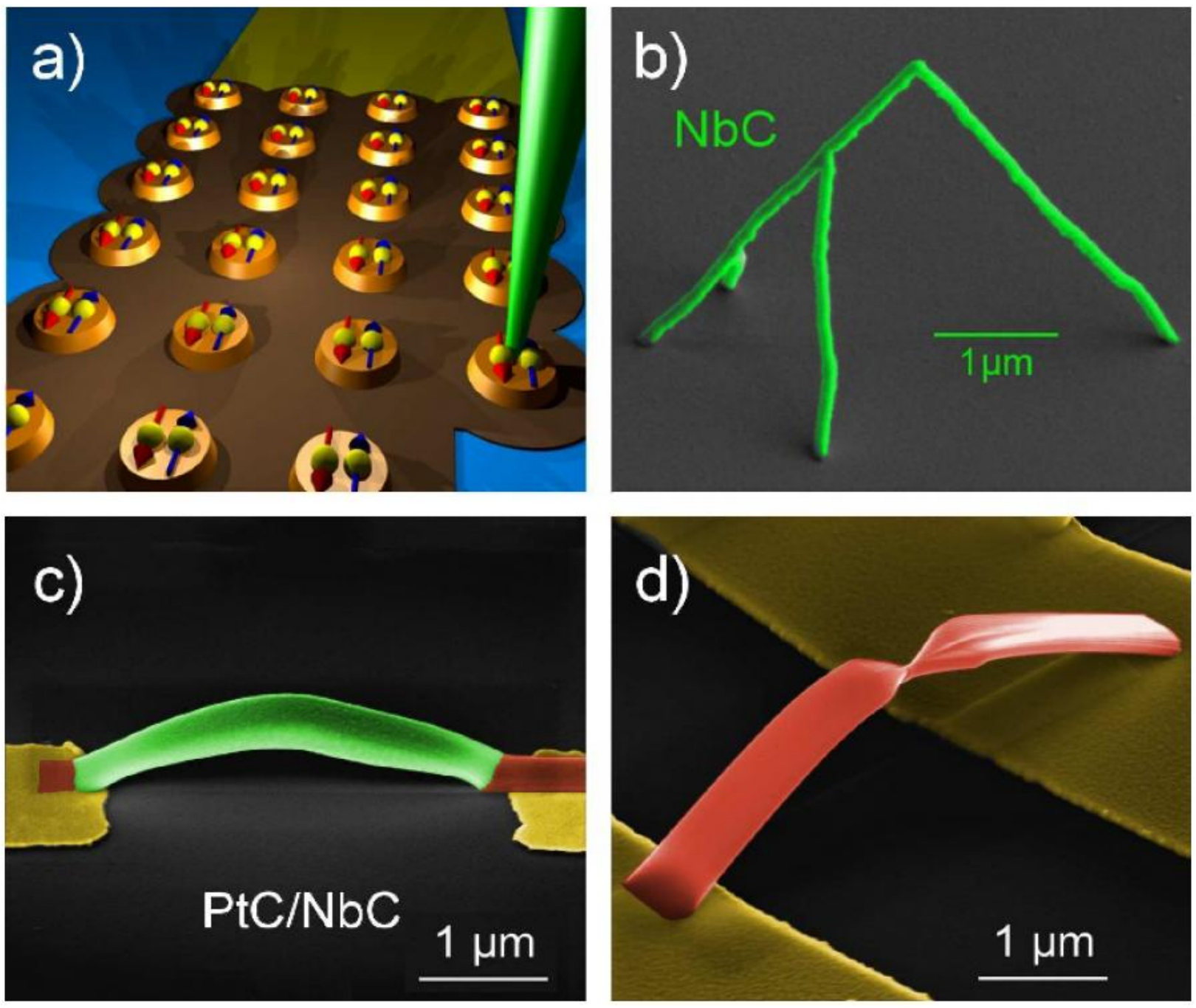

**Figure 9.2.** Sketch of a Josephson junction array printed by $Ga^+$ focus ion-beam induced deposition (FIBID). The array is made by NbC dots coupled through the proximity effect via a 2D granular metal layer [11]. b) SEM micrograph of a freestanding 3D NbC nanowire printed by FIBID. Reprinted with permission from [12] Copyright © 2019, American Chemical Society. c) 3D PtC/NbC heterostructure obtained by covering a PtC (orange) scaffold by selective CVD with NbC (green). Adapted with permission from [10] Copyright © 2023, American Chemical Society. d) SEM image of a 3D PtC nanobridge with constriction fabricated by focused electron beam induced deposition (FEBID) [13]. The PtC scaffold might be covered by selective CVD with NbC to form a tubular 3D Josephson junction.

### 9.4 Concluding Remarks

As we learn how to tame the complexity of atomically sharp and ultra-clean twisted interfaces between quantum materials with non-stochiometric light-element composition, a new perspective opens up for exploiting subtle emergent states in specific applications, mimicking the complex shapes and chirality observed in biology.

Furthermore, new states and functionalities emerge as we learn to control the matter and the geometry of superconductors at mesoscopic scale. This directly relates to the recent impressive advancements in nanofabrication as exemplified, e.g., by the study of the breakdown of superconductivity for periodic nanoscaled systems and the investigation of 3D superconductors and tubular nanostructures.

**Acknowledgements**

N.P. acknowledge Flavio Lo Sardo for useful feedback during the preparation of the manuscript and the partial funding support from the European Union, ERC-CoG 3DCuT, 101124606, by the Deutsche Forschungsgemeinschaft (DFG 512734967) and Terra Quantum AG. F.P. gratefully acknowledge the support of the Frankfurt Center of Electron Microscopy (FCEM). The work of P.Sz. was supported by Projects APVV-23–0624, VEGA 2/0073/24 and SAS Project IMPULZ IM-2021–42.

## 10. Magnetic nanoelements and their arrays

*Gleb Kakazei*

University of Porto, Portugal

### 10.1 Status

With the advancement of nanofabrication methods in the last decade of $20^{th}$ century, periodic arrays of magnetic nanoelements have attracted attention due to their unique magnetic properties at the nanoscale, which enable high-density storage, sensitive detection capabilities, and novel biomedical applications, such as targeted drug delivery and magnetic resonance imaging enhancement. Novel lithography techniques have been developed or improved to allow fabrication of ordered magnetic nanostructures with desirable features, including arrays of elements with reproducible sizes in the range down to 10 nm, which can be extended over large areas. Many parameters (material, crystallinity, number of layers, array geometry, etc.) were varied to tune the properties of these systems. Also, magnetic nanostructures were prepared with very controllable shape: from the simplest dots, bars and lines, to rectangles, triangles or zigzag paths, and used as reconfigurable vortex pinning sites (see also Section 5) and magnetic counterparts in superconducting spintronics (see Sections 11 and 12). The extensive studies on the behavior of individual dots have revealed that the main magnetic properties of these nanostructures present important differences with respect to continuous films [1].

As magnetic elements are reduced from micrometer to nanometer dimensions, the interplay between exchange, anisotropy, and magnetostatic energies changes profoundly. Microsized elements usually form multiple domains, while below a few hundred nanometers single-domain or flux-closure states become favored. Below ~100 nm, magnetization tends to remain uniform. Meanwhile, reducing the element thickness to a few nanometers enhances surface and interfacial contributions, such as perpendicular anisotropy, exchange bias, and Dzyaloshinskii–Moriya interactions. These size-dependent effects govern switching behavior, thermal stability, and dynamic responses in nanoscale magnets.

Remarkably, one of the possible realizations of such flux-closure configurations is the vortex state, which often occurs in circular or square soft-magnetic elements a few hundred nanometers wide and a few tens of nanometers thick, where the aspect ratio supports flux closure, with the vortex core located at the element's center (Fig. 10.1). Vortices are considered iconic structures in nanomagnetism and serve as ideal candidates for exploring new physical concepts and experimental techniques due to their high frequency oscillations when displaced from the center [2]. Recent advancements in nanostencil lithography have enabled the extension of these typically two-dimensional (2D) vortex configurations into 3rd dimension [3], significantly increasing their oscillation frequencies to approximately 5 GHz—well above the typical sub-GHz range of planar vortex oscillators. This advancement opened new avenues for optimizing nanoscale magnetic devices, allowing for enhanced performance in frequency-dependent applications.

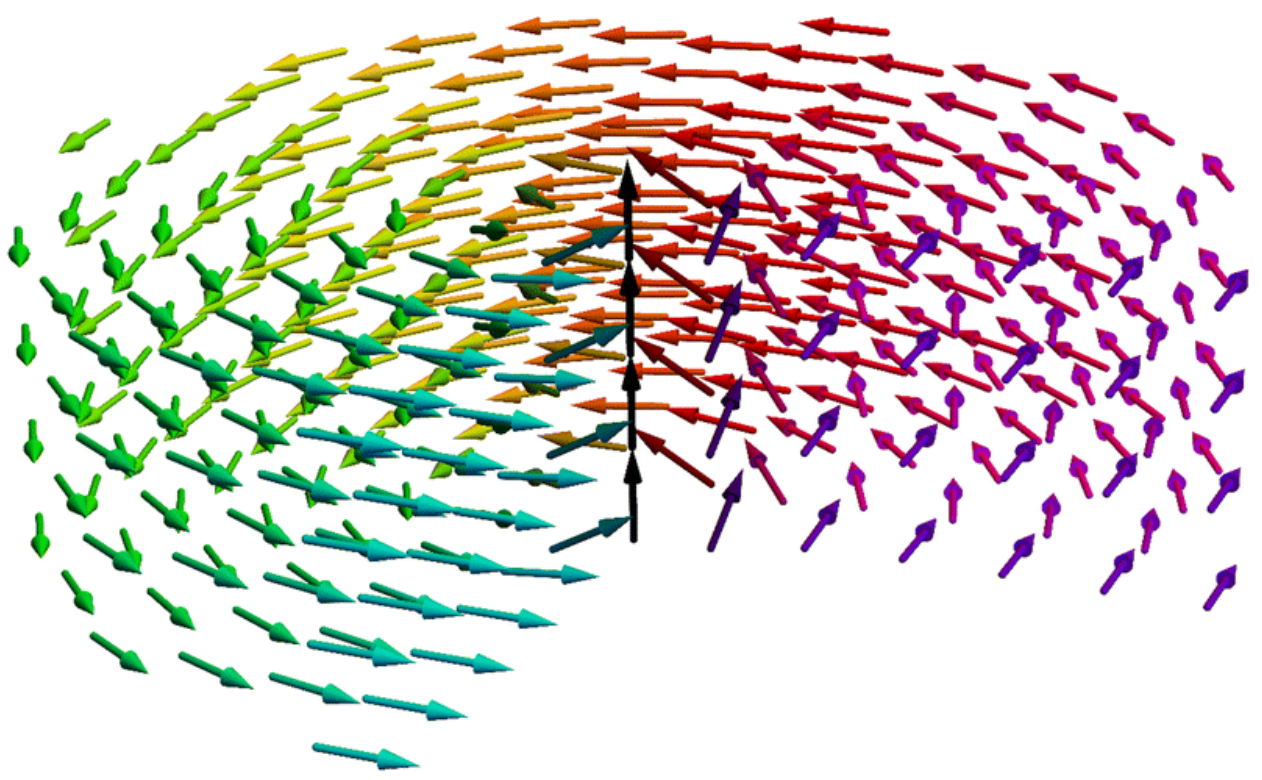

**Figure 10.1**. Cylindrical magnetic dot in the vortex state, arrows represent spin momenta directions of individual spins. The color of the spins corresponds to the direction of the planar component of the spin, spins on the axis of the particle point out of plane and form the core of the magnetic vortex. Reproduced from Ref. [3] with permission from the Royal Society of Chemistry.

The arrays of magnetic nanoelements are not only interesting for their intrinsic magnetic properties, but also due to their interaction with other systems, particularly with superconductors. Such arrays are known to constitute effective ordered pinning centers for the vortex lattice when they interact with type II superconducting films. Particularly, it was revealed in [4] that the pinning of superconducting vortices is strongly enhanced for the magnetic vortex state. This enhanced pinning was attributed to vertical stray magnetic fields produced by the magnetic vortex cores. Very recently, the upper frequency limits for guided and rectified net motion of superconducting vortices in epitaxial Nb films decorated with ferromagnetic nanostripes were explored, revealing the vortex guiding and ratchet effect arise on the background of the competition between the intrinsic weak pinning in the Nb films and the strong periodic pinning induced by the Co nanostripe array [5].

In magnetic data storage, continuous media employ uniform magnetic films where bits are encoded as magnetic domains, achieving areal densities up to around 2 Tb/in$^2$. However, further scaling faces superparamagnetic instabilities and write field limitations. In contrast, patterned media consist of a highly ordered array of magnetic nanoelements, where each nanodot stores a single bit, theoretically allowing densities up to 10 Tb/in$^2$.

Finally, magnetic nanoelements are very important for magnonics - an emerging field of research focused on the study and manipulation of magnetic excitations known as magnons, which are quantized spin waves propagating through magnetic materials [6]. Magnons carry information in the form of spin, making them potential candidates for developing new information processing technologies. In Ref. [7], a new type of magnonic crystal – NiFe films deposited on top of periodic 2D array of circular NiFe dots – was proposed. In this structure, the continuous film can be considered as a magnonic waveguide, when the role of dot array is to create periodic perturbations of internal fields in the neighbor regions of the film which can be controlled by magnetic field and film thickness.

### 10.2 Current and Future Challenges

Nanomagnetism, while a rapidly evolving field with significant advancements, faces several challenges that impact both research and application development. One of the primary current challenges is achieving consistency in manufacturing. Ensuring uniformity and reproducibility in the fabrication of nanostructures remains a problem, as variability in size, shape, and material properties can lead to inconsistent magnetic behaviors, complicating the development of reliable applications. Additionally, advanced characterization techniques are needed to probe the magnetic properties of nanostructures at the nanoscale, with many current methods lacking the necessary spatial resolution or sensitivity to fully understand the complex behaviors of small magnetic systems.

Traditionally, periodic arrays of magnetic nanoelements have been confined to 2D single- or multi-layered planar structures, where the elements are typically arranged in a flat geometry over a substrate. This planar configuration offers simplicity in design and manufacturing but inherently limits certain magnetic behaviors and functionalities to the two-dimensional plane. Therefore, one of the main challenges is to enable the expansion of magnetic nanoelements into third dimension, opening new possibilities for manipulating spin configurations and achieving functionalities beyond the limits of the 2D plane. In three-dimensional (3D) structures (see also Section 8) the magnetization orientation can extend vertically, allowing for richer and more complex spin configurations This development paves the way for diverse applications, including spin-wave computing, which utilizes the propagation of spin waves in magnetic materials for information processing with minimal energy consumption [8].

Regarding the future, scalability remains a critical challenge. While current lithography methods can produce nanostructures on relatively large areas, scaling these processes for mass production while maintaining quality and precision will be essential for commercial applications. Energy efficiency will also be a significant concern as devices become smaller; developing nanomagnetic systems that operate efficiently while providing high performance will be an ongoing challenge. Additionally, as quantum effects become increasingly significant at the nanoscale, further research will be required to understand how these behaviors impact the magnetic properties of nanostructures and to leverage them for new applications.

Thermal stability is another significant challenge; as the size of magnetic elements decreases, their stability can diminish, making them more susceptible to thermal fluctuations that can adversely affect performance in real-world applications such as data storage devices. Furthermore, as arrays of magnetic nanoelements become densely packed, the dipolar and exchange interactions between elements can introduce complexities that alter their collective magnetic behavior. Understanding and controlling these interactions are essential for practical applications.

### 10.3 Advances in Science and Technology to Meet Challenges

To address challenges in magnetic nanoelements, significant advancements in science and technology are being developed. Precision fabrication techniques, such as nanoimprint lithography and extreme ultraviolet lithography, are improving manufacturing consistency by producing high-resolution nanostructures with uniform magnetic properties. Enhanced characterization techniques, including scanning tunneling microscopy, atomic force microscopy, and electron holography, provide better spatial resolution and sensitivity for studying magnetic behaviors at the nanoscale.

One particularly impactful technique in 3D nanofabrication is Focused Electron Beam-Induced Deposition (FEBID), a maskless approach that allows for the construction of complex 3D nanoelements with lateral resolutions below 10 nm (see also Section 8). FEBID enables precise control over the geometry of nanoelements by using a focused electron beam to deposit materials from a gaseous precursor. Importantly for magnetic applications, material composition can be tuned in situ via the e-beam waiting time and post-growth irradiation with Ga ions [9]. Recently, this technique has been used to fabricate magnetic "nanovolcanoes"—innovative 3D architectures where nanodisks are overlaid by nanorings (see Fig. 10.2) [10]. Spin-wave resonance measurements revealed that the rings surrounding the volcano craters harbor high-frequency eigenmodes, while lower-frequency modes are concentrated within the crater. By varying the crater diameter, the ability to fine-tune the higher-frequency eigenmodes without affecting the lowest-frequency mode was demonstrated. These nanovolcanoes can serve as building blocks for next-generation nanomagnonic devices.

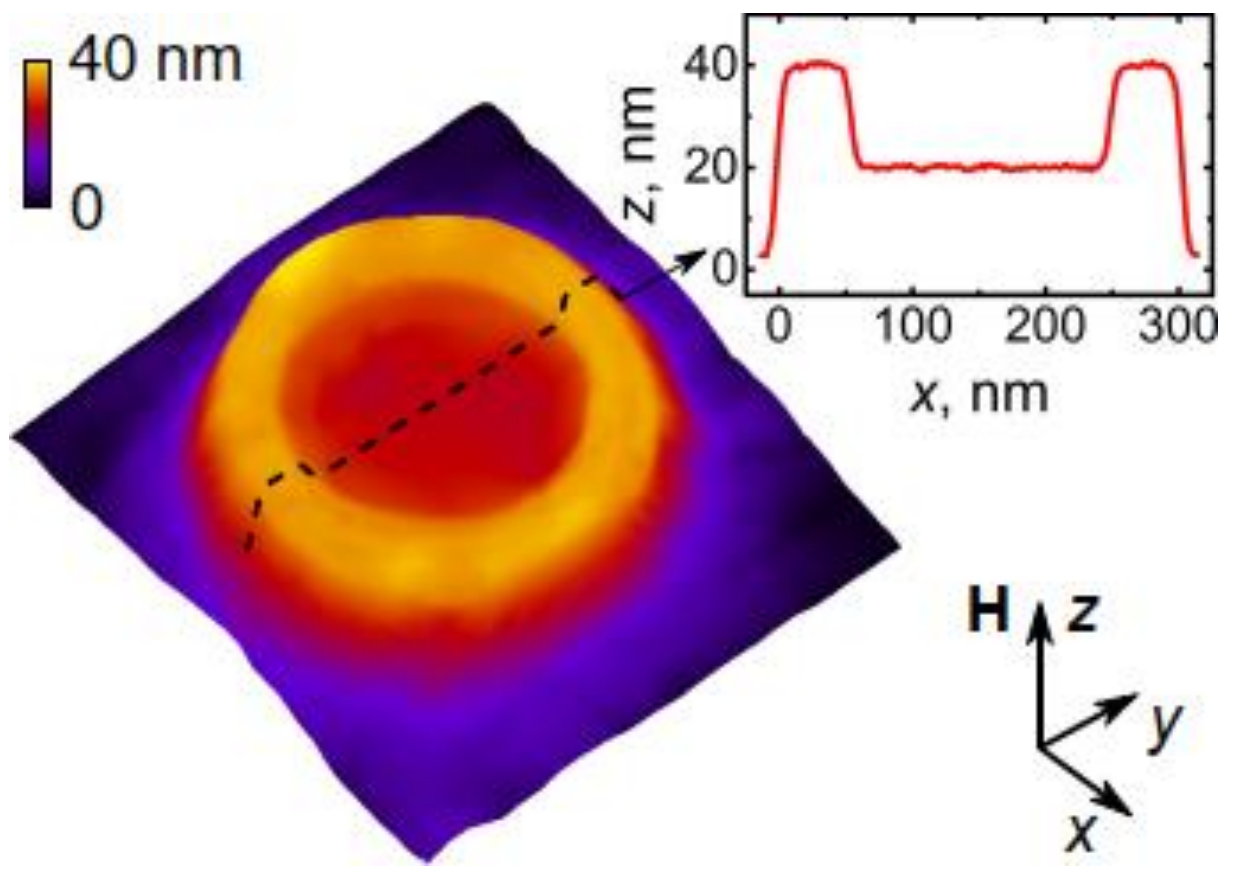

**Figure 10.2.** Atomic force microscopy image of the nanovolcano with the outer diameter of 300 nm and the crater diameter of 200 nm. Inset: cross-sectional line scan, as indicated. Adapted with permission from [10], CC BY.

Advances in magnetic simulation and modeling techniques help manage inter-element interactions by predicting collective magnetic behaviors within densely packed nanostructures. Integration with existing technologies is facilitated by hybrid systems that combine magnetic nanostructures with compatible semiconductor materials, enhancing their application in electronic and photonic devices. For energy efficiency, spintronic devices are being explored to reduce power consumption, alongside the development of novel ferromagnetic materials for low power switching.

### 10.4 Concluding remarks

Magnetic nanoelements and their arrays play a crucial role in moving forward modern technology by offering pioneering solutions across various domains, including data storage, sensing, spintronics, fluxonics, and magnonics. There is a growing demand for magnetic nanomaterials that serve multiple purposes, such as acting both as storage media and sensors. The ability to create and control magnetic nanostructures with smaller dimensions opens up new opportunities, enabling faster data processing and improved energy efficiency. Additionally, nanomagnets hold promise for the development of next-generation quantum computing technologies that rely on spin-based phenomena. Tailoring nanomagnetic systems for these applications will require extensive research to improve our understanding of underlying physical mechanisms and to enhance materials accordingly. Addressing these challenges will be crucial for further advancements in magnetic nanoelements, enabling to realize their potential in next generation technologies and applications.

**Acknowledgements**

G.K. acknowledges financial support from FCT – Portuguese Foundation for Science and Technology through the projects LA/P/0095/2020 (LaPMET), UIDB/04968/2025; from FEDER – European Regional Development Fund through the project no. 17142|COMPETE2030-FEDER-00854500 (SynRoLoD); and from the European Cooperation in Science and Technology COST Action CA21144 (SuperQuMap).

## 11. Superconductor-ferromagnet hybrids

*Jan Aarts*[1], *Jason Robinson*[2], *Javier E. Villegas*[3]

[1]Huygens-Kamerlingh Onnes Laboratory, Leiden University, Leiden, The Netherlands
[2]Department of Materials Science & Metallurgy, University of Cambridge, Cambridge, United Kingdom
[3]Laboratoire Albert Fert, CNRS, Thales, Université Paris-Saclay, Paris, France

### 11.1 Status

The claim to fame of superconductor/magnet hybrids is that the magnetic layer can either couple two superconductors in an unconventional way, with the resulting Josephson junction possessing special features; or the magnet can carry a spin-polarized (spin-triplet) supercurrent, with the promise of dissipationless spintronic applications. A number of ways of fabricating Josephson $\pi$-junctions are estalished, as well as generic methods of generating spin-triplet Cooper pairs, based on the spin-mixing / spin rotation principle, utilizing two or more ferromagnetic layers. At present, research focuses on a number of follow-up issues. An important one, and actively researched, is the role of *spin-orbit coupling* (SOC) [1], which results in the breaking of inversion symmetry. Combining this with time-reversal symmetry breaking, such as from the magnetic exchange field of a ferromagnet or an external magnetic field, this can produce a junction with a built-in phase difference $\varphi_0$ that, for instance, can act as phase battery. Similarly, SOC is researched as an alternative for magnetic inhomogeneity in generating spin-triplet Cooper pairs (see Fig. 11.1a). Incidentally, it is also a key ingredient in supercurrent rectification through the *superconducting diode* effect [2]. A different way of generating triplet supercurrents is by creating and controlling certain spin textures in the ferromagnet (see Fig. 1b), enabling memory effects [3]. Another area of interest concerns the dynamic, rather than the static, coupling between the magnetization and the superconducting state. Ferromagnetic resonance (FMR) has been used to study spin transport through spin pumping by generation of spin-triplet pairs and their transmission through a superconducting layer via a SOC interface [4]. There is also a growing interest in combining superconductivity with *magnonics*, for instance by using magnons excited in ferrimagnetic insulators to create a quasiparticle spin-split state in a thin superconductor [5], or to couple magnons and vortices [6]. Finally, there is steady progress in the use of less conventional materials than simple transition or rare earth metals. For example, using half-metallic ferromagnetic oxides (see Fig. 11.2b), very long-range proximity effects have been demonstrated with relatively high critical current values [7], [8]. Furthermore, spin-triplet supercurrents have been demonstrated in the Kagome (chiral) antiferromagnet $Mn_3Ge$ [9] and long-range Josephson coupling has been induced by a van der Waals superconductor into a van der Waals ferromagnet [10], see Fig. 11.2a.

### 11.2 Current and Future Challenges

One of the central challenges in the field is realizing the potential for novel superconducting electronics and spintronics (see also Section 12). Beyond analogues of classical devices such as diodes [2] and memories [3], a challenge is to demonstrate and use spin transfer torque (STT) effects in the same way as is done in conventional spintronics. By placing a thin ferromagnetic layer in the path of a spin-polarized supercurrent or quasiparticle current, its magnetization can be flipped, or brought into motion, resulting in a nano-oscillator. Similarly, a non-equilibrium triplet supercurrent might be able to apply torque to a magnetic domain or a skyrmion. The fundamental question is whether the accompanying dissipation is detrimental to the superconductivity, although, analogous to vortex motion, this does not appear to be unavoidably the case. On the practical side, the demonstrated supercurrents are not as high as used in spintronics, and writing or moving bits appears hard to achieve via the conventional angular momentum transfer. Skyrmion motion requires far less current density, but few experiments have yet been performed involving skyrmions and superconductors. For memory applications, a challenge is that the 'writing' of a bit, when not by the STT mechanism, is mostly done with a not-too-small, and therefore impractical, magnetic field. A potentially interesting approach is to create novel functionalities, in particular for magnetic switching, by electrostatic control. Superconducting devices involving SOC-based phenomena including diode effects and/or spin-triplet supercurrents are a promising direction since anisotropic (Rashba) SOC is tunable with an electric field. Such electric control of SOC may enable electric control of spin-charge coupling and conversion though the tuning of triplet channels within the superconducting density of states of superconductor [5].

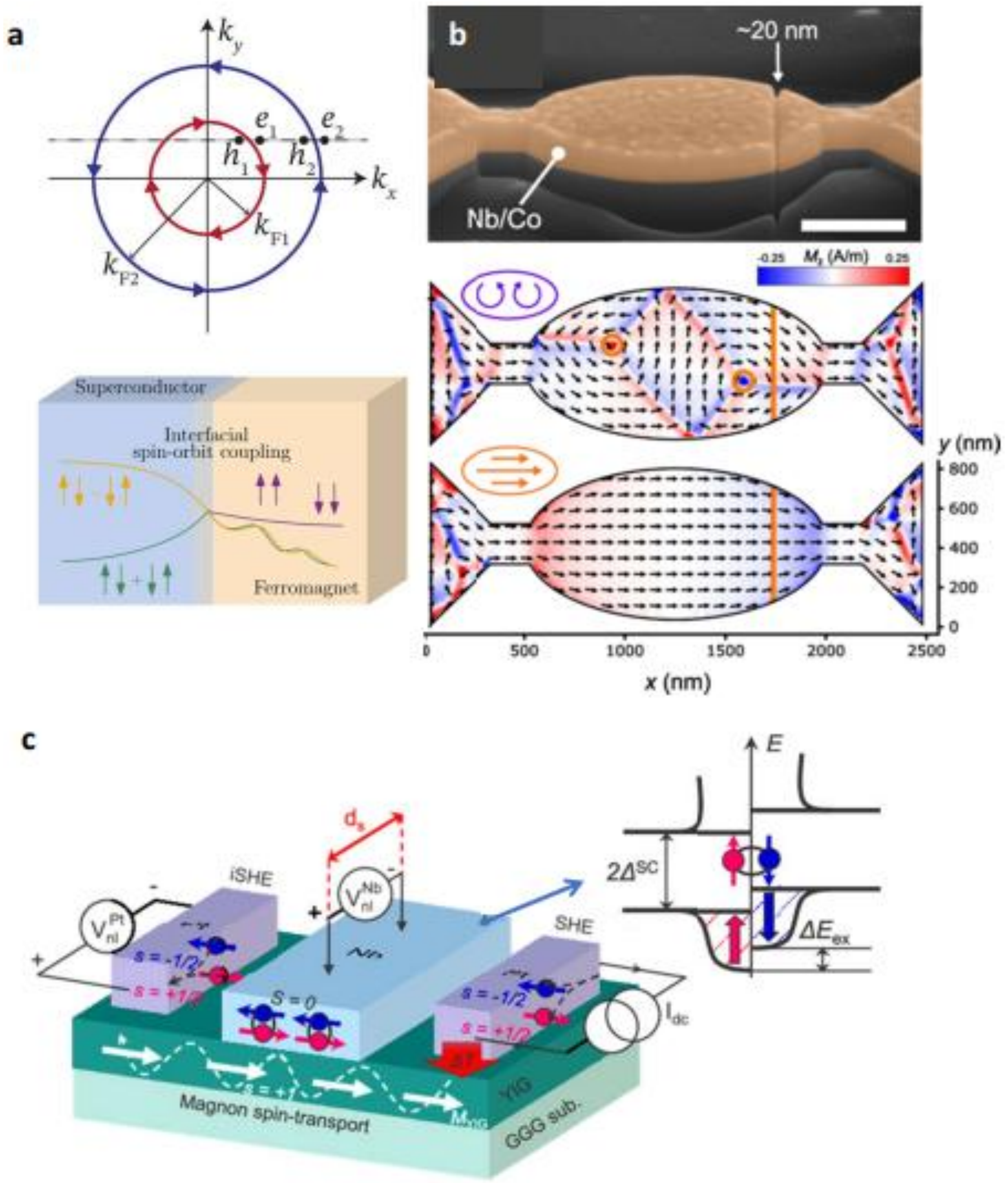

**Figure 11.1. (a)** Spin-orbit coupling effects are a key ingredient in the generation of triplet superconductivity. Adapted from [1]. **(b)** Mesoscopic superconducting memory based on as Nb/Co/Nb Josephson junction in which the switching between bistable magnetic textures is controlled by the magnetic history, leading to a switching of supercurrents. Adapted from [3]. (c) Layout of the device used to detect the giant inverse quasiparticle spin Hall effect by nonlocal magnon transport. The exchange splitting of the superconductor density of states plays a crucial role. Adapted from [5].

Another unexplored direction is to develop optical responses of a S/F hybrid system, allowing manipulation in the ultrafast time domain. For example, one could envision ultrafast optical manipulation of supercurrents in S/F devices by exploiting the already well-mastered excitation of ultrafast magnetization dynamics. Finally, there are the van der Waals materials, that offer a completely new palette of possibilities. For instance, Moiré engineering can create magnetic (bi)layers where ferromagnetic and antiferromagnetic regions are mixed. Also, it is the one class of materials where magnetism can be influenced by electrostatic gating. At the same time, this will require new theoretical efforts. The interfaces between superconducting and ferromagnetic layers are different than for metallic layers, and proximity effects or Andreev reflections will have to be reconsidered. Topological effects may come into play as well.

**11.3 Advances in Science and Technology to Meet Challenges**

Material science challenges are ubiquitous in S/F heterostructures [2-4,6-8]. That is so because many relevant mechanisms are extremely sensitive to the structural, magnetic and superconducting properties of the interfaces within devices. For example, the transparency for electron transport across an S/F interface or any interface, is dependent not only on the Fermi surface mismatch but also on variables such as disorder and impurity scattering at the interface, determining probabilities for Andreev reflection and thus the strength of the superconductor proximity effect. Likewise, properties such as the interfacial magnetic texture [3,9], anisotropy, and spin polarization [7,8], affect triplet correlations, spin transport, and magnetic exchange effects [5]. The challenge in the control over all of those properties lies in combining materials whose growth conditions and handling constraints can vary significantly. At present, S/F hybrids are mostly considered within a single class of materials, such as simple transition metals, oxides (see Fig. 11.2b), or van der Waals materials (see Fig. 11.2a), where, in all cases, different functionalities can be found and combined (with one notable exception: the absence of multiferroics in simple metals).

There is an urgent need to find proper interfacing that would allow combinations of different families of materials, such as an oxide with a transition metal - the ubiquitous problem being that the oxide oxidizes the metal, thereby complicating the interface. For van der Waals materials [10], the additional challenge is to either develop large-scale stamping techniques, or to develop growth techniques, with Atomic Layer Epitaxy (ALE) prominent among them. The progress in ALE when it comes to, for instance, transition metal dichalcogenides (TMDCs) such as $WSe_2$, that are of great interest as semiconductors in (opto)electronics and photovoltaics, have not yet reached the S/F front.

Theoretical efforts will be required to understand some of the emergent research directions outlined above. This is true in particular when it comes to combining knowledge and advances in different subcommunities. A prototypical example would be the ultrafast manipulation of S/F interactions using circularly polarized light. Because optical excitations are usually very energetic, especially as compared to the superconducting gap, non-equilibrium effects are expected to interplay with magnetization dynamics and proximity mechanisms. These complex interactions will likely demand further attention from theory, to guide or interpret experiments. Other examples are magnonic or optical mechanisms to couple spin without a net charge current. Last but not least, in van der Waals materials, theoretical (Density Functional Theory) calculations of the electronic properties of few-layer materials and the properties of interfaces are much needed. The equivalent of Fermi velocity mismatch when it comes to transparency of the interface has yet to be determined.

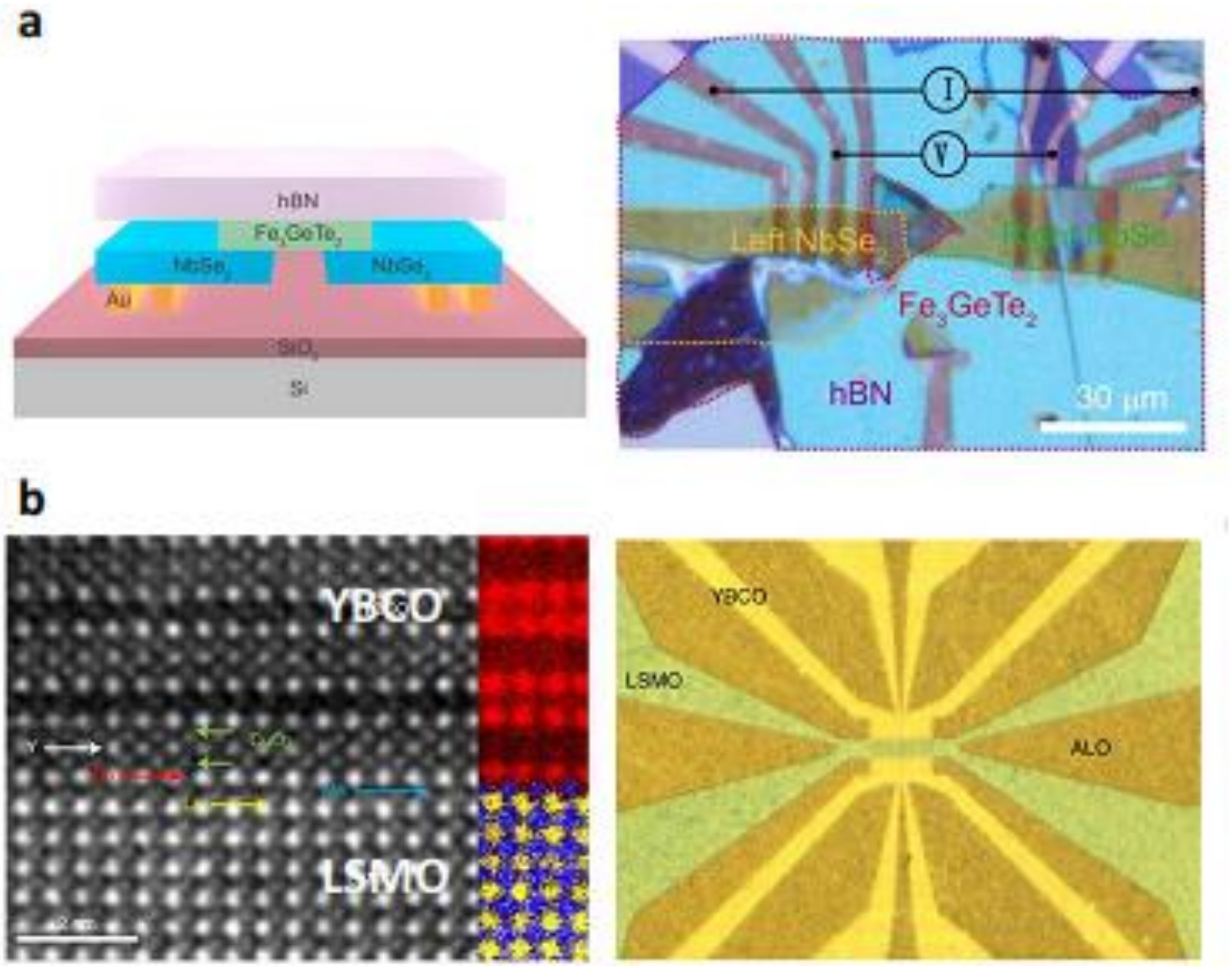

**Figure 11.2. (a)** Layout (left) and actual microscopy image (right) of a triplet Josephson junction based on the vdW materials $NbSe_2$ and $Fe_3GeTe_2$ Acdapted from [10]. **(b)** (left) Scanning transmission electron microscopy (dark field and electron energy loss spectroscopy) of the interface between a high $T_C$ superconductor ($YBa_2Cu_3O_7$) and a half-metallic ferromagnet ($La_{0.7}Sr_{0.3}MnO_3$) together with a micrograph of a planar Josephson S/F/S device based on those materials, in which long range triplet supercurrents have been observed. Adapted from [8].

### 11.4 Concluding Remarks

Superconductor-ferromagnet hybrids are fertile ground for novel physics and technological prospects. Despite the considerable research efforts, and the steady progress, during the last two decades, covering a plethora of topics that span from flux dynamics manipulation, unconventional proximity and Josephson effects to spin-dynamics in superconductors, the area offers an extraordinary range of tantalizing opportunities for basic and applied science areas. These can be hardly summarized here, but research directions such as developing sensitivity to optical stimuli, searching for topological effects, exploiting the interaction between supercurrents and magnetic textures and realizing the potential for spintronics applications are specially promising. Incorporation of novel materials, such as van der Waals materials, may be a key ingredient, which is accompanied by non-negligible material science and theoretical challenges.

### Acknowledgements

Work at Laboratoire Albert Fert supported by French ANR through grants ANR22-CE30-00020-01 “SUPERFAST”, ANR-22-EXSP-0007 PEPR SPIN “SPINMAT”, the European Union’s EIC pathfinder grant 101130224 “JOSEPHINE”. Work at Leiden University is supported through grant OCENW.XS23.4.079 from the Dutch Research Council (NWO). JV, JA, and JWAR acknowledge support from the COST action “SUPERQUMAP”. JWAR also acknowledges funding through the EPSRC (No. EP/P026311/1 and No. EP/N017242/1)

# 12. Superconducting spin currents in hybrid structures

*M. Althammer[1], H. Huebl[1,2], A. Kamra[3], and M. Weiler[3]*

[1]Walther-Meißner-Institut, Bayerische Akademie der Wissenschaften and Technical University of Munich, TUM School of Natural Sciences, Physics Department, Garching, Germany

[2]Munich Center for Quantum Science and Technology (MCQST), Munich, Germany

[3]Fachbereich Physik and Landesforschungszentrum OPTIMAS, RPTU Kaiserslautern-Landau, Kaiserslautern, Germany

### 12.1 Status

Superconducting currents and magnetic excitations can be readily coupled via electromagnetic radiation giving rise to hybrid (quantum) devices, which are intensely researched in the field of spin cavitronics. Typical settings rely on superconducting resonators loaded with magnetic media [1]. Alternative schemes rely on a direct interface between superconducting and magnetic layers in thin-film heterostructures, which allow to study the direct transfer of angular momentum between spin dynamics and superconducting currents. This provides the exciting perspective for dissipation-less manipulation of magnetic order and spin dynamics in superconductor/magnet hybrid structures.

The concept of angular momentum transfer is studied in spin pumping experiments [2,3] that focus on the injection of spin currents generated in the magnetic layer into the superconductor by externally driving spin dynamics. A pioneering work [2] showed that adding a heavy metal layer to the thin film heterostructure can enhance spin-orbit coupling effects in the superconductor via proximity effects. This, in turn, can give rise to spin triplet currents in the superconductor, improving the effectiveness of angular momentum transfer. In ac spin pumping experiments illustrated in Fig. 12.1(a), the ac component of the spin current flowing in the superconductor is transformed into a charge current by spin-orbit coupling and can then be inductively detected in microwave spectroscopy experiments. Thereby, the ac in- and out-of-phase components of the generated charge currents are identified as damping-like and field-like terms [3]. In these experiments, Nb and NbN were employed as the superconducting layers and the observed damping-like inverse spin-orbit torques $\sigma_d$ below the critical temperature $T_c$ of the superconductor were attributed to quasiparticle spin transport effects in the superconductor (see Fig. 12.1(b), (c)). In addition, it is observed that the spin singlet Cooper pairs effectively block any spin current flow in these heterostructures below $T_c$. Moreover, the quantitative analysis of inductively detected spin-orbit torque effects in superconductor/ferromagnet hybrids yields an enhancement of field-like torque contributions below $T_c$ [3].

Recent theoretical proposals suggest generating and utilizing spinful triplet Cooper pairs in structures that interface a conventional superconductor with two magnets. In such schemes, spinful Cooper pairs are generated in the first magnet by exploiting the spatial or temporal non-collinearity between the magnetic moments stemming from uniform precession [4], spin waves [5], or antiferromagnetic canting [6]. These propagate to and interact with the second magnet, providing a reactive coupling between the two magnets and serving as an application and confirmation of spinful Cooper pairs.

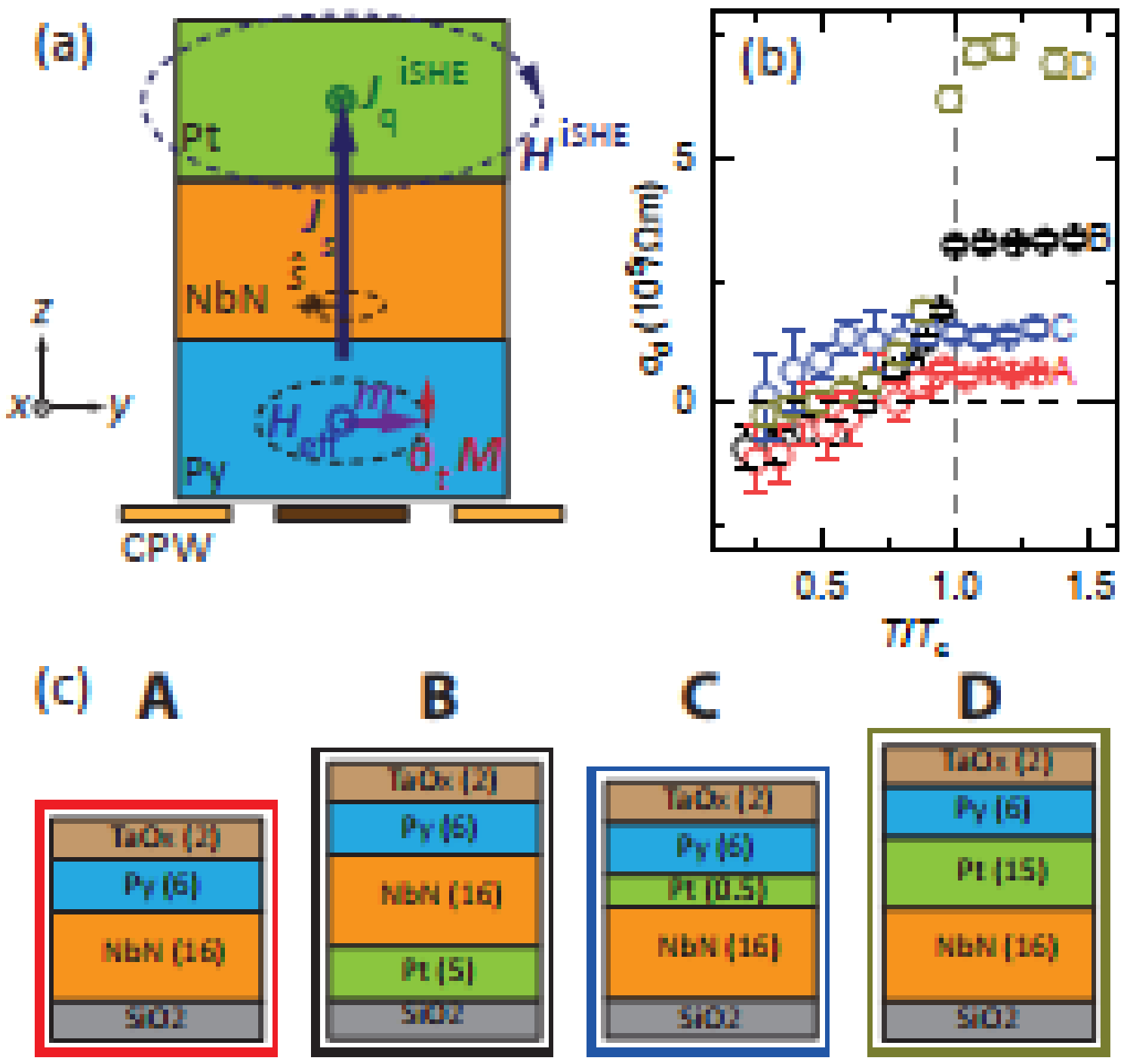

**Figure 12.1.** Inductive detection of inverse spin-orbit torque effects in superconductor/ferromagnet heterostructures. (a) Illustration of the measurement geometry. The sample is mounted on a coplanar waveguide (CPW), where the generated ac magnetic field drives the magnetization dynamics in the magnetic layer. The precessing magnetization $\mathbf{m}$ pumps a spin current $\mathbf{J}_s$ across the interface into the superconducting layer. The spin current is then transformed into a charge current $\mathbf{J}_q^{\mathrm{iSHE}}$ via spin-orbit coupling, which inductively couples to the CPW. (b) Extracted damping-like inverse spin-orbit torques as a function of the reduced temperature for different superconductor/ferromagnet heterostructures. (c) Illustration of the multilayer stacks for the samples A, B, C, D used in panel (b). Adopted with permission from Ref. [3].

## 12.2 Current and Future Challenges

At the current state, intriguing results have been obtained by spin pumping experiments in superconducting / ferromagnet heterostructures, but we lack a complete picture of the delicate interplay between angular momentum transfer, superconducting currents and excitations in superconductors. The impact of experimental parameters like for example the layer thicknesses and interface properties need to be explored further. For example, flux vortices in superconductors can also mediate angular momentum transfer [Ref. 6 of Sec. 11] or the topology of superconductors (see Section 1) could also provide novel functionality. Here, theoretical models need to provide guidance to disentangle the interplay of multiple contributions. Experiments utilize complex multilayers containing more than one magnetic layer or interlayers from heavy metals [2,3,7 and Ref. 1,4,5,7 of Sec. 11]. In such structures the influence on angular momentum transport, especially at the interface, is not well understood. Moreover, it remains unclear experimentally if antiferromagnetic layers admit the formation of spinful triplet Cooper pairs (see also Section 11). An important step is the experimental demonstration of spin transfer toque effects induced by a supercurrent acting on the magnetic layer in thin film heterostructures. As discussed in Section 11, this is challenging due to the available critical current densities in typical employed superconductors. Thus, cleverly designed experiments are required to gauge the potential of future applications.

The application potential of superconducting / magnetic heterostructures needs to be explored with a view beyond the generation of superconducting spin currents. For instance, the strong variation of magnetic damping at the superconducting transition temperature due to suppression of spin pumping [3] might be exploited for magnonic power limiters or in other non-linear magnonic devices. This requires further exploration of the superconductor / ferromagnet hybrid devices based on low-damping materials such as yttrium iron garnet.

The equal-spin triplet Cooper pairs carry spin supercurrents and are generated in conventional superconductors utilizing two or more noncollinear spin-active elements, such as magnets or spin-orbit coupling [2, 7, and Ref. 1 Sec. 12]. Their experimental observation has most often relied on long-range Josephson coupling through magnetic materials (e.g., [7]), which is expected to be infeasible via spin-singlet Cooper pairs. A more direct observation of their spinful nature continues to evade experiments, although fresh theoretical proposals on, for example, magnonic directional couplers mediated by superconductors [5] raise hopes for new avenues in this endeavor.

### 12.3 Advances in Science and Technology to Meet Challenges

From a fabrication point of view, it is desirable to learn more about the quality of interfaces and how this affects the current experiments. In addition to the improvements in fabrication techniques discussed in Section 11, this requires a further careful optimization of fabrication techniques to better control the interface properties for reliable sample fabrication and a critical assessment of defects within these interfaces via advances in spatially resolved defect spectroscopy techniques. This approach enables us to identify successful strategies to enhance spin supercurrent effects and pave the way toward future applications.

To quantify the transfer or angular momentum from ferromagnets to superconductors, studies have mostly focused on studying linewidth contributions due to spin pumping or dc voltages from inverse spin Hall effects [2,3 and Refs. 4, 5 of Sec. 11]. Both measures are not sufficiently developed in the presence of superconducting layers, which, e.g., short-circuit spin Hall voltages and change the electromagnetic environment at the superconducting transition. However, a full quantitative understanding of magnetic resonances in the presence of superconducting circuits is required to disentangle evidence for spin injection from spurious signal contributions. In light of the complexity of the microwave frequency response of the heterostructures [Refs. 2,3 of Sec. 11], spatial mapping of the resulting ac charge currents, like shielding Meißner currents in the superconductor, represents an important guidance for a full quantitative modelling to obtain a deeper understanding. This requires further progress in scanning probe techniques to enhance these methods' time and spatial resolution and to minimize the perturbation of the heterostructures by the probe.

A deeper understanding of these complex physics should allow the optimization of the hybrid structures towards applications, like magnetic memories, spin-torque oscillators (and their coupling to superconducting microwave circuits), and Josephson π-junctions for flux qubits based on spinful triplet supercurrents in conventional superconductors.

Recent experiments observe long-range Josephson coupling via an antiferromagnet in its non-collinear ground state, indicative of spin-triplet Cooper pairs. From a theory perspective, it can be understood in terms of the recently proposed Néel triplets [8] whose pairing amplitude oscillates on an atomic scale similar to and commensurate with the magnetic order in an antiferromagnet. This could enable a niche for memory schemes based on non-collinear antiferromagnets.

**12.4 Concluding Remarks**

The field of spin currents in superconductor hybrid structures has shown great potential, and further significant breakthroughs are expected in the coming years. The ultimate goal to control magnetic order and spin dynamics in the magnetic material remains fundamentally possible, with the promise of considerably reduced power dissipation. Here, the integration of magnetic insulators and superconductors is expected to obtain the best performance. Non-collinear spin structures will be in the spotlight of theory and experiment in the next years as they provide an exciting perspective for the realization and investigation of spinful supercurrents in hybrid structures.

In addition, the angular momentum character is not exclusive to spins, and hence the orbital angular momentum has similar potential, e.g., showing effects of enhanced amplitude [9]. This could also allow for an engineered conversion or interfacing between spin and orbital angular momentum using superconductors.

The rich interplay between the unconventional (d-wave etc.) pairing in superconductors and magnetism in altermagnets may lead to unprecedented phenomena, e.g., a perfect superconducting diode [10].

This short overview shows that many uncharted territories for superconductor/magnetic layer hybrids can and should be explored for future applications. We are excited to see how the field will prosper in the future.

**Acknowledgements**

MW acknowledges financial support by the Deutsche Forschungsgemeinschaft (DFG, German Research Foundation) via TRR 173/3–268565370 “Spin +X” (Projects B13 and B15). HH and MA acknowledge financial support by the Deutsche Forschungsgemeinschaft (DFG, German Research Foundation) via Germany’s Excellence Strategy EXC-2111-390814868, and TRR360 (Project ID 492547816). HH acknowledges the Munich Quantum Valley, which is supported by the Bavarian state government with funds from the Hightech Agenda Bayern Plus.

## 13. Angular resolved photoemission

*J. Hugo Dil*[1,2] *Daniil Evtushinsky*[1]

[1]EPFL, Switzerland

[2]Paul Scherrer Institut, Switzerland

### 13.1 Status

For more than 30 years by now, the progress in the condensed matter physics and material research has gone hand-in-hand with the developments in the technique of angle-resolved photoemission spectroscopy (ARPES). ARPES is an experimental tool unique in its ability to map the electronic structure of a system of an arbitrary level of complexity with energy and momentum resolution, allowing to see every single electronic state independently. Primary examples of this co-development are the inclusion of full angular separation of the photoelectrons and the rise of highly anisotropic superconductors, and the addition of spin resolution (SARPES) and the discovery of topological materials. One of the highlights in experiments on superconductors was the observation of the d-wave gap distribution over the Fermi surface in cuprates [Shen 1993], contrasting the behaviour of well-known conventional superconductors. Overall, in the promising research areas of topological electronics and unconventional superconductivity, ARPES has provided large amounts of self-evident data, thereby establishing many fundamental physical aspects [Lv review 2019, Sobota 2021, Lim review 2024].

Currently we are witnessing the next revolution in the technical development of ARPES, which will again have significant influence on the condensed matter community. For a long time, photoemission setups were relying on collecting the signal from a mm-sized area. Recent advance in the photon beam focussing and sample positioning, enabled scanning of the sample with spatial resolution as good as 1 µm and better. This, in turn, has opened a way towards in-situ tuning of the electronic properties by means of electric gating, imposing a magnetic field, and introducing controllable stress for a wide variety of systems. With all that at hand, one can aim at the detailed studies of the superconducting nanodevices, both in a sense of figuring out the microscopic origin of the electron pairing, and in a sense of imaging the behaviour of the superconducting condensate.

First SARPES experiments on high temperature superconductors (see also Section 5) have shown that there is no direct link between the measured spin signal and the superconducting state [Fanciulli 2017]. However, the inclusion of the spin as an additional observable can give access to hidden order if combined with other parameters such as photon energy or polarisation. For example, SARPES at resonant photon energies can help distinguishing the symmetry of the electron pairing [Tjeng 1997]. With the large number of observables accessible via ARPES (energy, momentum, spin, orbital order, bosonic coupling, gap size etc.) one can be assured that the inclusion of further control parameters, see Fig. 13.1, will open up a whole realm of opportunities, where the limits might be imposed only by the imagination of the researcher.

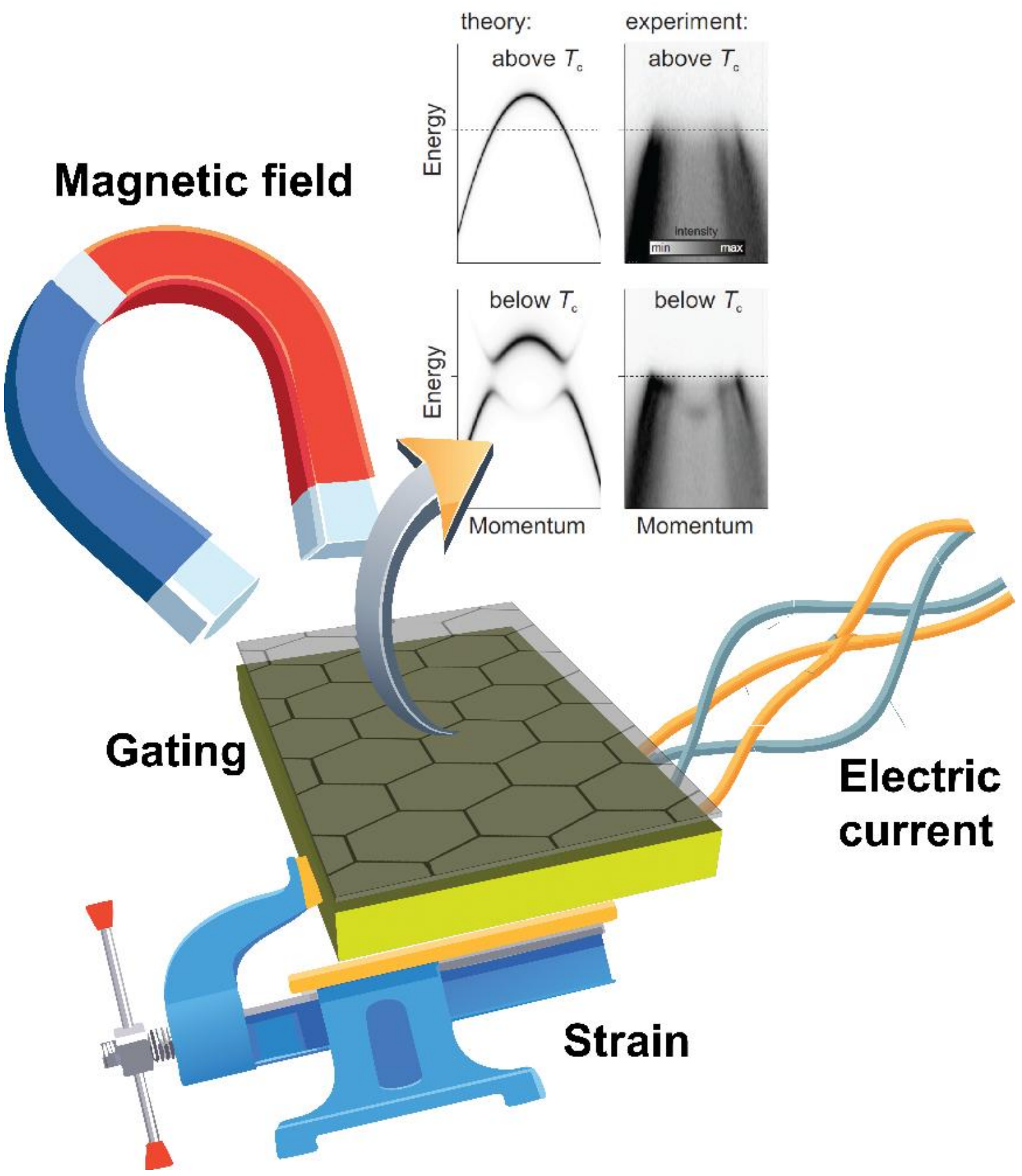

**Figure 13.1**. ARPES experiment under *in situ* application of electric gating, magnetic field, mechanical strain, and/or electric current. The top right shows theoretical and experimental electronic spectra below and above the critical temperature of the superconducting transition.

## 13.2 Current and Future Challenges

The main challenge in any photoemission experiment is to find a balance between the different types of resolution, while still maintaining a level of efficiency that keeps the measurement feasible. A clear example comes from the use of ultrashort laser pulses to achieve time resolution. Due to the Fourier uncertainty principle for time and wavelength (energy), shorter pulses impose a worse energy resolution and for a 100 fs pulse duration the best achievable energy resolution is around 100 meV.

For a reduced spot size, there is no fundamental relation to the achievable energy resolution, but there is a strong influence due to so-called space charge effects. If several electrons are emitted around the same time from a small area, they will repel each other resulting in an energy broadening. Therefore, the photon flux needs to be low enough, which will have the additional advantage that it reduces damage to the sample. Consequently, every electron that is emitted should detected, placing stringent demands on the efficiency of electron energy analysers. On the other hand, the more point-like character of the source generally enhances the momentum resolution and reduces the incoherent background.

On the scientific side, the challenges are different. In order to build a comprehensive model of superconductivity, one needs to know the pairing interactions that enter the gap equation. The solution of the gap equation is very sensitive to the input parameters, so that practically one needs to know the underlying electron dispersion and scattering rates rather precisely. ARPES measurements can provide this information when they are performed with care and taking selection rules into account.

Furthermore, based on the spectrum obtained by ARPES, one can quantify the interactions of an electron with phonons, magnons, polarons, other electrons and in general with any excitation present in the system. Those interactions form a basis for numerous effects important both from the point of view of fundamental research and applications, but separating the different bosonic interactions typically requires an additional observable or very high energy and momentum resolution combined with theory support.

Often for the complex system the sought-after extraordinary properties only occur when the parameters are tuned to the needed values. This is where the small spot size plays an important role because it allows for advanced ARPES experiments in combination with electric gating, applied magnetic field, or strain as ways to tune the system.

**13.3 Advances in Science and Technology to Meet Challenges**

The technological capabilities to focus a photon beam down to micrometer size or even lower, are currently well developed, although there is still room for improvement with regard to user friendliness. Depending on the photon energy range, the focussing is achieved by transmissive, reflective, or diffractive optics, partially relying on interference effects such as in Fresnel zone plates. Of course, the advantage of such a small spot size is lost if the sample stage does not have a better stability. State-of-the-art ARPES compatible sample stages now achieve a stability of tens of nanometers and a reproducibility in the micrometer range, while achieving temperatures below 10K. There is still a rapid development in these aspects and we consider these technical challenges solved and rather shift towards what is needed to develop new science based on such a small photon spot.

Pioneering experiments that integrated electrostatic gating in the ARPES environment [Krempasky 2018, Nguyen 2019] clearly showed that the many free charges excited by the photon beam, impose a different strategy as in conventional device operation. For sample thicknesses beyond a few atomic layers, either a top gate or laterally well positioned gates should be used, highlighting the need for a small spot size to account for any inhomogeneities in the field. At the same time, these studies showed that it is possible to alter the material properties by an electric field and directly image the response by ARPES. Similarly, the application of an electric current in an ARPES experiment is far from trivial, but has recently been achieved [Suen 2024], again relying on a small spot size. Technical developments in placing electrodes, preferentially in UHV, are desperately required to push these applications forward. But the award is worth the effort as it will become possible to directly follow the properties of the superconducting gap in reciprocal space as a function of these parameters and thus to finally define the pairing interaction. Furthermore, one can imagine that we will be able to fine tune the magnetic order via electromagnetic coupling or directly by a current, or to move the superconducting vortices through the beam and measure the contrast of the electronic properties as compared to its normal-state core.

Another important dimension for tuning the system is given by the possibility to apply the mechanical strain. As it turns out, there are many examples where the lattice and electronic degrees of freedom are strongly coupled. Subjecting unconventional superconductors to the strain [Lim review 2024] reveals interesting details helping in unravelling the entangled nature of the superconducting state. This way, by applying the uniaxial pressure, it was possible to see in finest details how the van Hove singularity is driven right through the Fermi level in $Sr_2RuO_4$, [Sunko 2019], or to find direct evidence that the unusual nematic electron order has a common origin with the effective electron pairing in iron superconductors [Rhodes 2020].

The next logical step is the inclusion of a magnetic field in a ARPES experiment. For decades one of the main concerns has been to keep the magnetic field out of the measuring chamber, including the stray magnetic field of Earth. The operation of any photoemission setup relies upon the fact that photoelectrons move in straight and predictable trajectories after leaving the sample. Even in the presence of a relatively small field (0.1 Tesla), the trajectories of electrons would start to wind in tight helixes around the field lines, with a width of only several tens of micrometers. This would result in a complete loss of the energy and momentum resolution, and even in the electrons not reaching the detector at all. There are several ideas regarding how one could solve this problem, with one of the most viable ones being to drastically restrict the volume of space where the magnetic field is present again relying on a small spot size [Ryu 2023]. Much work is still needed, but the prospect of controlling the phase of the superconducting condensate, or driving the system through a topological transition, merits the effort.

### 13.4 Concluding Remarks

Although ARPES will not achieve the types of spatial resolutions described in Sections 14-16, the available small spot sizes open up a realm of new experiments. Beyond the examples given above, one can imagine to measure individual Qubits as described in Section 23 to determine relevant properties of the SC condensate.

Going even further, it would be extremely interesting to perform phase sensitive (S)ARPES experiments as a function of the position within a Josephson junction. Combined with time-resolved capabilities one can then envision to locally decohere the superconducting condensate and measure how the systems respond, and in what time scale it recovers. In general, superconductivity is a perfect example of a non-trivial phenomenon with far-reaching technological use where ARPES, and its many derivatives, can provide an excellent tool to explore the underlying fundamental phenomena.

**Acknowledgements**

We gratefully acknowledge all our current and previous collaborators and the many fruitful discussions concerning ARPES over the last decades.

# 14. Probing Superconductors by Scanning SQUID Microscopy

*Beena Kalisky*[1] *and Yonathan Anahory*[2]

[1]Bar Ilan University, Ramat Gan, Israel

[2]The Hebrew University of Jerusalem, Jerusalem, Israel

**14.1 Status**

Superconducting nanodevices have become foundational elements in the pursuit of quantum computing and advanced quantum materials. Understanding superconductors in confined geometries is critical to unlocking their potential for coherent manipulation in quantum systems. Reaching this level of understanding requires developing tools with relevant spatial resolution and sensitivity to probe superconductivity at submicron scales. This section discusses the role of scanning superconducting quantum interference device (SQUID) microscopy in probing superconducting materials at microscopic scales, focusing on vortex physics, superfluid stiffness, and unconventional superconductors [KIR99, KOL17, PER22, CHR24].

The SQUID is a highly sensitive magnetic flux to voltage converter. This flux sensor uses a superconducting loop with two Josephson junctions or weak links, as illustrated in Fig. 14.1, (see also Sections 19-22) to detect minute magnetic flux changes with extreme sensitivity, down to micro-$\Phi_0$ levels [KIR99, KOL17, PER22], where $\Phi_0=h/2e$ is the flux quantum. Scanning SQUID microscopy (SSM) consists of scanning the SQUID loop over the sample surface, allowing spatial mapping of the magnetic flux distribution above the surface of (superconducting) samples.

A wide variety of SSM designs provide versatile access to a wide range of problems including spatial distribution of electrical current (Fig. 14.2(c) and (d)), vortex physics (Fig. 14.2(a)), and magnetism (Fig. 14.2(b)). The main design concepts are shown in Fig. 14.1 and described below.

- **SQUID on Chip (SOC)**: SOCs are fabricated in a planar geometry as illustrated in Fig. 14.1a. The SQUID loop contains a small (100 nm to 5 µm) sensing area, the pickup loop, which is integrated directly on the chip. SOCs offer the highest field sensitivity and functionality, with spatial resolution limited by the pickup loop's distance from the sample (~200 nm). The layered device allows for integrating more electronics like modulation and excitation coils, enabling simultaneous measurements of additional magnetic properties such as susceptibility, and strain response. [KIR99, HUB08, HIL03,CUI17].

- **SQUID on Tip (SOT)** and **SQUID on Lever (SOL)**: The SOT design reduces the tip-to-sample distance to ~10 nm and the loop diameter to ~40 nm, enabling resolutions of a few tens of nm [FIN10,VAS13,ANA20]. The SOT configuration places the SQUID loop at the apex of a hollow quartz tube (Fig. 14.1b). Recently, the SQUID loop was integrated into an atomic force microscope (AFM) lever, forming a SOL [WYS22] (Fig. 14.1c). Both configurations offer the advantage of scanning closer to the sample, which enhances spatial resolution but sacrifices the magnetic field sensitivity.

Each configuration has unique benefits and limitations in terms of resolution, sensitivity, and compatibility with magnetic field ranges, allowing researchers to select the optimal setup for specific superconducting applications [KOL17, PER22, CHR24].

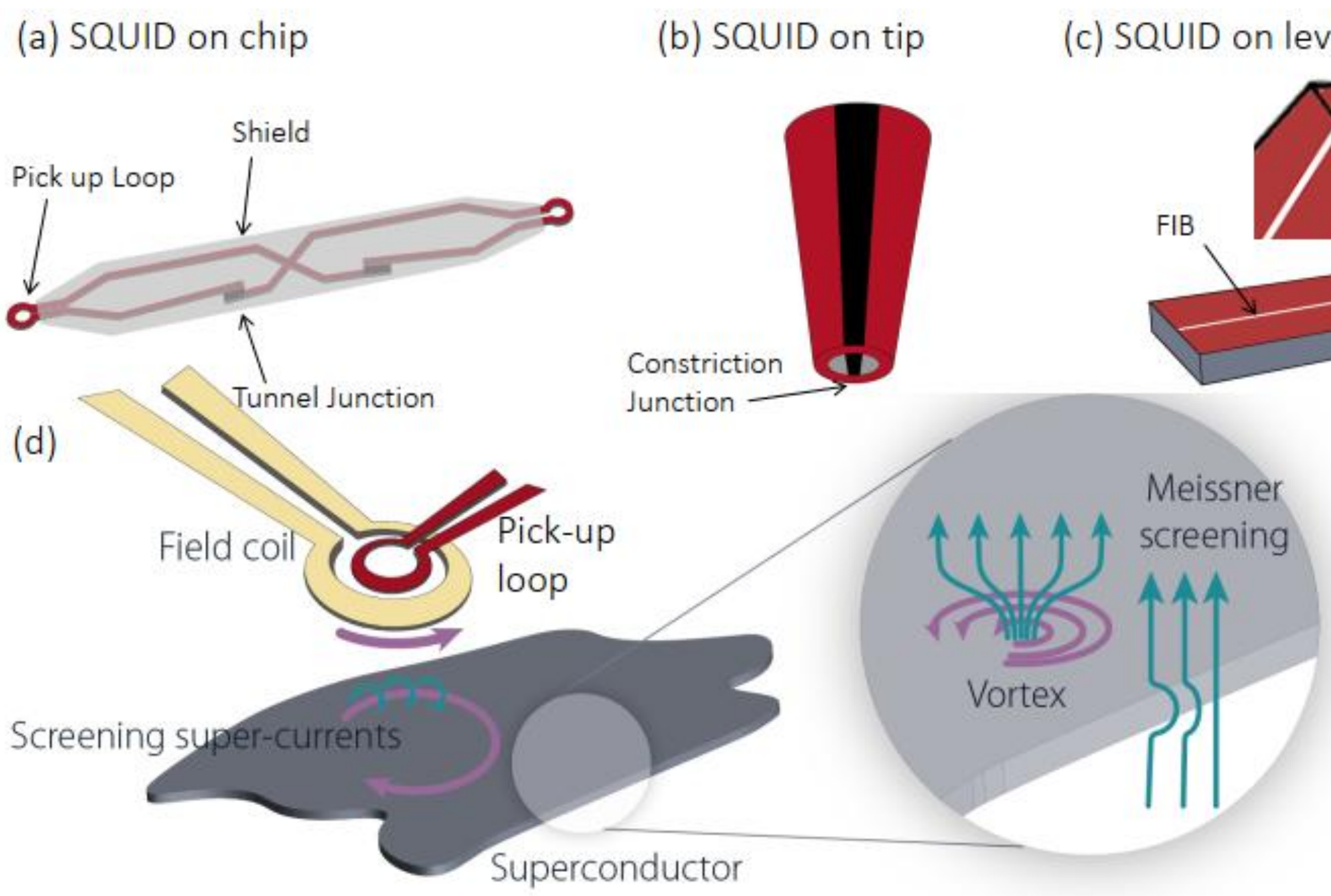

**Figure 14.1: Scanning SQUID sensors and applications to superconductivity.** SSM sensors and operation modes. (a)–(c) Illustration of three sensor designs: (a) SQUID on chip (SOC) (b) SQUID on tip (SOT) (c) SQUID on lever (SOL). Illustration from Ref. [CHR24] (d) Scanning SQUID imaging of superconductors. (top left) Local mutual inductance measurements applied to a superconductor. Current in a loop concentric to the SQUID's pickup coil (SOC design) generates magnetic fields at the sample. The superconductor then generates screening currents, which produce fields detectable by the SQUID. (inset) Other magnetic field profiles typical in superconductors: distorted field lines near the edge, due to the Meissner effect, and a vortex. These are imaged by DC magnetism mode of all SQUID designs. Illustration from Ref [PER22].

### 14.2 Current and Future Challenges

One of the main applications of SSM over the years has been the study of superconductors. For example, in the Meissner state, the presence of screening currents can be detected by SSM, providing direct access to the London penetration depth. See illustration in Fig. 14.1(d) and measurement in Fig. 14.2(a). Additionally, SSM reveals the positions and dynamics of superconducting vortices. These vortices can be observed as they move, interact, or get pinned by defects, particularly under changing temperature or magnetic field conditions. Imaging these phenomena enhances our understanding of flux pinning, a critical factor in improving the performance of superconducting materials in applications like quantum computing. In what follows we discuss three topics where SSM has played an important role. Citations to the original studies can be found in the reviews we cite [KOL17, PER22, CHR23].

Superfluid stiffness, or the resistance of a superconductor to phase disturbances, can be directly detected by SSM, as shown in Fig. 14.2(a). By mapping the superfluid stiffness, the SSM can detect local variations that stem from defects or inhomogeneities. These measurements can be conducted at various temperatures, tracking the evolution of the penetration depth towards the critical temperature, Tc. These measurements provide local identification of Tc, and crucial information about the structure of the gap or the presence of multiple superconducting gaps. The gap structure, which expresses itself in the temperature-dependent data, can, for example, resolve between a full and a nodal gap, and identify 2D superconductors by the shape of the Berezinskii-Kosterlitz-Thouless transition.

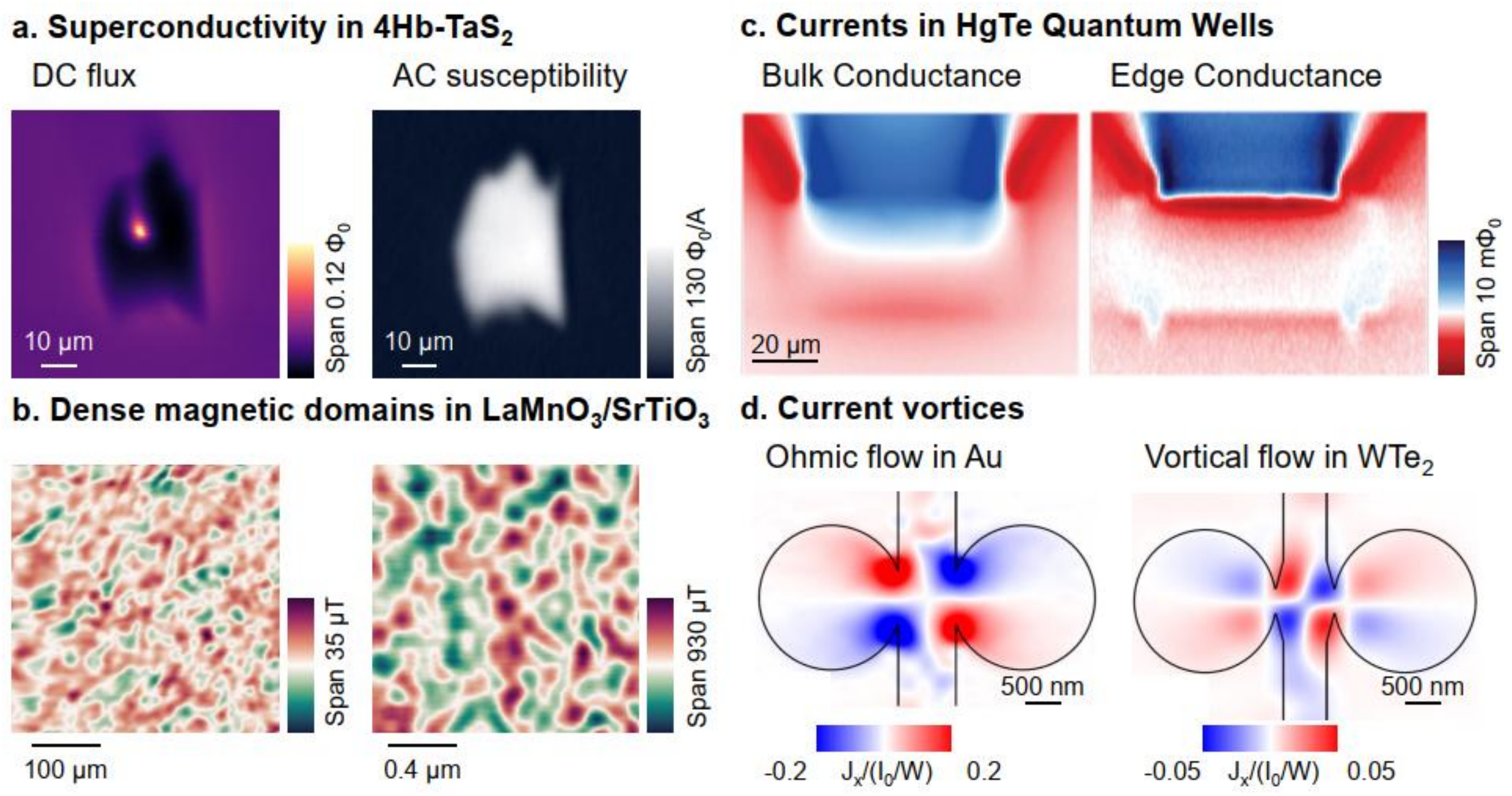

**Figure 14.2: Examples of SQUID imaging**. (a) SSM on a superconducting flake of 4Hb–$TaS_2$. The DC flux image shows the Meissner effect and a vortex in a superconducting flake. The AC susceptibility images the local diamagnetic magnetic response. (b) Magnetic domains in $LaMnO_3/SrTiO_3$ heterostructure imaged by an SOC (left) and SOT (right). (c) SSM imaging of magnetic fields produced by a transport current flowing in an HgTe quantum well Hall bar. The images show conduction through the bulk (left) when the system is in a metallic regime, and conduction along the edge (right) when the bulk is tuned into an insulating regime via electrostatic gating. (d) Current density $J_x(x,y)$ in Au (left) and $WTe_2$ (right). The current shows laminar flow in Au, and a closed loop vertical flow, indicative of a hydrodynamic nature, in $WTe_2$. Illustration from Ref [CHR24].

There are different ways to measure the superfluid density. One method consists of measuring the magnetic susceptibility as a function of the distance. This method uses an on-chip coil as a field source to apply small fields (typically 0.001-0.1 Gauss), the superconductor eliminates the locally applied field, and the SQUID detects the extent of this elimination, reflecting the local superfluid density (up to Pearl length of millimeters). By measuring this with a lock-in technique, this measurement is done in parallel to DC magnetometry, or AC current flow, images. Another method consists of measuring the decaying magnetic field profile around a vortex or directly imaging the Meissner shielding of the sample. In the bulk, the characteristic length scale of the exponential decay is a direct measure of the London penetration length from which the superfluid density can be calculated. For ultra-thin samples the Pearl length can be measured.

**Vortex physics.** A vortex consists of the local suppression of the superconducting order parameter over a characteristic length corresponding to the coherence length, which allows the magnetic field lines to thread the sample in that area generating a quantum of flux, as illustrated in Fig. 14.1d. The vortex matter has been a fertile field of investigation since their theoretical prediction nearly 70 years ago. In principle, vortices are expected solely in type-two superconductors. However, in practice, they are relevant in nearly all superconducting thin films, as well as any high-$T_c$ superconductor. Their role is dominant in many aspects, from dissipation in superconducting magnets and power transmission lines to decoherence in superconducting qubits.

SSM resolves the magnetic field distribution around a vortex emerging from circulating screening currents, as shown in Fig. 14.2(a). For example, the presence of vortices with non-integer number of $\Phi_0$ is evidence of a non-trivial current phase relation as discussed below. SSM is more adapted for imaging larger fields of view (~ tens of microns) or for larger bandwidth measurements [CUI17]. As a result, SSM can image flux flow with a bandwidth ranging from a few Hz to GHz over a length scale of many microns.

**Exploring unconventional superconductivity and exotic order parameters.** Scanning SQUID microscopy has proven invaluable in exploring unconventional superconductors and topological states, particularly through its ability to detect minute magnetic signatures and phase fluctuations. For unconventional superconductors, SSM enables the direct observation of exotic states like chiral superconductivity and spontaneous vortices that arise in materials where time-reversal symmetry is broken. This sensitivity allows tracking fluctuations near phase transitions and allows studying the coexistence or competition between superconductivity and other electronic orders.

SSM's current-phase relation (CPR) measurements are crucial for identifying unique pairing symmetries and unconventional behaviors in Josephson junctions. These measurements require a ring-shaped geometry, where a superconducting loop encloses the junction, enabling precise detection of phase variations. The scanning SQUID locally induces alternating current in the ring and tracks it sensitively. CPR analysis can detect phase shifts, such as π-junction behavior, or unusual periodicities indicative of nontrivial order parameters like chiral or p-wave symmetries, resolve between full and nodal gap, and examine two-component SC. These phase-sensitive observations are key to detecting broken time-reversal symmetry and uncovering exotic pairing mechanisms [PER22, HIL03].

Combining SQUID microscopy with stress measurements enables the study of strain-sensitive properties in superconductors, particularly regarding how mechanical distortions impact the superconducting order. By applying uniaxial strain, researchers can observe shifts in the critical temperature, alterations in vortex behavior, and changes in the CPR that might indicate a change in the superconducting order parameter. For instance, applying stress can help distinguish between different types of pairing symmetries, as certain unconventional superconductors exhibit strain-dependent characteristics, which can reveal chiral or other exotic orderings. Strain also affects other electronic orders visible to SSM, like magnetism, shown in Fig. 14.2(b).

### 14.3 Advances in Science and Technology to Meet Challenges

With recent advancements, SSM is positioned to push the boundaries of superconductivity research. As SSM continues to evolve, the following future directions highlight potential developments:

- **Improved spatial resolution and bandwidth**: Developing nanoscale SQUIDs with higher sensitivity and spatial resolution will allow for detailed imaging of superconducting states. Enhancing bandwidth will also enable real-time measurements of magnetic noise, capturing dynamic superconducting behavior.

● **Multifunctional probes**: Integrating SSM with other on-chip sensors and stimuli (e.g., local gates, stress, or microwave excitations) could enable simultaneous measurements of multiple material properties.

● **Correlated measurements with STM**: Combining SSM with atomic-scale imaging techniques, such as Scanning Tunneling Microscopy (STM), allows for direct correlation between local electronic structures given by the order parameter modulus in STM and the phase sensitive measurement given by the SSM, thereby opening new pathways in superconductivity research.

These advancements promise to expand the scope of SSM beyond traditional superconductivity studies, enabling in-depth examination of the physics behind quantum materials and providing robust tools for nanoscale device characterization.

### 14.4 Concluding Remarks

SSM has demonstrated itself as an invaluable tool for probing the behaviors of superconducting materials at the nanoscale. From imaging vortices and mapping superfluid density to revealing unconventional superconducting orders, SSM enables a deep understanding of superconductivity in nanodevices. Continued advancements in SSM technology are set to broaden its applications, making it accessible to a wider scientific community. As superconductivity research pushes the frontiers of quantum technology, SSM will remain a critical tool for exploring nanoscale superconductors' unique phenomena and functionalities.

## 15. Cryogenic Scanning Hall Microscopy

*Simon Bending*

University of Bath, United Kingdom

### 15.1 Status

The use of scanning Hall probes to image magnetic materials dates back over 60 years to when micrometer-based stages were first employed to map samples with thin film Bi or InSb sensors, achieving typical spatial resolutions of a few micrometers. The invention of high mobility modulation-doped semiconductor heterostructures and the development of piezoelectric nanoscale positioning systems in the late 70s and early 80s subsequently revolutionised the field. Present-day scanning Hall microscope systems have evolved to be able to precisely map the surface of a magnetic sample with nanoscale Hall effect sensors. By recording the calibrated Hall voltage at each pixel one can reconstruct a high-resolution image of the local magnetic induction. At cryogenic temperatures this approach has been widely used to map the intermediate state in type I superconductors and the mixed state in type II materials. Individual fluxons can readily be resolved in modern microscopes (c.f., the images of vortices in a Nb single crystal in Fig. 15.1(b)-(f)) allowing vortices and vortex patterns to be investigated in conventional and high-$T_c$ superconductors. Measured vortex profiles can be fitted with theoretical models to extract the 'local' magnetic penetration depth and flux content, revealing, for example, the signatures of unconventional electron pairing, non-monotonic vortex-vortex forces and strong crystalline anisotropy. The coherence length also plays an important role in determining vortex profiles but is extremely short in most high-$T_c$ superconductors and generally very challenging to extract from scanning Hall microscopy images. In addition, the interaction of vortices with antidots and ferromagnetic dots has been widely explored in nanostructured films, yielding key insights into microscopic flux pinning mechanisms in superconductors.

Surface tracking protocols based on the integration of sensors with scanning tunnelling microscopy tips as well as atomic force microscopy cantilevers and tuning forks have all been demonstrated. Scanning Hall probe imaging remains of enduring importance because the sensors are almost completely non-invasive, can be used over a wide range of temperatures (0.3–300 K) and magnetic fields (0–7 T) and yield quantitative maps of one component of the magnetic induction. In this way scanning Hall microscopy provides an excellent complement to other real space imaging technologies such as scanning SQUID microscopy (c.f., Section 14) and scanning tunnelling microscopy (c.f., Section 16). Furthermore, when operated in 'gating mode' with a large potential difference between sensor and sample, they can be used to generate maps of the surface topography which can be spatially correlated with magnetic field images. GaAs/$Al_xGa_{1-x}As$ two-dimensional electron gas (2DEG) sensors remain the workhorse for cryogenic imaging (c.f., Fig. 15.1(a)) due to their extremely large low temperature electron mobilities (>$10^7$ $cm^2/Vs$) and low carrier concentrations (typically ~$1.5x10^{11}$ $cm^{-2}$) and the operation of sub-micron sensors with minimum detectable field $B_{min}$ < 0.1 $\mu T/Hz^{0.5}$ has been demonstrated at cryogenic temperatures [1]. However, the noise level of sensors increases dramatically at room temperature when their performance becomes rather poor. Their ultimate spatial resolution is also limited to 100 nm due to the fact that the active 2DEG is approximately this distance below the surface of the wafer. Narrow gap semiconductor heterostructure sensors have also been widely explored due to their much higher 300K mobilities, but these have few advantages over GaAs/$Al_xGa_{1-x}As$ at cryogenic temperatures. In recent years, attention has turned to graphene due to the very low, tuneable carrier density of this atomically thin material combined with the extremely high mobilities arising from its unique band structure. While the performance of CVD graphene sensors is limited by scattering from charge centres in the adjacent substrate, this can be suppressed by the encapsulation of graphene in two layers of hexagonal boron nitride (hBN). This raises the exciting prospect of high sensitivity imaging with encapsulated graphene Hall sensors whose spatial resolution is ultimately limited by the ~10 nm thick hBN capping layers.

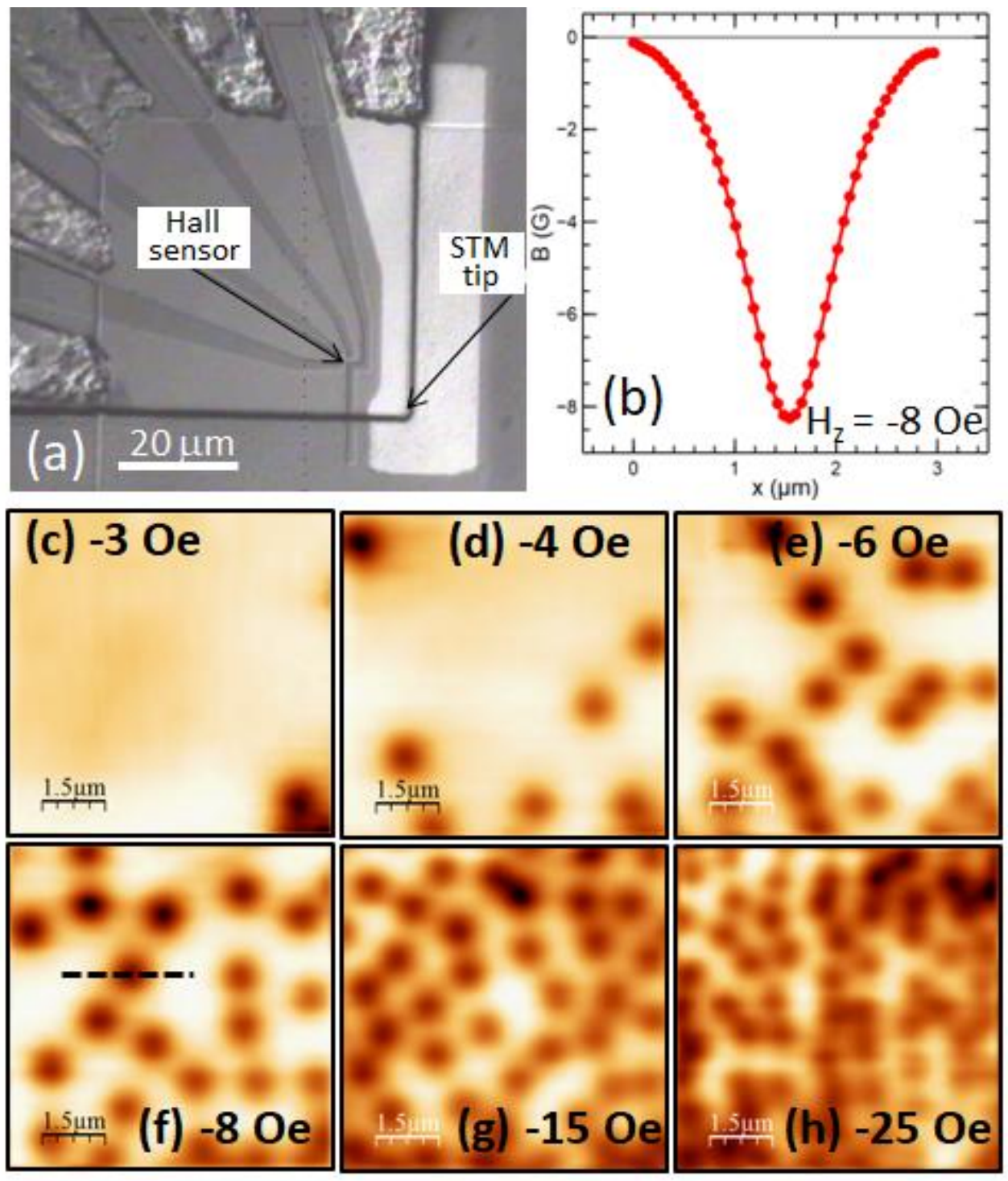

**Figure 15.1**: Optical micrograph of a sub-micron GaAs/$Al_xGa_{1-x}As$ Hall sensor with an integrated STM tip. (b) Field profile across a vortex in a Nb single crystal at T= 5K along the horizontal dashed line indicated in (f). (c)-(h) Scanning Hall images of vortices in a high quality Nb single crystal at various indicated negative applied fields and T= 5K.

## 15.2 Current and Future Challenges

Developing scanning Hall probe systems for different operation conditions is primarily a Hall sensor materials optimisation problem. The signal:noise ratio (*SNR*), which is limited by Johnson noise at frequencies well above the $1/f$ noise corner, is a key figure-of-merit that determines the minimum detectable field of the sensor:

$$\mathrm{SNR}=\frac{I_{\max}R_{\mathrm{H}}B}{\sqrt{4k_{\mathrm{B}}TR_{\mathrm{V}}\Delta f}} \propto \sqrt{\frac{\mu}{n_{2\mathrm{d}}}}I_{\max}, \qquad (15.1)$$

where $R_H$ is the Hall coefficient, $R_V$ is the output resistance at the voltage contacts, $\Delta f$ is the measurement bandwidth and $\mu$ and $n_{2d}$ are the two-dimensional carrier mobility and concentration respectively. $I_{max}$ is a maximum empirical operation current which, at high temperatures, is generally limited by the saturation drift velocity. At low temperatures electron transport in the sensor can be quasiballistic and $I_{max}$ is associated with a rapid increase in $1/f$ noise which can have several different origins, e.g., random telegraph-like events due to the trapping and emission of `hot' electrons at deep traps. Preserving the high carrier mobility after pattering nanoscale Hall sensors is a particular challenge, as microscopic noise processes linked to edge disorder can greatly increase 1/f noise levels. The ultimate spatial resolution of a sensor is governed by a combination of the width of the active Hall cross ($w$) and the sample-sensor spacing ($h$), which can be approximately combined in an effective resolution $\approx(w^2+h^2)^{1/2}$. In semiconductor heterostuctures the minimum sensor width is typically limited to ~100 nm by sidewall depletion after etching, while the minimum scan height is also set to ~100 nm by the required remote doping and capping layers above the active 2DEG in the heterostructure stack. This is an area where 2D semiconductors, e.g., graphene, can play a very important role as the vertical height of these atomically thin structures allows extremely small (≥10 nm) sample-sensor spacings to be achieved. In this case the spatial resolution is limited by the narrowest wire widths one can pattern without significant reduction of the carrier mobility. One final important criterion for a Hall sensor is that the `offset' resistance, which notionally arises due to slight misalignment of the opposing Hall contacts, be very small. In reality this is nearly always due to microscopic inhomogeneities in the sensor material which can lead to large 'offset' voltages' that must be electronically compensated during imaging. This offset resistance can significantly degrade minimum detectable fields beyond predictions based on equation (15.1) and must be kept to a minimum by optimising the spatial homogeneity of the carrier system.

### 15.3 Advances in Science and Technology to Meet Challenges

Recent developments in graphene heterostructures have the potential to significantly impact cryogenic imaging, combining extremely high spatial resolution with excellent minimum detectable fields. The highest carrier mobilities have been achieved when mechanically exfoliated graphene is encapsulated between two thin (~10 nm) layers of hexagonal boron nitride (hBN) [2] since this suppresses scattering at charge centres in adjacent oxide layers on, e.g., Si/$SiO_2$ substrates. Low temperature mobilities in excess of 140,000 $cm^2$/Vs have been demonstrated [3] with $B_{min}$ estimated to be as low as 80nT/$Hz^{0.5}$ at 4.2K in a Hall cross based on the intersection of two 1µm wide leads [4]. Moreover, these devices can be tuned with a back gate allowing $B_{min}$ to be optimised just either side of the charge neutrality point. To date magnetic imaging has only been demonstrated with semi-encapsulated (top side encapsulated) nanoscale sensors at room temperature as illustrated in Fig. 15.2 [5]. Fig. 15.2(a) shows an optical image of a micron-sized scanning Hall sensor patterned in a monolayer graphene flake deposited onto pre-patterned Au contacts and excapsulated with hBN. Fig. 15.2(b)-(e) demonstrate the use of the scanning sensor to image stripe domains in a ferrimagnetic YIG film at 295K, something that would not be possible with GaAs/AlGaAs 2DEG Hall sensors which have very much higher noise levels at room temperature.

From a fabrication perspective, graphene Hall sensor sizes comparable to the 10 nm thickness of hBN capping layers should be readily achievable, in stark contrast to the minimum size of GaAs/AlGaAs probes which is limited to about 100 nm by sidewall depletion. This will allow vortex images to be captured with very much higher spatial resolution and vortex patterns to be resolved up to much higher magnetic fields. An ultimate limit on the spatial resolution will be set by the opening up of a significant band gap in graphene nanoribbons (GNRs) with line-widths below 10 nm [6]. However, there are reports of a substantial reduction in the carrier mobility of GNRs for line widths below 60 nm due to the onset of strong edge scattering [7] and It remains to be seen if this can be suppressed in encapsulated graphene devices. Scanning Hall microscopy requires the integration of a second sensor to control the scan height during imaging and map the surface topography, and the combination of the two complementary data sets yields new information about the sample. The ultimate spatial resolution can be obtained by patterning Hall sensors directly on the end of AFM tips, and there have been several attempts to achieve this [8]. While this approach has proved to be very challenging, one can anticipate future developments in this area in order to optimise the imaging spatial resolution. A major challenge to the widespread adoption of encapsulated graphene Hall sensors is the labour-intensive device fabrication, involving several mechanical exfoliation and dry transfer processes [9]. However, recent advances in CVD thin fim growth have enabled the realisation of both graphene/hBN heterostructures [10] as well as graphene/hBN/graphene trilayers [11], raising the possibility that the direct CVD growth of hBN/graphene/hBN trilayers may become possible in the near future. This would allow the ready wafer scale processing of encapsulated graphene sensors.

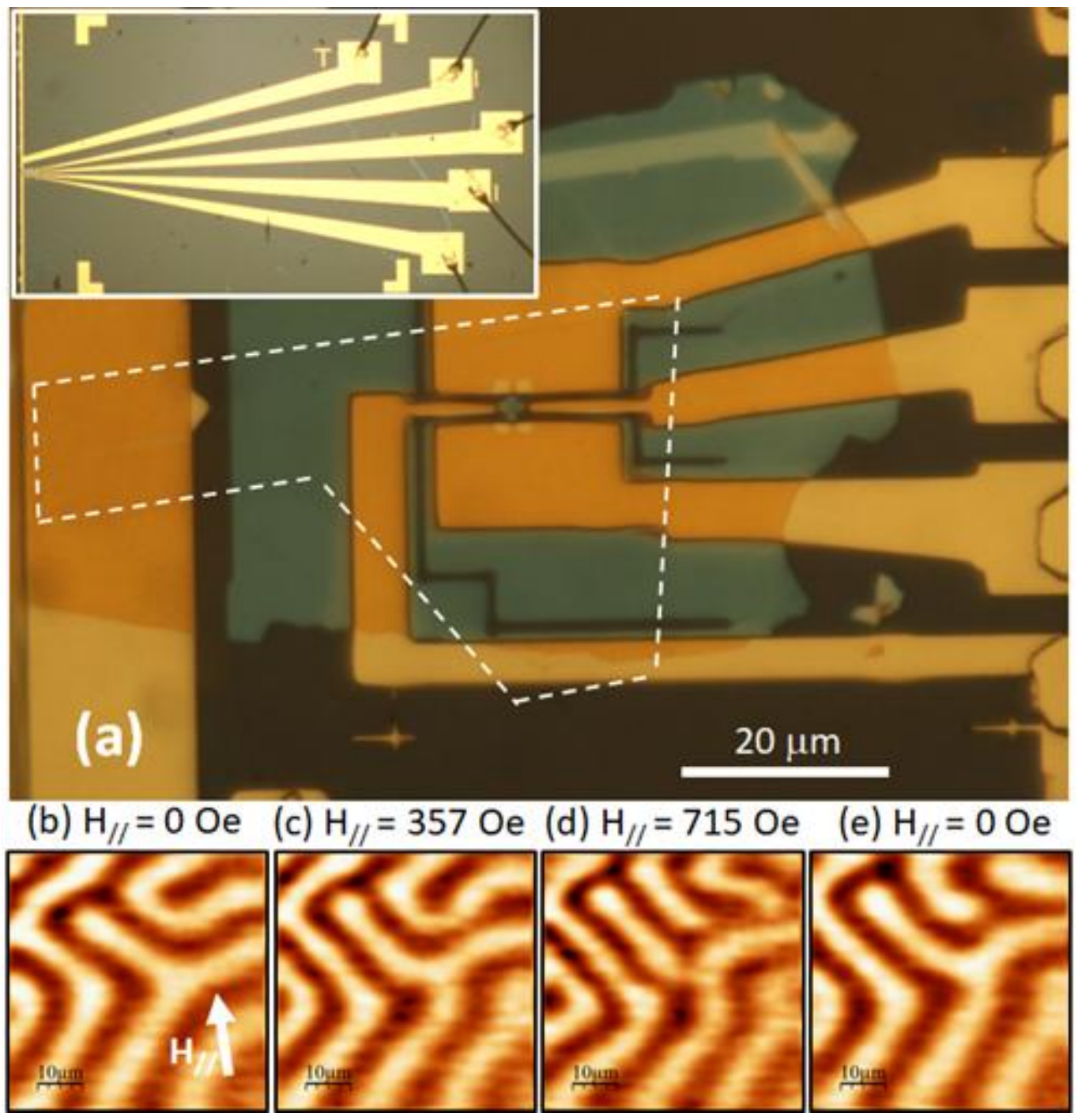

**Figure 15.2**: Optical micrograph of semi-encapsulated graphene Hall sensor based on the intersection of two 1 µm wide wires. The inset shows an image of the diced and wire bonded sensor chip. (b) – (e) Scanning Hall images of a ferrimagnetic YIG film showing the characteristic labyrinth magnetic domain captured at 295K after the in-plane field [parallel to the arrow in (b)] was increased from zero to 715 Oe and back to zero again [5].

### 15.4 Concluding Remarks

The current state-of-the-art of cryogenic scanning Hall probe microscopy has been described. The technique provides a valuable complement to other scanned probe magnetic imaging techniques; it is non-invasive, quantitative and can be used over a wide range of magnetic fields and temperatures. Considerable efforts are being devoted to the exploration of new Hall sensor materials for system optimisation under different experimental conditions, e.g. at room temperature and for high spatial resolution. Of particular current interest is the use of encapsulated graphene sensors which combine very high carrier mobilities with strong potential for deep nanoscale pattering. However, many challenges remain to be solved including the development of wafer scale fabrication approaches and the preservation of high carrier mobilities in devices based on leads a few nanometers wide.

**Acknowledgements**
S.J.B. acknowledge financial support from EPSRC in the UK under grant numbers EP/X015033/1 & EP/W022680/1 and the Superqumap COST Action CA-21144.

## 16. Scanning tunnelling microscopy and spectroscopy in superconductors

*H. Suderow[1], P. Liljeroth[2], A. Hassanien[3], I. Guillamón[1], E. Herrera[1]*

[1]Universidad Autónoma de Madrid, Spain

[2]Aalto University School of Science, Finland,

[3]Jozef Stefan Institute, Ljubljana, Slovenia

### 16.1 Status

Traditionally, superconductivity has been viewed as a bulk phenomenon, characterized by a bulk order parameter which is given by the Cooper pair wavefunction in Ginzburg-Landau theory. Below the critical temperature $T_c$, electrons condense into Cooper pairs and a gap opens in the density of states close to the Fermi level. The superconducting gap is related to the amplitude of the Cooper pair wavefunction. In a Scanning Tunnelling Microscope (STM) a metallic tip is brought into tunnelling distance to a sample. The tunnelling current depends on the density of states of the sample and can be used to trace changes in the electronic properties at very small length scales, visualizing atoms and other atomic scale changes in the electronic properties. The tunnelling conductance can be measured as a function of the bias voltage and provides at sufficiently low temperatures a direct measurement of the electronic density of states as a function of energy. Thus, very soon after its invention, STM was used to observe the energy dependence of the superconducting density of states and trace the spatial dependence of the superconducting gap.

In a superconductor under a magnetic field, vortices enter the sample. The gap decreases inside the vortex cores, making the STM a powerful tool to obtain vivid images of the superconducting vortex lattice [1-4]. These possibilities defined for almost four decades the objectives of many STM studies in contributing to understand the bulk properties of superconductors. The STM has been very successful, determining the superconducting gap structure of dozens if not hundreds of compounds, the properties of superconducting vortex lattices and of vortex cores, and showing how superconductivity is suppressed at magnetic impurities down to atomic scale [3-5]. However, STM probes the local density of states (LDOS) of top atomic layer of the sample and does not directly access the bulk properties. The surface is important by itself and is in principle a two-dimensional electronic system presenting sometimes distinct features from the bulk.

The distinct features between surface and bulk properties are due to the abrupt changes in the atomic potential when ending the solid at the surface. Often, semiconductors present a crystalline structure at the surface which is different from the bulk. Dangling bonds re-arrange at the surface giving another atomic arrangement than in the bulk [6]. By contrast, bonding in metals is fundamentally different. Thus, quite often the surface only shows slight changes in the atomic positions, consisting for example in a slight extension or compression of the lattice constant along an axis perpendicular to the surface. The in-plane symmetry often remains the same as the bulk one. This leads to a surface electronic band structure similar to the bulk electronic band structure, partly justifying the extended trend of viewing STM as indicative of the bulk behaviour in superconductors. However, metals also present surface states in gaps of the bulk electronic band structure. Electronic surface states build a two-dimensional electron gas which is, to some extent, decoupled from the bulk.

For a long time, these surface states have been considered to have single electron particle properties and not able to present correlated electron states. Very recently, advances in surface and sample preparation and improved STM techniques have shown that two-dimensional correlated electrons, 2DCE, exist in different systems and can be superconducting [7-10]. These 2DCE can be visualized successfully and studied precisely using the STM.

### 16.2 Current and Future Challenges

Whereas bulk and surface of a conventional BCS superconductor are expected to be very similar, this is no longer the case for unconventional and topological superconductors. A case known since long is the one of d-wave superconductors. Andreev reflection of BCS quasiparticles at a flat surface mixes order parameter components of different sign and is pair breaking. Therefore, an in-gap state is expected at the surface of a d-wave superconductor. This is rarely observed, as it requires flat surfaces oriented in such a way as to produce sign change in the underlying superconducting order parameter. As explained in Section 1, the bulk boundary correspondence of topological materials provides a surface state which is protected by topology. For example, the edge states of the Quantum Hall effect lead to edge conduction channels that remain ballistic, independent of the level of disorder, due to the chiral edge conduction. Contrary to the states at the surface of a d-wave superconductor, in a topological superconductor, edge modes have distinct Cooper pair and spin conduction properties which are protected by topology.

There are several routes to obtain such a topological superconductor (see also Section 1). One is by taking a usual superconductor to induce by proximity superconductivity in a topological insulator (Fig. 16.1a). Another one is by taking an unconventional bulk superconductor and studying its boundary modes (Fig. 16.1b). In both cases, STM is the most adequate technique to unveil topologically protected boundary modes, and both cases require careful consideration of the surface electronic band structure. For a topological insulator, it is crucial to isolate surface behaviour, for example in the form of crystalline reconstructions, from bulk boundary modes. In an unconventional bulk superconductor, it is key to demonstrate that the strong electronic correlations leading to unconventional superconductivity also govern the properties of the two-dimensional surface states. On the other hand, in van der Waals materials, the interlayer bonding is very weak. Thinning a van der Waals material down to the monolayer limit leads to a two-dimensional electronic system (Fig. 16.1c). The observation of strong correlations in single layers of transition metal dichalcogenides, giving heavy quasiparticles, opens the way to address a third path towards unconventional superconductivity in two dimensions [9,10].

### 16.3 Advances in Science and Technology to Meet Challenges

A key aspect allowing to understand the electronic properties of two-dimensional electronic systems comes with the unique possibility of creating quantum dot states. Because the surface behaves as a two-dimensional electron gas, lateral quantization leads to the formation of discrete states in between steps or in confined geometries [9,10]. We can distinguish two situations. In the first, discrete states are formed by lateral quantization of a two-dimensional electron gas at the surface of on an uncorrelated metal. Using typical lateral distances, the separation between energy levels (several tens of meV) is usually much larger than the superconducting gap (of the order of a meV, Fig. 16.1.d).

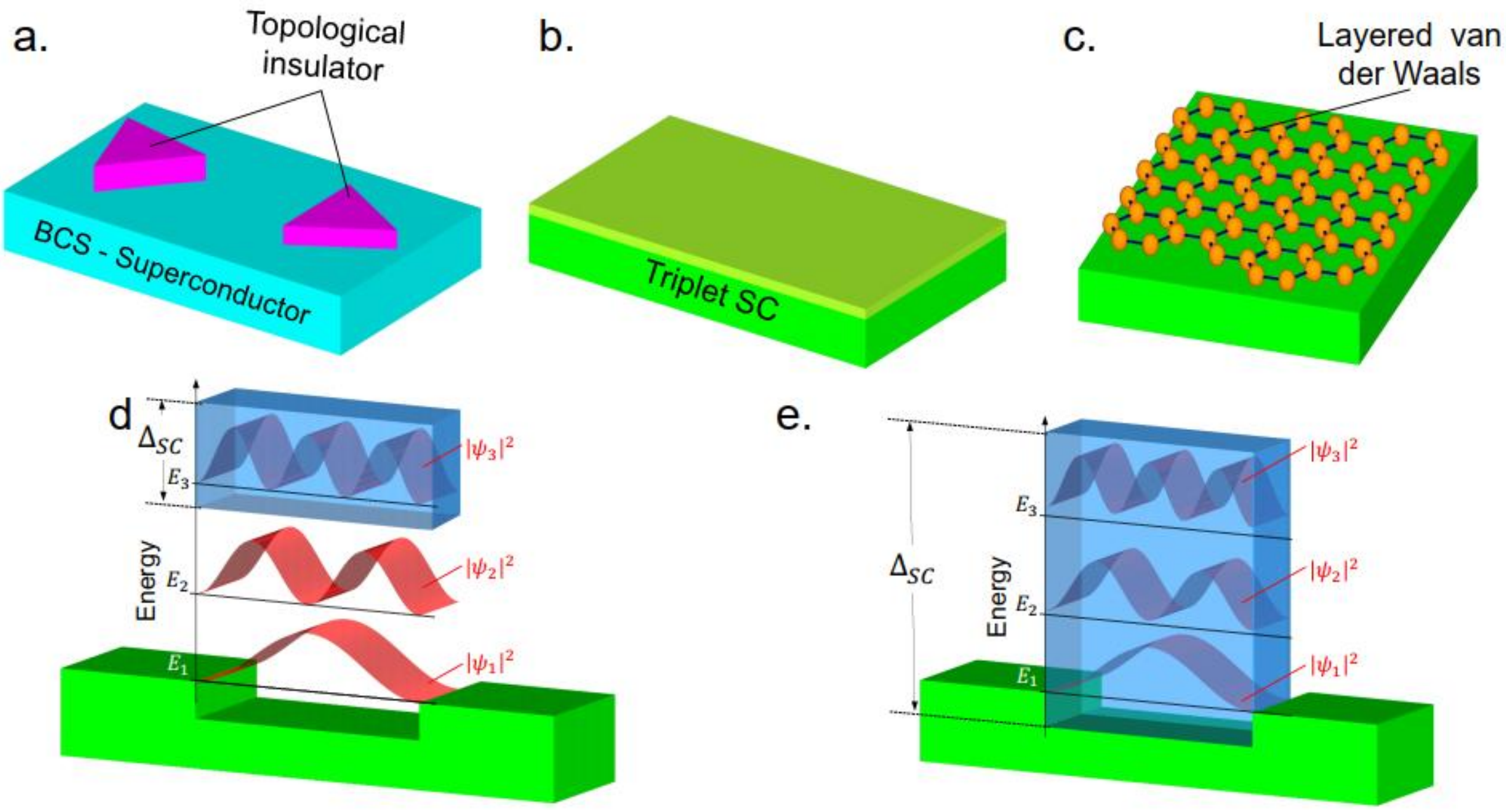

**Figure 16.1** (a-c) Possible routes to topological superconductivity. (a) Cartoon of a topological insulator (violet) grown on an atomically flat s-wave BCS superconductor (blue). (b) Triplet superconductor (green) hosting topologically protected boundary states at the surface (light green). (c) Strongly correlated two-dimensional layered material on an insulating substrate (green). (d) Quantum well states of a normal metal at the surface form by lateral confinement on a step. The step at the surface is shown in green. The quantum well states schematically in red. In blue we highlight the superconducting gap opening. Eventually, a single quantum well state might be present. (e) Quantum well states in a highly correlated metal with heavy fermions. The separation in energy is much smaller, so that there is a much stronger interaction between quantum confinement and superconductivity, shown schematically by the gap opening in blue.

However, an energy level positioned inside the superconducting gap creates a spin degenerate Andreev in-gap state, showing that the two-dimensional electron gas can indeed carry superconducting correlations [9]. On the other hand, a surface state of a heavy fermion system (Section 7) leads to energy level separation which is comparable to the superconducting gap (in both cases a fraction of a meV, Fig. 16.1e), giving a 2DCE. Here, the finite lifetime of bound states modifies the proximity induced superconducting gap (Fig. 16.2a-c) [10]. Furthermore, edges show electronic symmetry breaking at the surface [10]. These results provide new ways to explore the connection between bulk and surface properties in unconventional superconductors.

Naturally occurring heavy fermion materials are a rich playground for unconventional superconductivity. However, two-dimensional materials open completely new opportunities to engineer systems to exhibit heavy quasiparticles and potentially unconventional superconductivity. As illustrated in Fig. 16.2d-g, it is possible to create heterostructures that combine the key ingredients of heavy fermion materials, namely mobile conduction electrons and a lattice of localized magnetic moments [8]. Kondo coupling between these creates a Kondo lattice that is essentially responsible for the physics of heavy fermion systems. In contrast to traditional materials, the Kondo lattice and the heavy fermion hybridization gap can be probed separately in these systems. When the layer with the magnetic moments is on the top (in this case Mott insulator 1T-$TaS_2$), the STM probes the LDOS of this

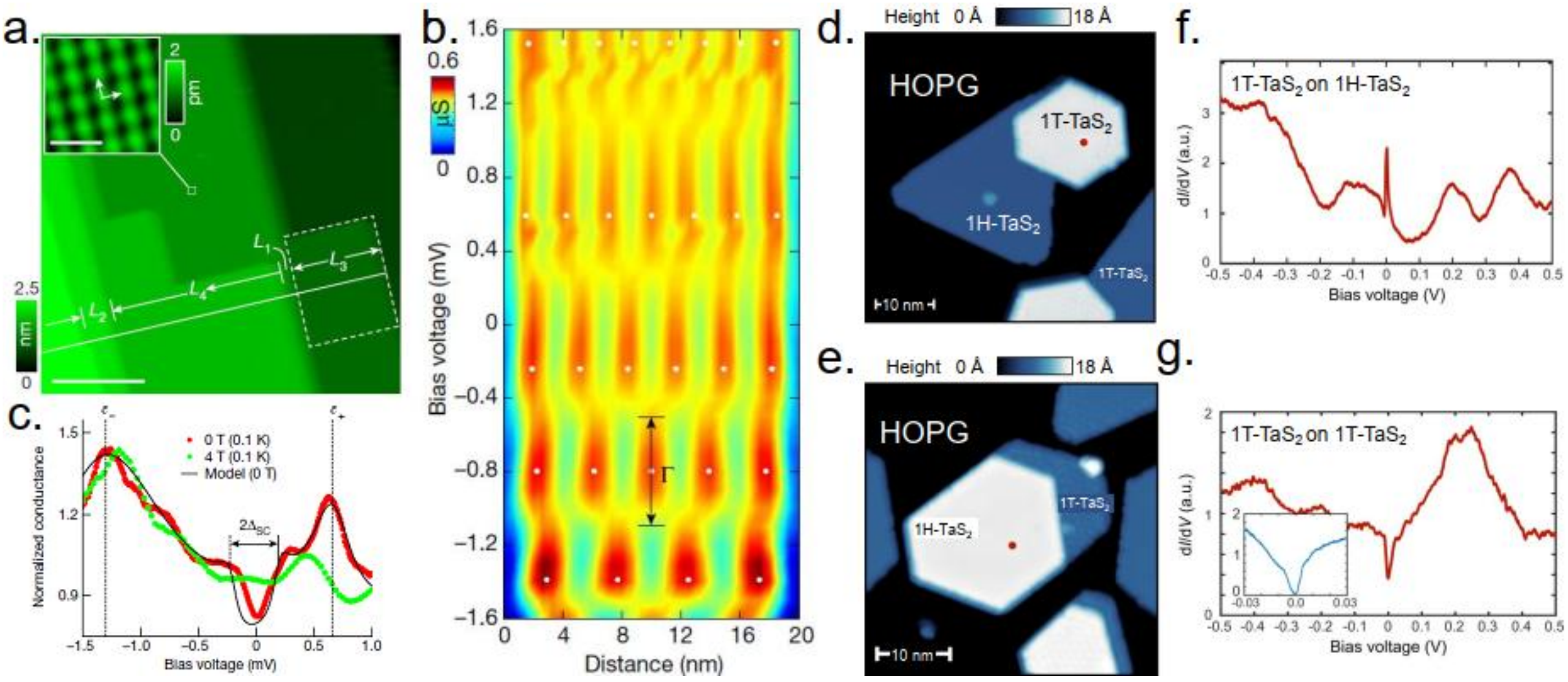

**Figure 16.2** (a) Topographic image of URu2Si2, showing atomically flat terraces marked by $L_{1\text{-}4}$. The inset shows an atomic resolution image of the U square lattice. White scale bar is 10 nm long. (b). Tunnelling conductance is shown as a colour scale, as a function of bias voltage and distance, on terrace $L_3$. White dots show the position of quantum well states formed by heavy fermion two-dimensional quasiparticles. We schematically show the broadening by interactions as Γ in black. (c) Tunnelling conductance as a function of the bias voltage at very low voltages (zero field in red and 4 T in green). The proximity induced calculated superconducting density of states is shown as a black line. (d,e) Epitaxially grown layers of 1T-TaS2 and 1H-TaS2. (f,g) Tunnelling conductance vs bias voltage obtained at the red points shown in (d,e). The inset in (g) shows a zoom into small voltage range. Reproduced from [7,10].

layer and spectroscopy shows a Kondo resonance at the Fermi level (Fig. 16.2f). Alternatively, when the layer with the itinerant electrons is the top layer (in this case metallic 1H-$TaS_2$), we can observe the heavy fermion hybridization gap at the Fermi level. The goal of the future research in this direction is to probe the heavy fermion phase diagram (transition between heavy fermion behaviour and magnetic order, see [11]) and trying to realize superconductivity in doped artificial heavy fermion systems.

Studying 2DCE requires advances in the possibilities of STM. The position and location of atomic size structures can be manipulated with the tip to study the spatial dependence of proximity induced superconductivity [12]. The spatial resolution can be enhanced by functionalized tips [13]. The energy resolution is typically enhanced using superconducting tips [5]. On the other hand, while the usual density of states measurement provides an indirect coupling to the Cooper pair wavefunction through the density of states, the Josephson critical current is directly proportional to the Cooper pair wavefunction. However, the Josephson effect between two superconductors is strongly smeared by thermal fluctuations in atomic scale STM junctions. Recently, it was shown that a feedback driven junction sustains spontaneous oscillations of the Cooper pair current, notably enhancing the possibilities of atomic scale Josephson measurements [14] in view of the detection of modulated superconducting states as pair density waves [15].

### 16.4 Concluding Remarks

The advent of 2DCE is key to identify and study topological superconductivity. A notable remaining issue is the observation of vortices in superconducting 2DCE. Under some conditions, these might carry Majorana modes at the core. More importantly, however, is that the uniqueness of the value of the Cooper pair wavefunction at a single location is the defining characteristic of a superconductor, and more generally of a macroscopic quantum coherent system. Observing a superconducting gap or a peak in the conductance at low bias similar to the one expected from the Josephson effect in a 2DCE does not demonstrate quantum coherence. Instead, the observation of a vortex or a vortex lattice does so, as the flux quantization requires a unique Cooper pair wavefunction. Vortices in 2DCE continue to be very elusive and efforts to observe these are very difficult using magnetic imaging techniques, because the magnetic field on a two-dimensional superconducting system is nearly homogeneous (the penetration depth being very large, see Section 14). STM can probe the density of states and the Josephson current, being much more adequate to identify vortices in low dimensions.

On the other hand, a significant challenge is to unite efforts made at atomic scale with making devices with transport properties linked to the 2DCE. The fabrication of contacts that allow driving a current through a 2DCE is a relevant challenge. Recent advances in two-dimensional interfaces (Sections 2 and 9), in materials (Sections 3 and 6) and in devices (Sections 17 and 22) show that the control over interfaces is leading to surprising new insight which requires interfaces that are close to perfection. Applying these techniques to surface states is certainly not easy, but it could lead to obtain devices based on 2DCE that are integrable into quantum circuits.

**Acknowledgements**

Support by the Spanish Research State Agency (PID2023-150148OB-I00, PDC2021-121086-I00, TED2021-130546B-I00 and CEX2023-001316-M), the European Research Council PNICTEYES grant agreement 679080 and GETREAL grant agreement 101142364 and by the Comunidad de Madrid through program Mag4TIC-CM (Program No. TEC-2024/TEC-380). We have benefitted from collaborations through EU program Cost CA21144 (superqumap.eu).

# III. Superconducting quantum devices

## 17. Electrical tuning of superconducting devices

*Alejandro V. Silhanek[1], Joris Van de Vondel[2], and Anna Palau[3]*

[1]Université de Liège, Liège, Belgium

[2]KU Leuven, Leuven, Belgium

[3]ICMAB-CSIC, Barcelona, Spain

**17.1 Status**

Since the early 1960s, the idea of controlling the superconducting state with an electric field, like its use in semiconductors, has intrigued the scientific community due to its vast potential for technological applications, including transition-edge sensors, memristive devices, superconducting filters, interferometers, and resonators. All-electrical actuation of a superconducting device can be obtained through various approaches depending on the specifics of the target material and the exact nature of the mechanisms at play. As schematically shown in Fig. 17.1, the electric field could be used to (i) modulate the density of carriers (field-effect band bending, panel (a)), (ii) induce a redox reaction at the interface (chemical effect, panel (b)), or (iii) stimulate displacement of the atoms (composition/ stoichiometry effect) by using either a static electrical field (panel (c)) or momentum transfer from the flowing charge carriers (panel (d)).

Unlike semiconductors, conventional (superconducting) materials screen the electric field over a very short distance, i.e. Thomas-Fermi screening lengths $r_{TF} < 0.1$ nm. As a result, the impact on the charge density in films with a thickness exceeding a few nanometers, will be negligible. In these materials, electrical tunability can be achieved by dragging atoms via momentum transfer from flowing charge carriers, thus modifying the system's response in targeted regions with high current density [1]. In order to explore actuation schemes based on charge density modulations, it is necessary to drastically boost the overall impact of the electric field by using ultrathin films [2] or even 2D layered structures [3], apply extremely high electric field intensities [4] or explore materials with a reduced number of charge carriers or a combination of these approaches.

High-critical temperature superconductors (HTS) have a tremendous advantage over conventional superconducting thin films due to their low carrier density and compatibility with high dielectric constant (high-κ) materials. The former implies a relatively large Thomas-Fermi screening length $r_{TF}$ over which band bending occurs. For instance, for $YBa_2Cu_3O_{7-x}$ (YBCO) $r_{TF}$ ~ 1 nm, comparable to the coherence length ~1.5 nm and spanning one unit cell of the material. This fact permitted to expand the working region for achieving effective actuation based on charge density modulations and led to several important breakthroughs in electric field effect methods, such as ion-liquid gating of HTS [5], polymer electrolyte gating [4], ferroelectric gating [6], and direct metallic gating [7] or using intermediate buffer layers to facilitate redox reactions [8]. In these systems, being able to tune the carrier density by 0.2 electrons per unit cell permits to switch between an insulating state and a superconducting state [3]. This is a general feature of correlated electron systems, in which the interplay between several electronic phases can give rise to abrupt changes in the material's properties as a consequence of slight modifications in the strain, chemical composition, and disorder, among many other knobs. In this vein, selective electromigration of oxygen atoms in HTS has been demonstrated to be a viable route to modify the ordering and stoichiometry of the material [9,10].

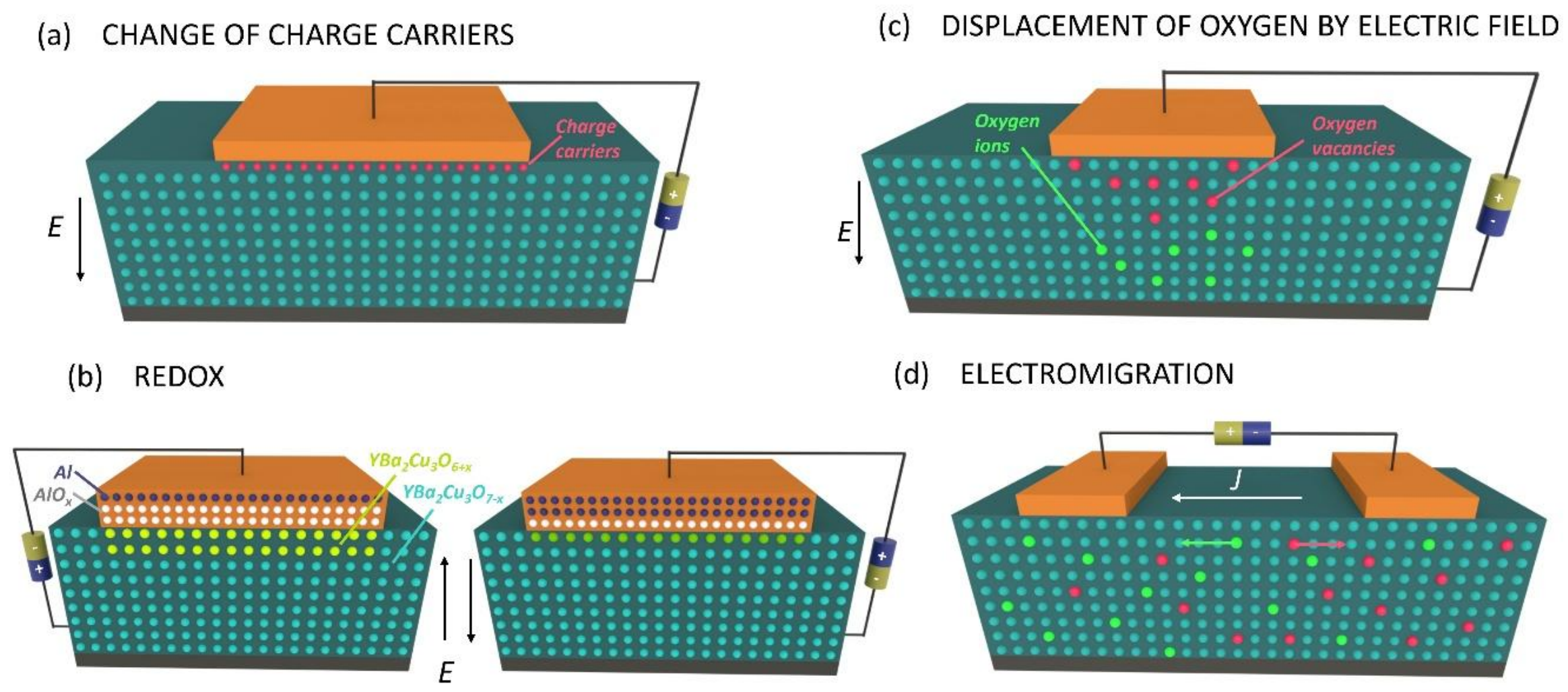

**Figure 17.1**. Different approaches implemented to induce electrical tunning of superconducting copper oxides. (a) electrostatic gating leads to a change of the charge carrier density. (b) When Al and YBCO come into direct contact, a layer of alumina oxide naturally grows at the interface by pumping oxygen from the YBCO structure and thus generating a few nanometers of underdoped YBCO insulating barrier. The volume of the barrier can be controlled by applying a voltage greater than the difference between the reduction potentials of Al and Cu. (c) Electric field induces displacement of oxygen atoms generating oxygen vacancies. (d) Current-induced counterflow of oxygen vacancies and oxygen atoms.

## 17.2 Current and Future Challenges

Electrostatic fields in HTS affecting the carrier density are generally reversible, fast, and permit large switching amplitude. In addition, this method enables the controllable change of the charge carrier concentration without affecting the level of disorder. A non-volatile control of the carrier density modulation may be achieved through ferroelectric polarisation. The challenge, however, lies in the limitation of these materials in reaching a multiplicity of states (beyond binary setting). Electrochemical field-induced doping offers good retention and speed while consuming less power than the resistance switching obtained through electromigration. Concerning device fabrication, electrostatic gating promoting substantial modification of the density of states requires adapted substrates with high dielectric constant, very thin and therefore excluding single crystals, and an excellent surface quality favoring the growth of high-quality epitaxial films.

Electromigration involves a long-distance diffusive process, naturally leading to irreversibilities and high-power dissipation unless very low resistance changes are induced. In addition, there is an inherent stochasticity in the process (similar to dielectric breakdown), giving rise to highly inhomogeneous stoichiometry, which requires extreme miniaturisation to achieve uniformity of the voltage switching response. The advantage of electromigration is the possibility of affecting the stoichiometry in a large portion of the sample volume with a certain degree of reversibility and control, which is appealing for exploring the phase diagram as a function of the atomic species under control [10] but it has limited potential for technological applications. Thus far, the vast majority of the studies focus on p-type copper oxides such as YBCO, and it would be interesting to extend this study to explore the response of n-type oxides such as $Nd_{2-x}Ce_xCuO_4$ (NCCO). In NCCO, superconductivity arises from electron doping of the $CuO_2$ planes, while its crystal structure, characterized by the absence of apical oxygen, modifies the local electronic environment. Consequently, the directionality of the electromigration process is expected to be reversed compared to that observed in hole-doped cuprates.

The current challenge is to apply these techniques to the design of electrically addressable superconducting devices with enhanced functionalities. This will require precise control of the material's switching at low temperatures, where ion and charge mobility may be significantly reduced. In addition, it has been recently reported the possibility of using electric field effect even in conventional superconducting devices [11]. However, the underlying mechanism is still under scrutiny and clearly deserves further investigation.

### 17.3 Advances in Science and Technology to Meet Challenges

An essential advancement fuelling the development of technologically ready devices based on the electric field effect on superconductors should involve their CMOS compatibility. This is a common challenge for complex-oxide functional materials, which possess a wide range of functional properties that depend on the crystal structure. Significant efforts have been made in this area, exploring various alternatives, including directly synthesising oxides on silicon or transferring exfoliated films. A concrete example is the growth of STO thin films on Si wafers. Nevertheless, much work remains to be done, particularly concerning high-temperature superconductors.

Another significant challenge that needs to be technologically addressed is device fabrication. The complexity of HTS materials, coupled with their sensitivity to nanometric defects and oxygen doping (see also Sections 5, 6, 19, 21, 22), places significant demands on micro- and nano-fabrication processes. Thus, nanoscale patterning of devices requires thorough optimization to precisely control the initial doping of the material, which will ultimately determine the device performance.

### 17.4 Concluding Remarks

All-electric gating of superconducting materials remains a promising and powerful approach for achieving reconfigurable electronic states. Energy-efficient and versatile components based on this method can find their way into a wide diversity of tunable superconducting devices (filters, SQUIDs, Josephson junctions, memories, etc.). Yet new exciting multifunctionality may still emerge when combining superconductors with voltage-controlled magnetic anisotropic in ferromagnetic materials. All in all, it is indispensable to improve the interface and material engineering aspects to extract the full potential of this phenomenon.

**Acknowledgements**
The authors acknowledge the COST (European Cooperation in Science and Technology) [www.cost.eu] program through the COST Action SUPERQUMAP (CA 21144) and support from the Fonds de la Recherche Scientifique - FNRS under the grant Weave-PDR T.0208.23. This work is supported by the Research Foundation Flanders (FWO) grant number G0D7723N and MCIN/ AEI PID2021-124680OB-I00 cofinanced by ERDF A way of making Europe.

## 18. Superconducting single-photon detectors

*Ilya Charaev*[1] *and Maria Sidorova*[2]

[1]University of Zurich, Switzerland

[2]Humboldt-Universität zu Berlin and DLR, Germany

### 18.1 Status

Superconducting nanowire single-photon detectors (SNSPDs) emerged in the early 2000s [1], building on niobium nitride (NbN) thin films, and demonstrated the potential for fast and precise detection with exceptionally low noise (dark counts). This significant potential is enabled by the physics behind nanowire operation [1] (Fig. 18.1). Upon photon absorption in a current-biased nanowire, the superconducting state locally collapses, forming a 'hot spot' that grows either through quasiparticle diffusion or vortex crossing until it blocks the current flow. This switches the nanowire to the resistive state, generating a measurable voltage signal, followed by an inherent reset mechanism via electron cooling.

Over the past two decades, the SNSPD technology has progressed significantly. Their ultra-low system timing jitter (3 ps)[2], low dark counts ($6\times10^{-6}$ Hz)[3], gigahertz photon count rate (1.5 Gcps)[4], and high system detection efficiency (98% at 1550 nm wavelength) [5] place SNSPDs at the forefront among existing single-photon detectors, such as photomultiplier tubes (PMTs), single-photon avalanche diodes (SPADs), transition-edge sensors (TESs). This combination of features also established SNSPDs as promising tools for single-photon detection in applications such as quantum optics, secure quantum and space communications.

In quantum key distribution (QKD), SNSPDs enable a readout of cryptographic keys encoded in photon states for secure data exchange. Additionally, their low dark count rates and high efficiency are indispensable in fundamental research experiments, such as dark matter search or test of quantum mechanics. SNSPDs also support advancements in high-resolution imaging and laser ranging (LiDAR), where precise photon detection is critical for applications ranging from autonomous driving to climate monitoring [7,8].

SNSPD technology is expanding well beyond telecom wavelengths; recent advancements now enable detection across a wide spectral range, from X-ray to mid-infrared wavelengths (up to 29 µm[6]). This flexibility results from tailored optimisation of both material composition and nanowire geometries. Mid-infrared detection is especially valuable for applications like exoplanet spectroscopy, infrared astrophysics, physical chemistry, remote sensing, and direct dark-matter detection.

For decades, the material platform of SNSPDs has been limited to low critical temperature ($T_C$) conventional superconductors such as NbN, TaN, MoSi , WSi, MoN, MoGe , Nb, NbSi, NbTiN and NbN/$\alpha W_5Si_3$ with operating temperature well below 4 K. Now, groundbreaking experiments have provided preliminary evidence for high-temperature operation up to 25 K [7] employing unconventional high-$T_C$ superconducting materials such as $Bi_2Sr_2CaCu_2O_8$, $La_{1.55}Sr_{0.45}CuO_4/La_2CuO_4$ and two-gaps $MgB_2$ with unknown detection mechanisms.

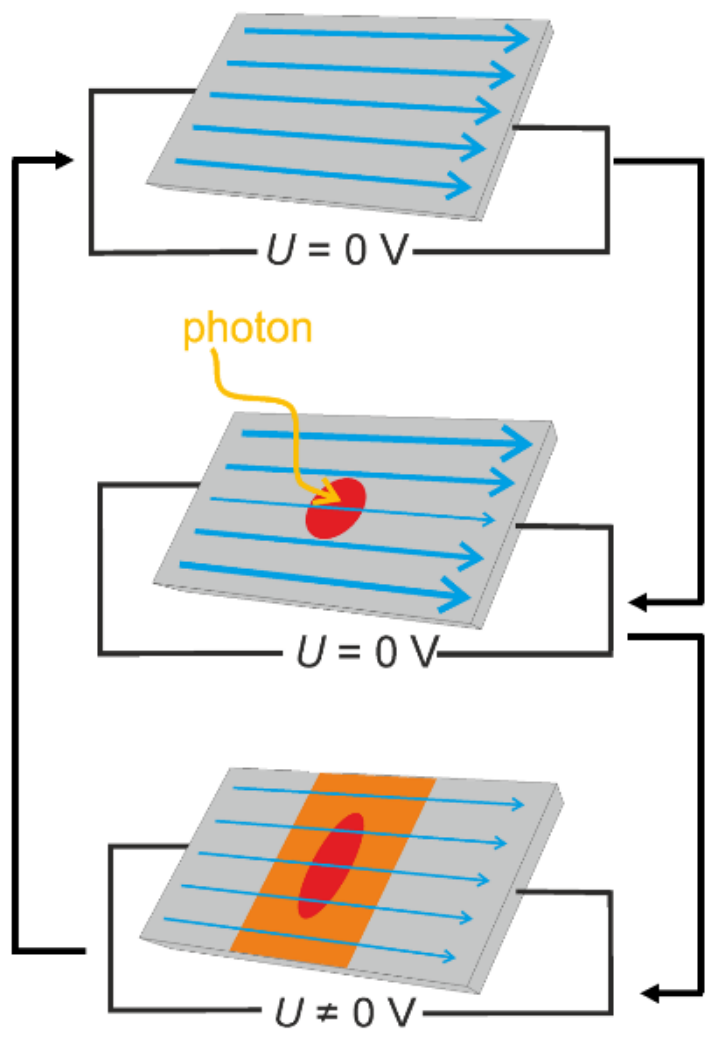

**Figure 18.1.** Principe of photon detection in nanowires.

### 18.2 Current and Future Challenges

Although superconducting nanowires with their non-linear properties have been successfully demonstrated as a key element for single-photon detectors, on-chip photonic integrated circuits, thermal switches (nTron and hTron), imagers, NASA cameras, artificial neurons and synapses, their potential has yet to reach the fundamental limit.

(i) Applications such as quantum imaging require a large active detection area, which necessitates the use of multiple SNSPDs arranged in arrays. Creating large arrays of SNSPDs with uniform performance across all detectors remains challenging. Variability in nanowire width, thickness, and material quality can lead to pixel-to-pixel variations in detection efficiency, timing resolution, and dark count rates. Improving fabrication precision and yield is essential to produce reliable, high-performing detector arrays suitable for imaging applications. As SNSPD arrays scale up, readout complexity and heat dissipation by each pixel increases significantly. Each pixel element requires a low-noise readout channel, often involving cryogenic amplifiers, which adds to the system bulk, cost, and power consumption. Developing scalable, multiplexed, or integrated readout electronics compatible with cryogenic environments is critical to making large SNSPD arrays practical.

(ii) While SNSPDs offer unprecedented performance compared to other single-photon detection technologies, their optimal operation requires cooling in complex and expensive cryogenic systems well below 4 K. Moreover, the unique properties of such low-temperature detection systems rapidly degrade as the temperature increases, significantly limiting their performance under higher thermal conditions. Research into high-$T_C$ temperature superconducting materials, such as $MgB_2$ and cuprates, aims to raise operating temperatures, which could reduce cooling requirements and extend SNSPD usability (see also Section 5). Although single-photon detection has been successfully demonstrated at elevated temperatures (up to 25 K), prototype's efficiencies remain low, and the detection mechanism is unclear.

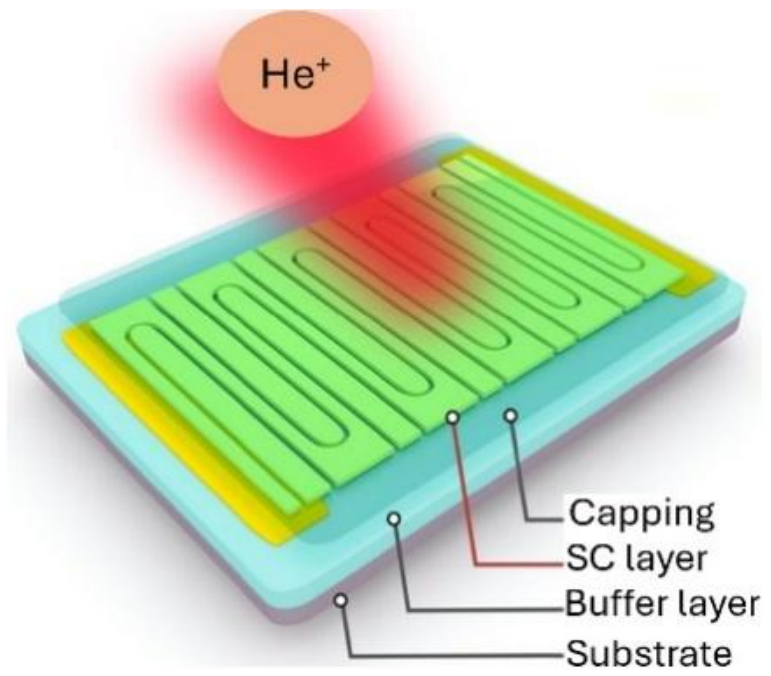

**Figure 18.2**: Schematic of high-$T_C$ device fabrication.

(iii) High efficiency in mid-infrared detection remains a challenge as detection efficiency in this range decreases exponentially with lower photon energy. Achieving consistent performance over these wavelengths is critical for applications such as infrared spectroscopy, astrophysics and remote sensing.

(iv) Spatially multiplexed SNSPD arrays enable photon-number resolution (PNR), which is in high demand for photonic quantum computing, characterization of heralded single-photon sources, and QKD. However, achieving high PNR fidelity requires the number of pixels significantly larger than the number of detected photons, as multiple photons may simultaneously hit the same pixel. This requirement combined with the need for high single-photon detection efficiency becomes challenging to meet even for three photons.

**18.3 Advances in Science and Technology to Meet Challenges**

(i) To overcome the limitations of array size and to achieve its high degree of multiplexing, a thermal row-column sensor element with thermally coupled imager readout can be effectively used, as was shown in the recent demonstration of a 400000-pixel camera [8]. In addition, the perpendicular orientation of the two nanowire layers enables polarisation insensitive optical cavity designs.

(ii) Recently, a new approach in fabrication has emerged, based on the introduction of dislocations into an otherwise orderly atomic arrangement in unconventional superconductors, known as precision dislocation engineering (Fig. 18.2). It is based on atomic level control in three dimensions by low ion dose and highly collimated helium-beam irradiation to achieve novel properties. By developing new fabrication methods using He+ irradiation, high critical temperature superconductors can be fabricated into nanowires without degrading their properties. Detection efficiency can be further enhanced by optical enhancements such as 3D lenses, optical cavities, and integration with waveguides.

(iii) To lower the energy threshold for mid-infrared detection, both the superconducting energy gap and the cross-sectional dimensions of the nanowires (width and thickness) must be reduced. However, this approach comes with trade-offs, including the need for even lower operating temperatures (below 1 K) and reduced readout currents (less than 2 μA). Further progress will require sophisticated materials and device engineering, e.g., broadening of energy threshold or stacking multilayers.

(iv) Recent experiments have shown that a SNSPD, made from a sufficiently long nanowire, can act as a cascade of thousands of pixels, allowing photon-number resolution of up to 5 photons with high efficiency [9]. This PNR information is encoded in the fast voltage signal and can be retrieved using advanced signal analysis. Enhancing PNR fidelity and extending it to higher photon numbers will require an understanding of the physics underlying the jitter in multi-photon detection and optimizing the nanowire geometry.

## 18.4 Concluding Remarks

Superconducting Nanowire Single Photon Detectors (SNSPDs) are critical to a wide range of advanced applications due to their exceptional sensitivity, fast response times and low dark count rates. These detectors are used in quantum optics and quantum information science, enabling the development of secure communication protocols through quantum key distribution. Their ability to detect single photons makes them invaluable for high-resolution imaging techniques, including biological and astrophysical imaging, where the detection of faint light signals is essential. In addition, SNSPDs play an important role in fundamental research, such as the study of quantum entanglement and the search for dark matter through direct detection experiments. As technology advances, the integration of SNSPDs into various systems promises to enhance capabilities in areas such as telecommunications, environmental monitoring and medical diagnostics, underlining their importance as a cornerstone of modern photonics.

# 19. Recent advances and perspectives for high-$T_C$ junctions

*Floriana Lombardi and Thilo Bauch*

Chalmers University of Technology, Gothenburg, Sweden

### 19.1 Status

Superconducting devices serve as essential tools across a wide spectrum of scientific and commercial applications, such as magnetoencephalography for neuroscience and geophysical surveys in resource exploration. At the core of these devices is the Josephson junction (JJ)—a weak link between superconducting electrodes— which enables precise control over superconducting current. Advances in high-$T_C$ materials (see also Section 5) have broadened the scope of JJ applications, as these materials operate at elevated temperatures, reducing the cooling complexity and creating pathways toward cost-effective, portable superconducting technologies.

The JJ, originally envisioned as a nanoscale constriction has driven research for decades, but the field has faced significant challenges. These include limited fabrication precision, material degradation during lithographic processes, and the absence of a comprehensive microscopic model that accurately addresses the local heating and non-equilibrium effects at these scales. As a result, current low- and high-$T_C$ superconducting electronics rely on complex interface engineering. For example, tunnel junctions incorporate inhomogeneous barriers that host fluctuating two-level systems, contributing to 1/f noise and decoherence[1], which hinders device reliability.

State-of-the-art high-$T_C$ Josephson devices are based on Grain Boundary Josephson Junctions (GBJJs) and are crucial to many superconducting applications. Despite progress, developments in high-$T_C$ technology over the past 30+ years have been incremental, with high-$T_C$ JJs still heavily reliant on grain boundaries, such as those formed on bicrystal or step-edge substrates[2]. Grain boundary characteristics can vary significantly, impacting reproducibility and performance, particularly in parameters like critical current density and normal state resistance, making it difficult to achieve the precision and consistency of low-Tc counterparts. As a result, achieving the performance and reproducibility of low-Tc Josephson junctions remains challenging.

A more reproducible process that also allows flexibility in circuit design would be transformative for high-$T_C$ superconducting sensors and electronics, significantly advancing the field.

### 19.2 Current and Future Challenges

In bicrystal-based grain boundary junctions, design flexibility is significantly constrained, as the junctions must align with the grain boundary of the substrate. Additionally, the variability in grain boundary properties across bicrystal substrates hinders the reproducibility of high-$T_C$ GBJJs. Step-edge junctions, which require multiple lithography steps and several epitaxial thin film depositions[3], face similar reproducibility challenges. Furthermore, the inherently low critical current density of GBJJs (typically below ~500 kA/cm$^2$ at 4.2 K for [001]-tilt angles larger than 10° [2]) restricts their performance in ultra-small junctions, limiting their use in applications that would otherwise benefit from high-$T_C$ materials, such as operation in strong magnetic fields.

Grain boundary-based $YBa_2Cu_3O_{7-\delta}$ (YBCO) junctions also suffer from considerable 1/f noise, largely due to charge trap defects within the barrier. Noise measurements on various types of GBJJs with different cross-sectional areas $A_j$ show that the scaled 1/f critical current noise at 1 Hz, $A_j S_I / I_c^2$, where $S_I$ is the critical current noise power spectral density and $I_C$ the critical current, is approximately given by a constant value $\sim 1 \cdot 10^{-8} \mu m^2/Hz$, [4]. This finding supports the charge trap model as an explanation for 1/f critical-current noise, suggesting that the product of trap density and Coulomb screening area remains similar across different grain boundary configurations. Compared to low-Tc tunnel junction based JJs the scaled 1/f noise in high-$T_C$ GBJJs is approximately 70 time larger. Therefore, the limitations of GBJJs, including high intrinsic noise, limited reproducibility, and challenges in achieving uniform critical current densities, continue to constrain their broader application in high-precision superconducting electronics.

A further challenge in broadening the application range of superconducting junctions is their fixed transport characteristics post-fabrication. Introducing tunable high-$T_C$ junctions, where transport properties can be dynamically adjusted, would greatly expand their functionality. This flexibility could pave the way for bio-inspired, neuromorphic superconducting hardware that operates with minimal power demands. Josephson junctions offer promising capabilities for neuromorphic computing due to their natural ability to emulate neuronal behaviors, such as spiking in response to input signals that surpass a threshold. Additionally, the tunability of critical current could serve as a means to mimic synaptic memory, allowing precise control over signal transmission between “neurons,” thus enabling more complex, brain-like computational architectures[5].

**19.3 Advances in Science and Technology to Meet Challenges**

New strategies to develop high-$T_C$ Josephson junctions, moving beyond traditional grain boundaries, include the use of Dayem bridges, high-energy ion irradiation of predefined wide YBCO bridges through a mask[6], and direct He focused ion beam (FIB) irradiation[7]. These techniques leverage bare epitaxial films grown on single crystals substrates, enabling more complex circuit designs and improved layout flexibility. High-$T_C$ nanoSQUIDs based on Dayem bridges exhibit low magnetic flux noise and benefit from a simpler fabrication process. However, the relatively high parasitic inductance of Dayem bridges limits their use in e.g. SQUID magnetometers at 77 K. Junctions defined with He FIB have demonstrated relatively low noise in SQUIDs operating up to 50 K, but their performance close to 77 K has yet to be fully validated.

A new type of junction based on the Dayem bridge is the Grooved Dayem bridge (GDB). GDBs can be defined anywhere on the chip and oriented at will within the film plane. Moreover, the bridge and weak link inside it are realized during a single lithography process, see Fig 19.1. SQUID magnetometers incorporating GDBs demonstrate low white magnetic flux noise, 6 $\mu\Phi_0/ Hz^{1/2}$, and magnetic field noise below 100 $fT/Hz^{1/2}$ at 77 K on a 5mm x 5mm chip[8], see Fig. 19.2. As such, GDB-based SQUIDs effectively combine the nanofabrication benefits and reproducibility typical of Dayem bridge structures with the high magnetic sensitivity of state-of-the-art SQUIDs based on grain boundary junctions [2].

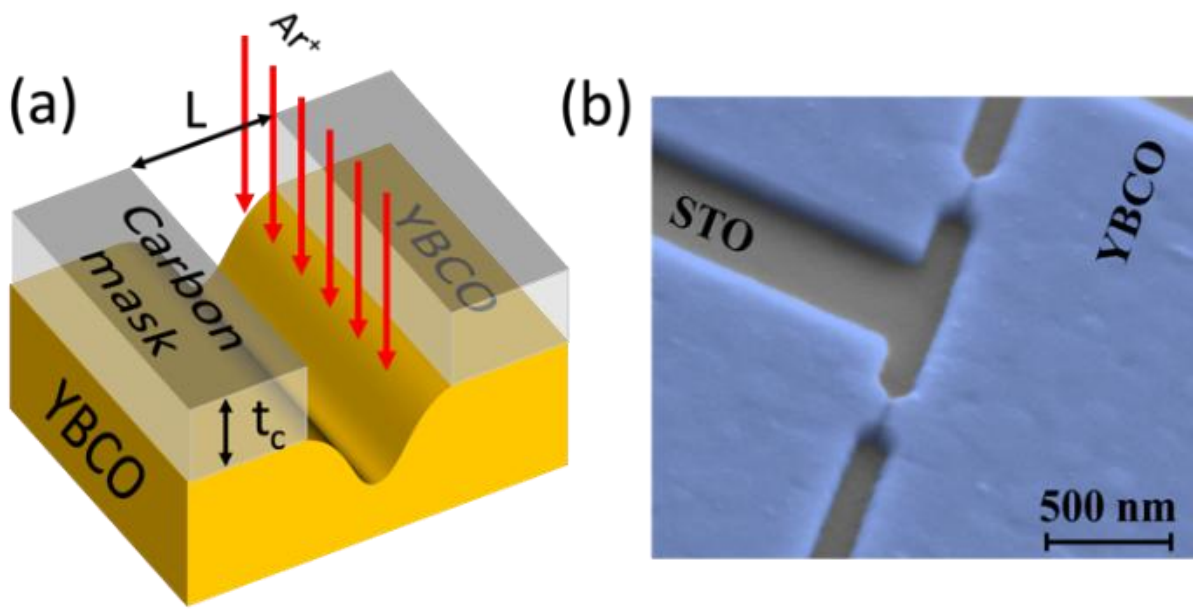

**Figure 19.1.** (a) Schematic of the Ar ion milling process inside the carbon mask gap. The width, W, and length, L, of the gap define the geometrical dimensions of the final grooved Dayem bridge (GDB) and can be varied to achieve different values of the critical current, IC. For aspect ratios of the gap in the carbon mask $t_C/L > 2$, the etching rate of YBCO during the Ar ion milling inside the gap is strongly reduced compared to the rest of the sample. This is the result of partial redeposition of the YBCO ablated by the Ar ions, which cannot be removed from the gap. (b) SEM image of two YBCO GDBs integrated in a SQUID loop.

However, reproducibility challenges in junction properties may occur between chips due to variations in mask characteristics, the ion milling process, and high-$T_C$ film quality. To address this, oxygen electromigration in YBCO nanowires can be a valuable tool for tuning electrical properties. Originally applied to modify hole doping in YBCO thin films and grain boundary junctions, oxygen electromigration has since been extended to YBCO micro-bridges and nanowires (see also Section 17). These advancements enable post-fabrication tuning of YBCO nanowires, and recent studies show that electromigration can significantly enhance SQUID performance[9].

While electromigration post-processing has shown promise for fine-tuning YBCO nanowires, its application to other types of junctions is yet to be fully explored. This technique could play a key role in achieving precise, ex-situ tuning of superconducting electronics, especially in applications where a large number of weak links with consistent, well-defined properties are essential.

As fabrication techniques advance, GBJJs could be integrated into hybrid superconducting structures that harness both the properties of high-$T_C$ materials and innovative materials, such as 2D materials. These hybrid systems have the potential to enable new quantum sensors and devices capable of operating across a broader range of environments. For example recent progress has led to the development of "vector substrates," using thin-film membrane technology[10]. In this method, a template layer is grown on a parent substrate with a sacrificial layer that can be easily dissolved leaving in solution the template layer that can be transferred onto a carrier substrate. This produces a vector substrate and allows in principle to create optimized substrates that would otherwise be challenging to obtain through traditional bulk single-crystal growth methods. One notable application of this approach involves fabricating nanometer-thick bicrystalline membranes of $SrTiO_3$, initially grown on a bicrystal substrate, and that are subsequently transferred onto sapphire substrates using the approach described above. The YBCO film is then deposited and patterned on this vector substrate, and the corresponding junctions exhibit excellent Josephson junction behavior. This technique would enable the creation of bicrystalline Josephson junctions of high-$T_C$ superconductors on a wide range of bulk substrates, offering greater flexibility in junction design and electronic properties.

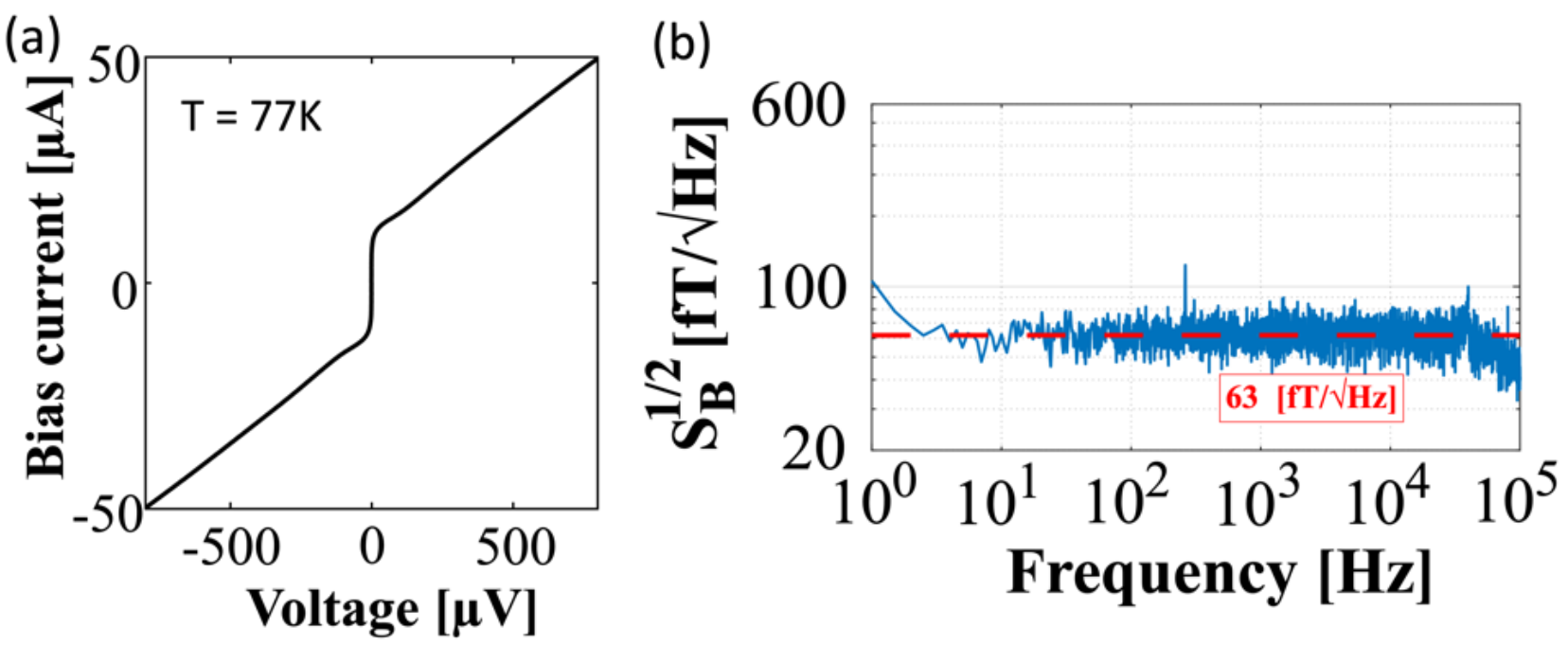

**Figure 19.2.** (a) Current voltage characteristic of a YBCO GDB-based SQUID measured at T=77 K. (b) Magnetic field noise measured at T=77 K of a YBCO magnetometer consisting of a GDB-based SQUID and a galvanically coupled pick-up loop on a 5 mm x 5 mm substrate.

In addition, this approach would allow to have different GBs angles on the same chip, by depositing an additional layer like CeO2 (or the same STO) which would make the region of the vector substrate, not covered by the membrane, compatible with the YBCO growth. In this way one could overcome the limitation of traditional bicristalline junctions and the possibility as anticipated to integrate them with other state of the art materials. Such hybrid systems could result in new types of quantum sensors and devices that can operate in a wider range of environments.

### 19.4 Concluding Remarks

In summary, advancements in high-$T_C$ Josephson junction (JJ) technology present both challenges and opportunities for future applications in superconducting devices. High-$T_C$ JJs have transformative potential across numerous fields, such as neuroimaging and resource exploration, due to their ability to operate at elevated temperatures, which reduces cooling requirements and expands application possibilities. However, reproducibility issues, such as variability in critical current densities inherent to grain boundary junctions (GBJJs) and the high 1/f noise, continue to limit their deployment in high-precision superconducting electronics. Recent innovations in Dayem bridge and grooved Dayem bridge (GDB) junction designs, alongside ion irradiation techniques and oxygen electromigration post-processing, represent promising alternatives to traditional grain boundary-based approaches. These new methods offer enhanced flexibility in circuit layout and show state of the art noise characteristics. Continued progress in developing reproducible, noise-minimized high-$T_C$ junctions will be essential for advancing superconducting technology and realizing its full potential in next-generation sensors, superconducting neuromorphic hardware, and potentially quantum circuit devices.

### Acknowledgements

This work was supported in part by the Knut and Alice Wallenberg Foundation (KAW) and in part by the Swedish Research Council under the project VR 2020-05184 and VR 2022- 04334. The authors would like to thank Edoardo Trabaldo for the fabrication and measurements of the GDB.

# 20. Josephson vortex in proximity junctions

Cheryl Feuillet-Palma[1], Vasily Stolyarov[1], Tristan Cren[2], Dimitri Roditchev[1]

[1]LPEM, ESPCI Paris, PSL, CNRS, Sorbonne University, Paris, France
[2]INSP, Sorbonne University, CNRS, Paris, France

### 20.1 Status

Josephson vortex in proximity Josephson junctions enable novel quantum functionalities [1,2], and can be used as a basis of numerous applications in superconducting technologies. Made of a thin non-superconducting metal (N) sandwiched by two superconductors (S), these SNS junctions mix up the intrinsic properties of N with the superconducting correlations induced from S by proximity. Electronic properties of these devices are governed by Andreev quasiparticles which are absent in conventional SIS junctions whose insulating barrier between the two S electrodes owns no electronic states. Similarly to Abrikosov vortices in superconductors - the $2\pi$-phase singularities of the superconducting wavefunction, Josephson vortex (JV) – are $2\pi$-phase singularities that can be created and manipulated inside N-regions by electric currents and magnetic fields [1] (see Fig. 20.1). These "proximity" JV, can be used as a basis of numerous applications in superconducting technology [2,3].

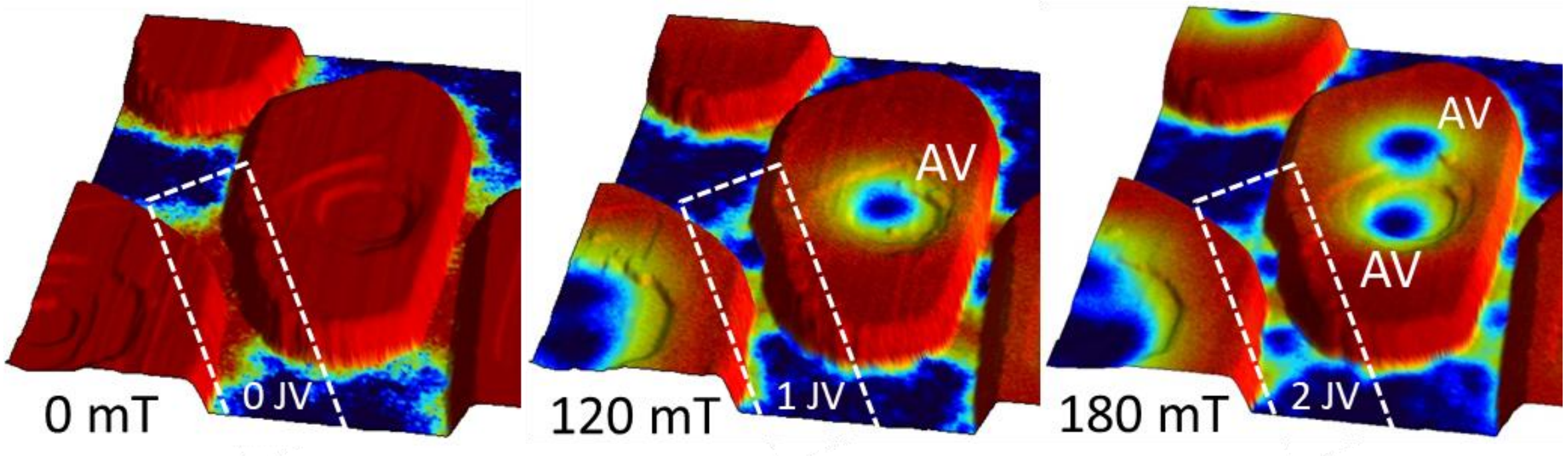

**Figure 20.1.** Josephson vortices (JV) appearing in lateral SNS junctions made of Pb-nano-islands linked by non-superconducting Pb-single atomic layers (STM experiment). Left panel: in zero-magnetic field, the superconducting gap (indicated by false red colour) is revealed in islands, but it is also induced between close islands by proximity (a region marked by white dashed rectangle). When a magnetic field is applied (middle and right panels), Abrikosov vortices (marked by "AV") enter the islands whereas Josephson vortices appear in the proximity regions.

### 20.2 Current and Future Challenges

The dynamics of JV could be extremely fast and low dissipating, since the Josephson vortex core has a significantly lower energy as compared to the Abrikosov vortex core. The JVs can be studied, on a local scale, by Scanning Tunneling (Section 16) and Magnetic Force Microscopies [1,2,4], as well as by transport experiments, on the macroscopic scale [4]. These experiments, along with precise numerical simulations, reveal the existence of several distinct dynamic regimes of the JV motion. One of them, identified as a fast hysteretic entry/escape below the critical value of Josephson current, is suggested for low-dissipative logic and memory elements whose states are encoded by the number of present Josephson vortices [6,7].

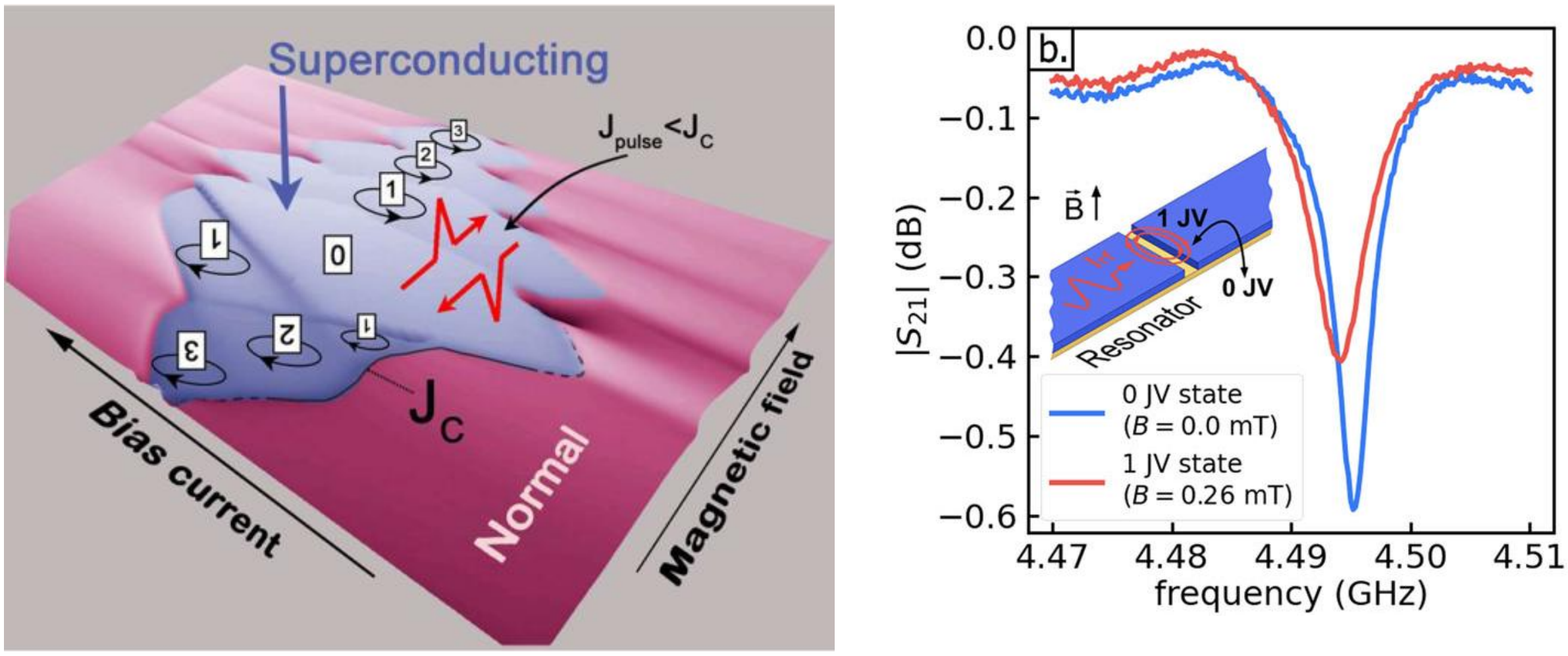

**Figure 20.2.** Single Josephson vortex memory [5]. Left panel: schematic phase diagram of a lateral Nb-Cu-Nb SNS junction (shown in the inset of the right panel) as a function of applied bias current and magnetic field. Grey-blue region corresponds to the dissipation-less regime below critical Josephson current Jc where JV of different number and polarity can be created by field and manipulated by short current pulses. Right panel: the resonance frequency and the quality factor of the microwave resonator coupled to the junction vary as a function of the number of JV present in the junction (from [5]).

### 20.3 Advances in Science and Technology to Meet Challenges

By integrating individual SNS junctions into a coplanar resonator and by applying a microwave excitation well below the critical current, it is possible to control the state of the system in an energy-efficient and non-destructive manner [8] (see Fig. 20.2). The remnant memory effect arises due to the presence of the natural edge barrier for Josephson vortices which can be precisely controlled by an appropriate choice of materials used to build the junctions and by precisely engineered junction geometry [9]. This opens new routes for non-destructive studies of spatially inhomogeneous superconducting materials and devices [6], and for creating scalable cryogenic memories directly compatible with superconducting microwave technologies.

### 20.4 Concluding Remarks

In the future, other functions and devices can be thought, such as quantum computers based on JV-qubits, control/correctors of quantum circuits, novel superconducting neural networks with information encoded in JV, reservoir computing based on superconducting electronics, superconducting digital and mixed-signal circuits, cryogenic memories and registers, among many others.

### Acknowledgements

This research was funded by the French ANR projects SUPERSTRIPES and CRYSTOP.

## 21. THz applications of superconducting Josephson junctions

*Vladimir M. Krasnov*[1] *and Benedikt Hampel*[2]

[1]Stockholm University, Sweden

[2]TU Braunschweig, Germany

### 21.1 Status

Electronic and photonic components capable of operation in the THz frequency range are essential for numerous applications. THz electronics is anticipated to play a crucial role in future generations of telecommunication systems and ultrafast computers. Advances in ultrafast electronics with radical innovations and novel post-CMOS solutions will be needed to overcome the "memory-wall" problem - a major technical challenge arising from the explosive growth in data volume. THz sources and detectors have potential uses in security, environmental monitoring, non-ionizing medical imaging, chemical analysis, and fundamental research. Despite significant progress made with semiconductor-based Quantum Cascade Lasers, developing efficient THz sources remains a technological challenge, often referred to as the "Terahertz gap." Generally, cryogenic cooling is required to build efficient sources and sensitive detectors in the low-THz range, creating suitable conditions for utilization of superconductors.

The quantum-mechanical nature of superconducting Josephson junctions (JJs) opens a possibility for building novel THz components with unique functionality. The ac-Josephson effect enables generation and detection of high-frequency electromagnetic signals. JJs can operate in a broad frequency range, from microwave to sub-THz for conventional low-$T_c$ junctions and in the full THz range (1-10 THz) for high-$T_c$ superconductor (HTS)-based JJs. Various types of Junction-on-cantilever sensors, enabling both high sensitivity and spatial resolution, were developed recently for scanning probe microscopy [1-3]. In particular, single JJs from $YBa_2Cu_3O_{7-d}$ were demonstrated as detectors in THz microscopy setups [1,3], see Fig. 21.1(a), allowing for spatial scanning of THz radiation from actively radiating or passively irradiated samples. Figure 21.1(b) shows a 2D scan of THz twisted light at 1.3 THz using such a THz microscopy setup. Antenna structures are required to efficiently couple THz radiation into a JJ and to evaluate both frequency and power [1,3]. Similar technology has been used to fabricate JJ arrays that serve as voltage standards [4].

Synchronization of multiple JJs in an array enables advanced functionalities, such as cascade amplification of detector readout and coherent superradiant emission with power proportional to the square of the JJ number [5-7], thus boosting radiation power efficiency (RPE). THz sources based on stacks of atomic-scale intrinsic Josephson junctions, naturally formed in the layered HTS $Bi_2Sr_2CaCu_2O_{8+d}$ (Bi-2212), have demonstrated up to 12% RPE at 4 THz [6], see Fig. 21.2. Additionally, wide-band frequency modulation of the Bi-2212 THz emitter has been shown [8], opening possibilities for ultra-high frequency signal processing suitable for next generation telecommunication systems.

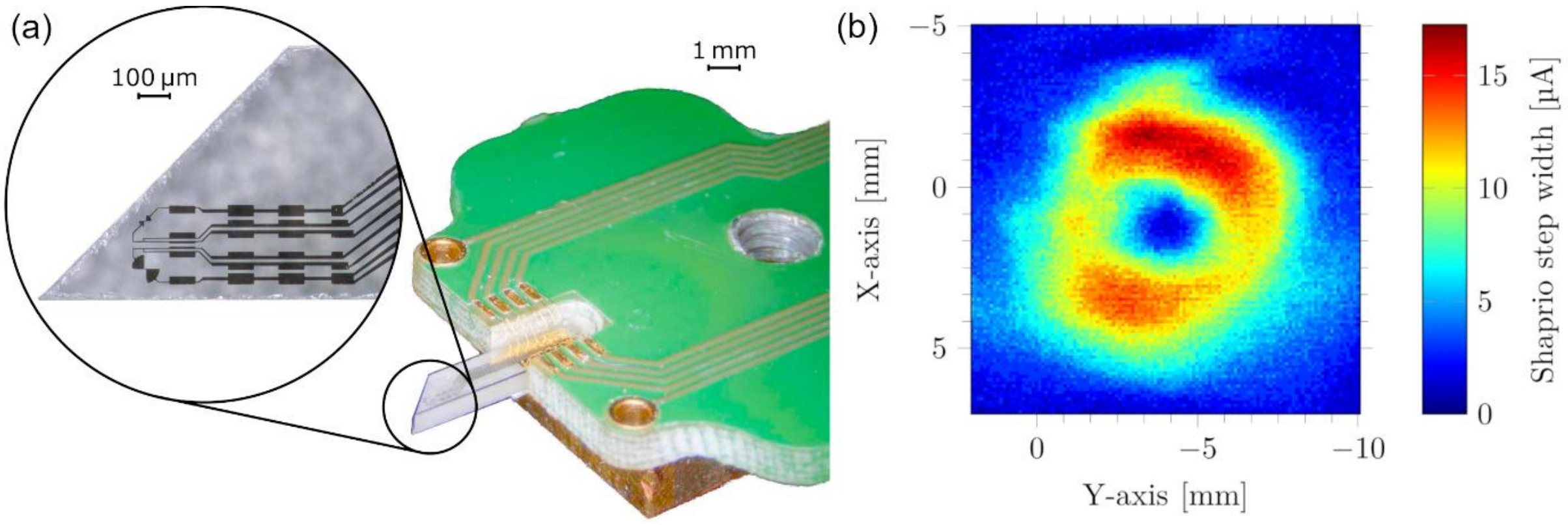

**Figure 21.1**. (a) Photomicrograph of a Josephson cantilever mounted on a PCB and copper carrier. The magnified inset shows its tip with two single JJs at the feed points of two antennas. Reprinted from Ref. [1], with the permission of AIP Publishing. (b) Measured beam profile of twisted light at a frequency of 1.302 THz generated by a far-infrared laser illuminating a spiral phase plate that was additively manufactured from cyclic olefin copolymer. The color-coded Shapiro step width is an indicator for the spatial distribution of laser power. Reprinted with permission from Ref. [3]. © 2023 IEEE

### 21.2 Current and Future Challenges

The key advantage of THz sources based on JJs is the inherent broad-range frequency tunability. Additionally, the high critical temperature (up to 95K for Bi-2212) allows for operation in liquid nitrogen, enabling simple, cheap, and compact monochromatic continuous-wave and frequency-tunable THz sources. The main challenge, however, is achieving high emission power in the far-field. The target value of $P$ = 1 mW has not yet been reached due to challenges with small RPE and large self-heating [6,7]. The typical RPE of Bi-2212 mesas is small, <1%, [5,7] while the thermal resistance is large, $R_{th}$ ~ 2-20 K/mW (depending on the mesa size). To reach $P$ = 1 mW, the mesa should need to dissipate $P$/RPE ~ 100 mW, which would cause a critical overheating, $R_{th}P$/RPE, well above $T_c$ thereby suppressing the Josephson effect [6].

For THz detector applications, single JJs and JJ arrays based on $YBa_2Cu_3O_{7-d}$ face fabrication challenges, as well as issues with coupling THz radiation into the JJs through antenna structures. The complex unit cell of $YBa_2Cu_3O_{7-d}$ requires lattice-matched substrate materials to ensure epitaxial growth. Pulsed laser deposition (PLD) using polycrystalline targets can produce high-quality thin films on single crystal substrates such as $(LaAlO_3)_{0.3}(Sr_2TaAlO_6)_{0.7}$, $SrTiO_3$, $LaAlO_3$, or MgO. Bicrystal substrates, a previously common approach to creating high-quality JJs at grain boundaries, are now less available, and the fixed grain boundary position on these substrates significantly limits circuit design. Other JJ fabrication technologies often exhibit greater parameter variability, posing challenges for the uniformity required in JJ arrays. Large JJ arrays for the Josephson voltage standard application require high uniformity of JJs across the entire array, and efficient coupling of THz radiation into each JJ is crucial. This presents challenges in both array fabrication and antenna design [1,3].

### 21.3 Advances in Science and Technology to Meet Challenges

There are several approaches to increase the emission power of Josephson THz sources:

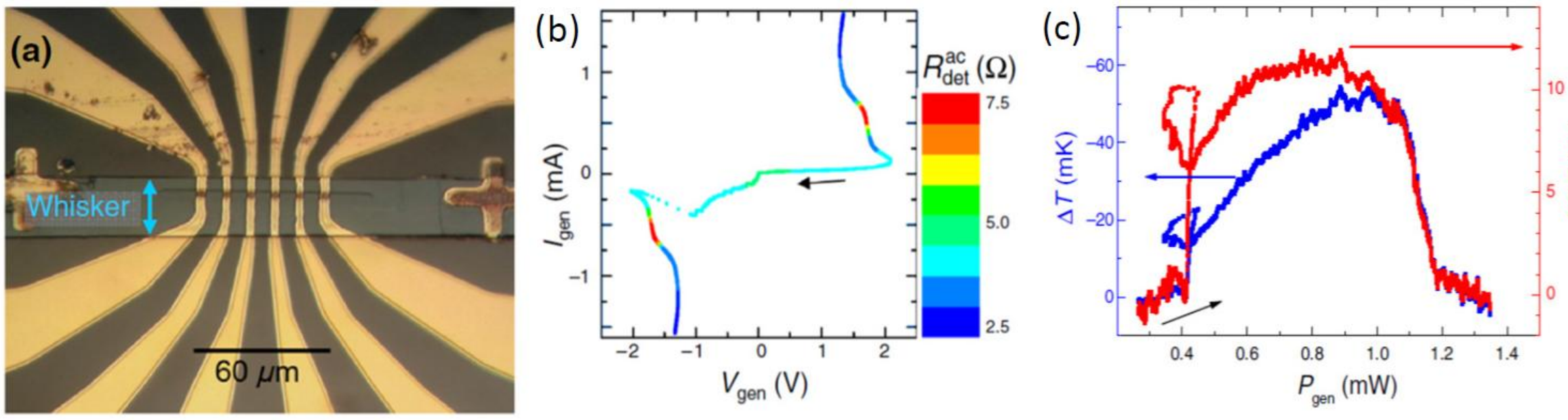

**Figure 21.2.** Coherent THz source based on Bi-2212 whisker. (a) An optical image of the device. (b) The Current-Voltage characteristics of the generating mesa recorded during a single bias swipe from positive to negative current. The color scale represents simultaneously measured detector response. The emission (red) occurs at $f \sim 4$ THz. (c) Radiative cooling as a function of dissipation power $P = IV$ (blue, left axis). The red line (right axis) shows the radiative power efficiency, reaching the record high value of 12%. Adapted from Ref. [6] under the Creative Commons Attribution 4.0 International license.

1) Introduction of Impedance-Matching Antenna Elements: The primary strategy should focus on enhancing RPE. The theoretical maximum RPE of 50% is achieved at a condition of impedance matching between the internal load resistance and the external (free-space) radiative resistance. In this case, the radiation power $P$ = 1 mW would be achieved at the net dissipation power of only 2 mW, thus precluding critical overheating of the device. Impedance matching involves incorporating properly designed antenna elements into the THz oscillator. Figure 21.2 shows an example of power-efficient THz sources based on Bi-2212 whisker crystal [6]. Figure 21.2(c) highlights a record-high RPE of 12%, deduced using the radiative cooling effect. This high RPE is attributed to the turnstile-antenna geometry, which enables effective impedance matching.

2) Reduction of Self-Heating: Thermal resistance can be reduced by employing stand-alone mesas with double-side cooling [7]. Self-heating at a given bias voltage (frequency) decreases linearly with the lateral size of the mesa, allowing for a high bias voltage and emission frequency in excess of 10 THz. However, the net power at a given voltage is proportional to the current and, therefore, decreases as a square of the lateral size. As a result, smaller mesas are less prone to self-heating but emit lower power.

3) Phased Arrays of Small Mesas: Coherent emission from multiple small mesas [7] could enable higher net power with little self-heating. Additionally, this approach could address another major issue for free-space emission: emission from Bi-2212 mesas involves high-order cavity modes, which produce complex radiation patterns with poor directivity. Phased antenna arrays could significantly enhance emission directivity. Technologically, the fabrication of such arrays has been demonstrated [9]. However, a key challenge would be in controlling phase shifts between individual mesas. Phase control can be achieved through surface electromagnetic waves propagating along metallic electrodes [7]. Accurate antenna design is essential in all cases, both for impedance matching and for improving directivity.

A newer approach to fabricating $YBa_2Cu_3O_{7-d}$ JJs involves using helium-focused ion beam irradiation [10]. This technique enables localized modification of the superconductor's properties, allowing for precise JJ placement and tunable properties based on the helium ion dose and other parameters. Combining this technology with antenna structures, potentially integrated with metamaterials in the future, is expected to advance the development of HTS Josephson voltage standards [4].

## 21.4 Concluding Remarks

The quantum-mechanical nature of the Josephson effect enables unique functionalities of high-frequency detectors and sources. In particular, it enables ultra-sensitive (quantum-limited) photon detectors with a spectroscopic resolution of both amplitude and frequency, and frequency-tunable oscillators. The utilization of Josephson junction arrays provides additional functionality, such as cascade-amplified detectors and superradiant amplification of radiation [5-7]. Recently, various types of Junction-on-cantilever sensors have been demonstrated for scanning probe microscopy [1-3], enabling high sensitivity and additional spatial resolution. High-temperature superconductivity expands the frequency range of Josephson electronics to the full THz range and facilitates comfortable operation at liquid nitrogen temperatures, opening possibilities for commercial applications.

Superconducting Josephson junctions will play a significant role in bridging the technological gap in the THz range by providing innovative solutions for generation and detection, thereby expanding the usage of THz technology across various fields for applications in imaging, spectroscopy, communications, security and ultrafast digital electronics (see also Section 23). New technological developments for the fabrication of JJs, the integration with well-matched antenna structures, and further optimization of current circuit designs will address present challenges to enable a broader use of Josephson THz technology in research and industry.

**Acknowledgements**

This work was supported in part by the Braunschweig International Graduate School of Metrology—B-IGSM, in part by the Laboratory for Emerging Nanometrology—LENA, in part by the Deutsche Forschungsgemeinschaft (DFG, German Research Foundation, under Germany’s Excellence Strategy—EXC-2123 QuantumFrontiers—390837967, in part by the Volkswagen Foundation and the Ministry of Science and Culture of Lower Saxony through “Quantum Valley Lower Saxony Q1” (QVLS-Q1).

## 22. Nanoscopic Josephson junctions and applications in quantum technologies

*María José Martínez-Pérez[1], Javier Sesé[1] and Dieter Koelle[2]*

[1]Instituto de Nanociencia y Materiales de Aragón (INMA), CSIC – Universidad de Zaragoza, Spain
[2]Physikalisches Institut, Center for Quantum Science (CQ) and LISA+, Universität Tübingen, Germany

### 22.1 Status

A SQUID (Superconducting Quantum Interference Device) consists of a superconducting loop interrupted by one (or more) Josephson junctions. Under the action of an externally applied magnetic field, a phase gradient arises that must sum to an integer multiple of 2π. If this condition is not fulfilled, a circulating current arises, producing sharp phase jumps at half-integer fractions of the magnetic flux quantum. This effect reduces the maximum critical current that the junctions can withstand, effectively turning the SQUID into a single junction with critical current periodic on the flux quantum. The smallness of the flux quantum turns SQUIDs into extremely sensitive devices, capable of transforming tiny flux changes into measurable electrical signals [1][2]. Additionally, this affects the SQUID's effective inductance, making Josephson junctions and SQUIDs a cornerstone of quantum technology, enabling highly tunable and nonlinear elements in quantum devices [3].

Already in the 1980's, reducing the dimensions of SQUIDs to the nanometric scale allowed for the fabrication of the most sensitive quantum flux detectors. This is achieved by two means. On the one side, reducing the dimensions of the nanoloop makes it more sensitive to tiny variations in the magnetization of nanoscopic magnetic structures placed nearby. On the other side, the device's flux noise is reduced with the reduction in the loop's geometrical inductance. NanoSQUIDs have been used to characterize nanoparticles (see Fig. 22.1a), nanowires or two-dimensional magnets [4]. Typically, the critical current of the SQUID is measured, but this process is stochastic, influenced by temperature and quantum tunneling. Alternatively, biasing a non-hysteretic SQUID with a current near its critical current generates a measurable voltage signal, periodic with the flux quantum. SQUIDs can operate in a nonlinear regime or within a feedback loop, which linearizes their flux-to-voltage response and increases their dynamic range up to many flux quanta.
Josephson junctions and SQUIDs are nowadays ubiquitous in the development of low-temperature solid-state quantum technologies like qubits, i.e., two-level systems used for quantum computing (see also Section 23). In addition, Josephson junctions are also fundamental building blocks for microwave superconducting circuits with tunable properties. In the most basic layout, the inductance of a LC resonator is modulated using the Josephson effect, with a flux-biased SQUID. This forms the basis of many circuit elements such as tunable couplers or quantum-limited amplifiers [5]. Josephson parametric amplifiers operate by modulating the resonance frequency of an oscillator with a pump tone. This process transfers photons from the pump to the incoming signal, amplifying it.

## 22.2 Current and Future Challenges

There are various practical methods to implement Josephson junctions. Tunnel junctions are commonly fabricated through shadow mask evaporation of aluminum combined with in-situ oxidation or by using niobium/aluminum oxide/niobium trilayers, which are patterned following several lithography steps. These junctions typically have critical current densities between 100 A/cm$^2$ and several kA/cm$^2$ at 4 K which is enough for building qubits and tunable microwave (mw) elements but totally impractical to build nanoscopic Josephson junctions for sensing. In addition, such junctions are very sensitive to external magnetic fields posing severe restrictions for magnetometry or applications in quantum magnonics.

Fabrication of nanoSQUIDs for quantum sensing is typically done by patterning a nanoconstriction in a superconductor with dimensions comparable to the coherence length [6]. Such devices can be patterned from a single lithography step but, unfortunately, are extremely hard to optimize. The effort can be worth if sensors must operate under the action of large magnetic fields. On the other hand, sandwiching a normal metal between two superconducting electrodes yields Josephson junctions with very high critical current densities of 10$^5$ A/cm$^2$ at 4 K, very attractive for the fabrication of nanoscopic devices [7]. Such junctions are, however, not well suited for operation in high fields. Finally, another way of observing the Josephson effect is to exploit the enormously anisotropic properties of high critical temperature superconductors like cuprates. Grain boundaries in YBCO ($YBa_2Cu_3O_{7-\delta}$) exhibit Josephson behavior with large critical current densities of 10$^5$ up to 10$^6$ A/cm$^2$ at 4 K. These barriers are e.g., fabricated from epitaxial growth over a bicrystal substrate of, e.g., STO ($SrTiO_3$), which yields high-quality thin YBCO films. However, this approach is impractical for building microwave YBCO circuits due to the significant microwave losses of STO.

## 22.3 Advances in Science and Technology to Meet Challenges

Integration of nanoSQUIDs into mw circuits is very promising for quantum sensing applications. In the past, SQUIDs have been operated as floating elements, inductively connected to a LC resonator with a characteristic frequency determined by the effective inductance of the SQUID, which depends on the flux threading the loop. In a newer version, the SQUID is integrated into the LC-resonator to detect small variations in its inductance using a reflectometry configuration, making it a sensitive flux-to-phase detector that operates in the non-dissipative regime, thus dissipating no power [8]. Such design allows one to directly realize Josephson parametric amplification, further improving the device’s sensitivity.

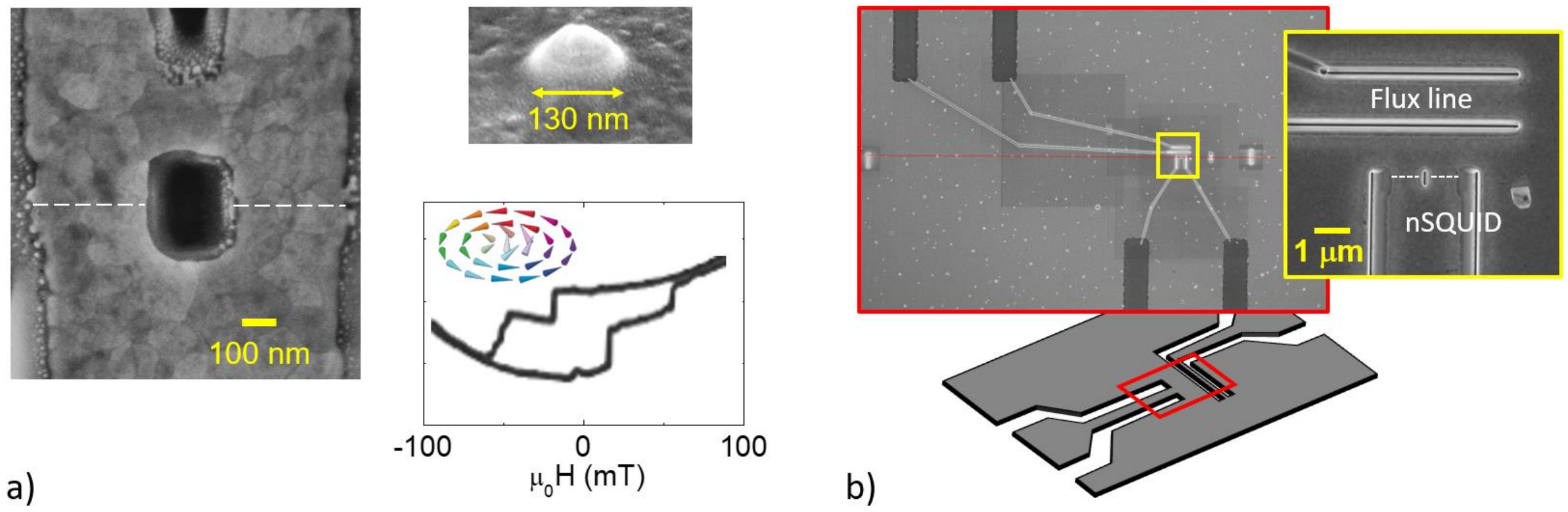

**Figure 22.1.** a) Left: SEM image of a YBCO nanoSQUID sensor on STO with the grain boundary junctions highlighted by dashed white lines. Top right: SEM image of a cobalt nanoparticle deposited on top of the nanoSQUID. Bottom right: Hysteresis curve measured with the nanoSQUID at 4 K. The inset shows a schematic of a vortex magnetic configuration responsible for the steps observed in the hysteresis loop. b) Bottom: Schematic of the YBCO nanoSQUID coupled to a transmission line. Top left: SEM image of the final device fabricated on MgO. Top right: Enlarged view of the nanoSQUID with the grain boundary Josephson junctions highlighted by dashed white lines (not published).

Significant effort is being devoted to improving YBCO nanoSQUIDs (see also Sections 5 and 19). One approach involves exploring substrates other than STO, such as MgO, which has a dielectric permittivity several orders of magnitude smaller than STO. This is promising as it enables the fabrication of non-hysteretic junctions from scratch, eliminating the need for resistive shunting and thereby reducing the intrinsic thermal flux noise of the device. Indeed, nanoSQUIDs based on MgO with unprecedentedly low noise characteristics have already been demonstrated [9]. Another promising route is the use of focused helium ion beam microscopes for directly patterning Josephson barriers. This technology enables precise control of the junction's critical current density by adjusting the irradiation dose. At high doses, the material becomes insulating, making it possible to directly write Josephson nanocircuits without milling the YBCO film [10]. Moreover, this technique paves the way for using alternative substrates such as LSAT or even sapphire. Together with MgO, the latter is enormously interesting for mw applications resilient to high magnetic fields including tunable superconducting cavities, parametric amplifiers or mw nanoSQUIDs. First devices have been fabricated showing the ability to tune the resonance frequency of mw cavities based on YBCO by several hundred of MHz or to directly couple nanoSQUIDs to transmission lines for mw readout (see Fig. 22.1b).

Additionally, the advent of ion microscopes has revolutionized the fabrication of constriction-like Josephson junctions. Previously extremely difficult to optimize, recently fabricated bridges using a combination of neon and helium milling have proven to be reproducible and allow for precise control of the resulting critical current [11]. These junctions start now to get combined with mature mw circuits for quantum sensing and signal processing, enabling their operation under larger applied magnetic fields.

## 22.4 Concluding Remarks

Josephson junctions are irreplaceable in the development of quantum sensors, quantum computers, and related technologies. The design of Josephson-based quantum circuits continues to evolve to mitigate decoherence mechanisms such as charge and flux noise, with new architectures like the fluxonium qubit demonstrating enhanced coherence times. Josephson parametric amplifiers (JPAs) enable near–quantum-limited amplification of microwave signals [5] and are now widely employed not only in superconducting quantum processors, but also across a variety of cryogenic microwave experiments, including axion dark-matter searches and other quantum sensing platforms. Traveling wave parametric amplifiers (TWPAs), which integrate Josephson elements into nonlinear microwave transmission lines, extend the operational bandwidth of conventional parametric amplifiers, albeit at the cost of increased design complexity and fabrication challenges. Furthermore, rapid single-flux-quantum (RSFQ) and related superconducting digital logic technologies, which utilize Josephson junctions to generate quantized voltage pulses, show considerable promise for high-speed qubit control and readout applications.

The fabrication of improved Josephson junctions is an active area of research, with many areas still to be explored and improved. Recent developments include the use of new focused ion beam techniques based on gas field ion sources, addressing critical challenges in the fabrication and reproducibility of Josephson devices. This is very promising for optimizing device performance under the application of high magnetic fields, opening the way for new applications in quantum sensing, cavity magnonics or spin-based quantum computing.

**Acknowledgements**

The authors acknowledge the European Research Council (948986 QFaST), Grant CEX2023-001286-S funded by MICIU/AEI/10.13039/501100011033, the Spanish MCIN and the European Union FEDER through project PID2022-140923NB-C21, MCIN for the Advanced Materials and the Quantum Communication programs with funding from European Union NextGenerationEU (PRTR-C17.I1) and the Government of Aragon, the European Union – NextGenerationEU (Regulation EU 2020/2094), through CSIC's Quantum Technologies Platform (QTEP), the Aragón Regional Government (QMAD E09_23R) and the COST Actions FIT4NANO (CA19140) and SUPERQUMAP (CA21144).

## 23. Superconducting qubits: challenges and innovations toward universal quantum computing

*Stefano Poletto[1], Alessandro Bruno[2], Davide Massarotti[3], Francesco Tafuri[3]*

[1] Rigetti Computing, USA

[2] QuantWare, The Netherlands

[3] Università degli Studi di Napoli Federico II, Italy

### 23.1 Status

The term "superconducting qubit" extends beyond individual qubits to include the broader field of multi-level artificial atoms fabricated on chip using superconducting material and following the principles of circuit quantum electrodynamics (cQED). The ability to fabricate artificial atoms with desired frequencies and coupling strengths, ranging from weak to ultra-strong, makes this approach an ideal platform for studying matter-light and qubit-qubit interactions. These artificial atoms are designed to operate in the low microwave spectrum, from several hundreds MHz to tens of GHz, and are cooled to 20 mK to minimize thermal noise and ensure the system remains in its ground state. Superconducting artificial atoms are carefully engineered with a delicate balance of capacitance, inductance, and a nonlinear element provided by the Josephson junction (JJ), see also Section 22. The Josephson effect enables the operation of both superconducting qubits and multi-level artificial atoms. Because of this flexibility and versatility, superconducting qubits represent one of the most promising approaches in quantum computing.

When a superconducting artificial atom is operated on its first two energy levels, it is referred to as a qubit. The first coherent oscillation in a superconducting qubit, the Cooper-pair box, was measured in 1999[1]. Although the system initially exhibited a short lifetime and significant noise, this breakthrough played a pivotal role in the development of superconducting qubits, quantum processors, and superconducting quantum devices. Over the past 25 years, the field has made remarkable progress. The lifetime of superconducting qubits has increased by several orders of magnitude, crossing the record value of 1 ms[2], thanks to improved materials, optimized fabrication processes, and improved qubit layouts. Single- and two-qubit gate fidelities now exceed 99.9%[3], while readout can be performed in as little as 40 ns with fidelity above 99%[4]. Multi-qubit quantum processors with hundreds of qubits are now fabricated and accessible to the public through quantum web services. Accordingly, there have been several successful demonstrations of how small-scale physical qubits can be efficiently manipulated and used to execute complex quantum algorithms. These include the watershed experiment demonstrating quantum computational supremacy[5] and the progress made recently towards fault-tolerant quantum computing with logical qubits surpassing the performance of physical qubits[6].

Quantum processors and technologies based on superconducting qubits are at an inflection point, where the quality of operations and the scale of devices have advanced to the stage where applications implemented on physical qubits can already deliver value. However, the path to building a universal quantum machine still presents significant challenges, requiring further advancement in both physics and engineering. It is impossible to give an exhaustive roadmap of current and future challenges on the superconducting quantum computing platform within the format and length limitations of this section. We will therefore limit to a representative section of some exemplary challenges without claiming to be exhaustive and direct the readers to the webpages of leading companies in the field for a timely update on their own roadmaps[7] and to a recent review[8].

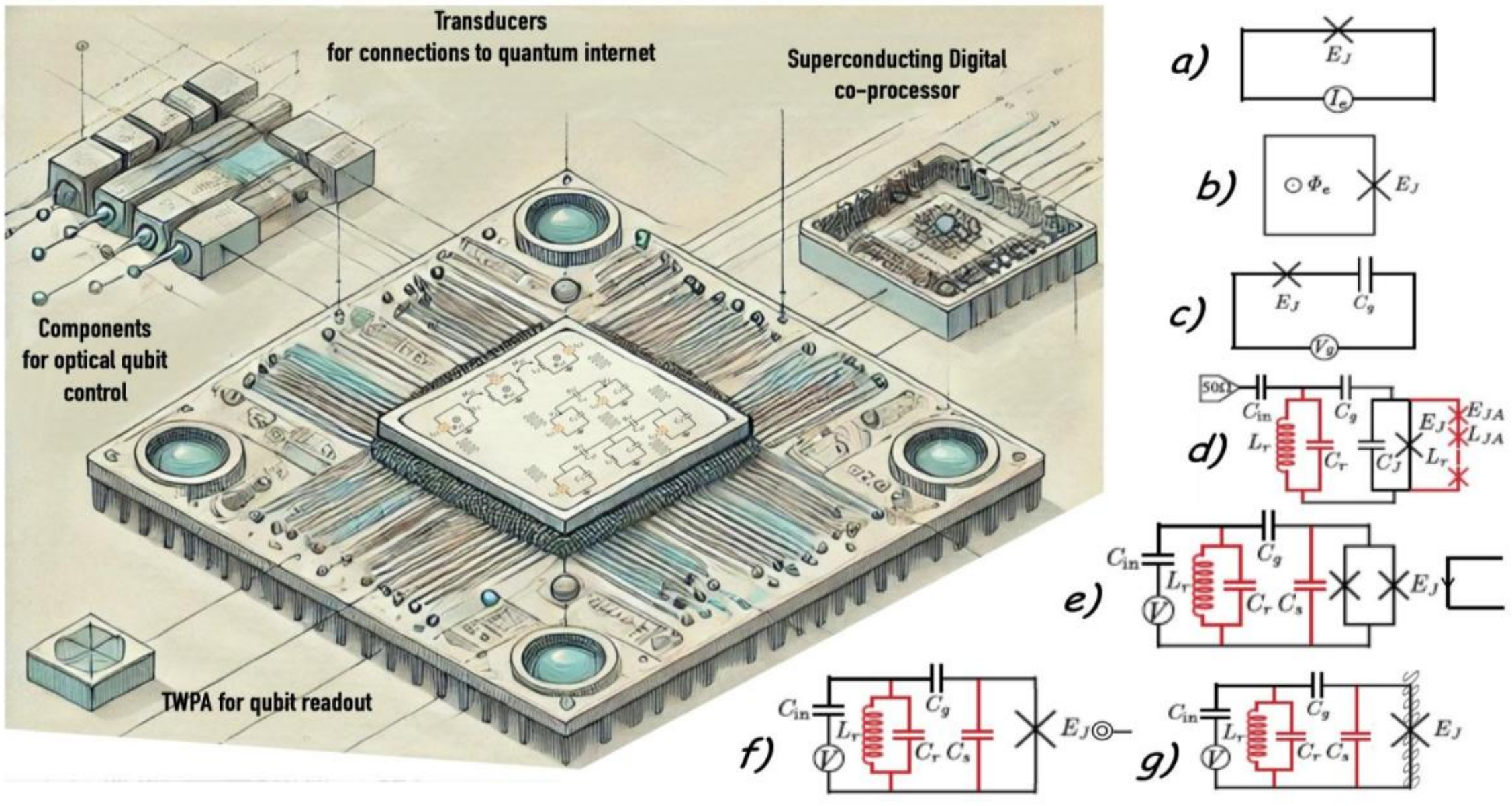

**Figure 23.1**. Cartoon of a future superconducting quantum processor with hybrid interfaces. In the insets, different types of superconducting qubits: a) phase qubit, b) flux qubit, c) charge qubit, d) fluxonium, e) transmon, f) gatemon, g) ferrotransmon.

## 23.2 Current and Future Challenges

Although the field of superconducting qubits has advanced rapidly, achieving the ultimate goal of building a universal quantum processor remains a significant challenge. Such a machine will depend on logical qubits, each constructed from hundreds of physical qubits, where errors are discretized, detected, and corrected using quantum error correction protocols. These protocols will encode quantum information in highly entangled quantum systems and implement quantum algorithms on logical qubits. Quantum error correction protocols are essential because even minimal errors can quickly accumulate leading to computational failure, but it comes at the cost of a larger number of qubits, low error rates for quantum gates and readout, and real-time data processing. Reaching this goal requires substantial improvements in multiple areas. These include achieving error rates in the range of 0.01–0.1%, a critical requirement for both noisy intermediate-scale quantum (NISQ) algorithms and fault-tolerant quantum computation, as well as fabricating and controlling processors with millions of qubits, extending relaxation and coherence times, reducing error rates for single- and multi-qubit gates, and enhancing readout fidelity while minimizing acquisition times. Careful systems engineering will also be crucial. For example, adjustable couplers, key components that enable dynamic control of qubit interactions, played a pivotal role in achieving several milestones. Further innovations, which may prioritize performance over simplicity, are needed to push error rates in large-scale systems into the $10^{-4}$ range and beyond, paving the way for robust and scalable quantum processors.

Large-scale Quantum Processing Units (QPUs) require not only higher-quality qubits and JJs but also a variety of quantum components, such as Traveling Wave Parametric Amplifiers (TWPAs) for qubit readout and more compact cryogenic electronics. From the perspective of classical control electronics, reducing its signal noise and intra-channel crosstalk, as well miniaturizing components like circulators and other microwave elements is essential. One potential approach is to leverage cryogenic electronics, such as Single Flux Quantum (SFQ) logic or other superconducting logic, though this is only one of several possible solutions being explored.

Scaling up quantum processors will also present significant challenges in packaging, wiring, and cryogenic systems. The largest superconducting quantum processors are already pushing the limits of commercially available cryogenic systems when fully wired with standard microwave frequency coaxial cables. High-density, small-footprint, and low thermal-budget wiring solutions are actively being explored and deployed in medium-scale devices. A comprehensive system engineering approach targeting the full stack must be taken in parallel, where one embraces the idea that many system parameters must be simultaneously optimized[9].

**23.3 Advances in Science and Technology to Meet Challenges**

One of the future challenges is to significantly reduce the footprint of superconducting qubits, enabling the implementation of processors with higher qubit counts within a relatively small chip size. The transmon qubit, one of the best-performing superconducting qubits, is also among the largest. Its size comes from an optimized layout to minimize the participation of its energy in lossy materials and interfaces. Reducing the size of superconducting qubits without compromising coherence and relaxation times will require advancements in material deposition, interface quality and the whole fabrication process, as well as innovations in design. High-quality angled-evaporation Aluminum-Aluminum Oxide-Aluminum (Al/AlOx/Al) JJs are essential for obtaining qubits with high coherence times; however, the yield of small JJs still needs further improvement. The "ideal" JJ requires an atomistic control of fabrication parameters, which in turn could allow suppression of two-level system defects. Process instability, device quality, and parameter variation hinder scaling to larger volumes and large-scale devices. The characteristics of these fabricated components and junctions typically are subject to considerable process variations, leading to a variance in qubit frequency of up to 10% in current QPUs, whereas less than 1% variation is needed. To address these challenges, post-fabrication JJ frequency trimming methods are explored[10].

As the number of qubits in processors increases, signal distribution will become a critical challenge. Current processors dedicate individual control lines to each qubit, with multiplexed readout limited to fewer than ten qubits simultaneously. This approach poses significant challenges in terms of cryogenic and packaging requirements. Developing high-density, low-thermal-conductivity wiring solutions and compact 3D-integrated packaging will be essential. Looking further ahead, multiplexed control techniques must be developed to manage more qubits than the number of control lines available in dilution refrigerators. Superconducting 3D integration methods are needed for complex high-coherence devices and systems as multi-chip modules, especially for large-scale QPUs, where 3D routing ensures minimal losses, crosstalk, and noise. These integrated packaging solutions may include multilayer superconducting metallization with superconducting Through-Silicon Vias (TSVs), air bridges, and flip-chip bonding. Large processors built from multi-chip modules is another active area of research and engineering integration[11].

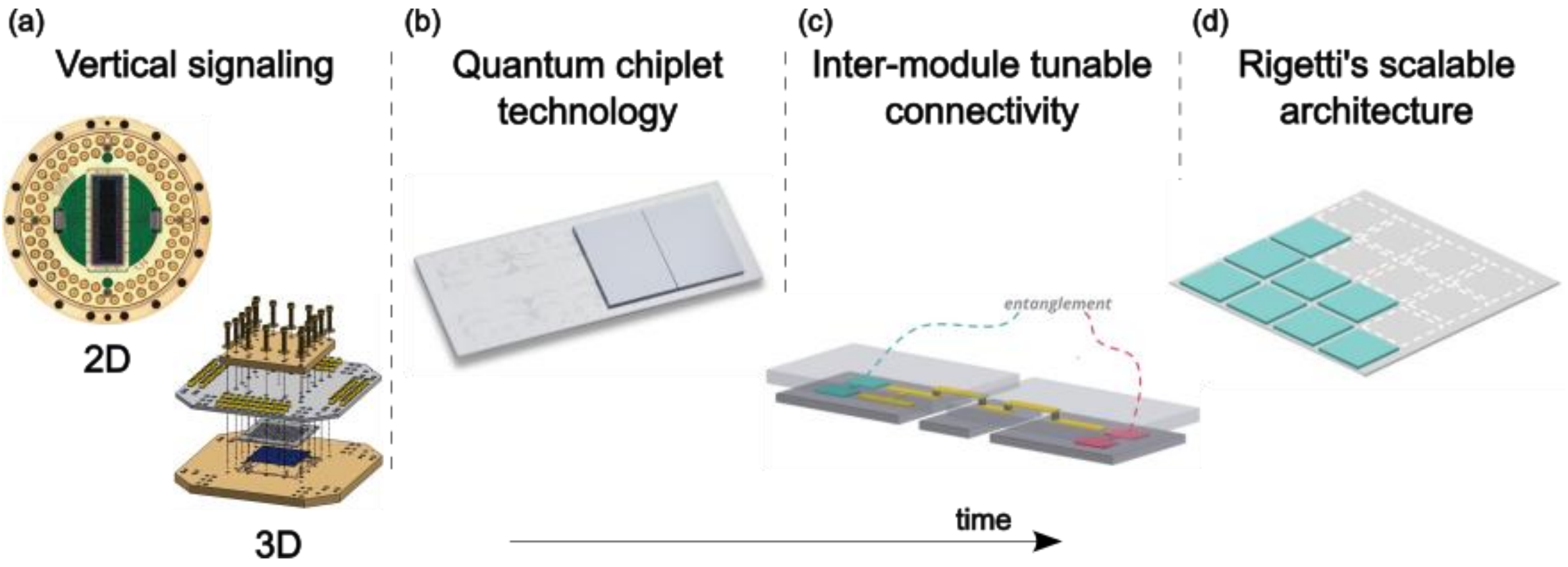

**Figure 23.2**: Envisioned Rigetti's scaling roadmap. (a) From laterally routed signals to 3D integration with IO signals delivered vertically to support large monolithic processors. (b) Modular assembly onto a carrier to enable high fabrication yield, improved performance and heterogeneous integration. (c) Inter-module tunable connectivity for high fidelity quantum entangling gates between modules. (d) Large scale processors build from identical tiles.

In the foreseeable future, optical-to-microwave transducers may be essential components for transmitting quantum information between quantum nodes housed in separate cryogenic systems, and they could potentially provide an alternative for controlling and reading out qubits on the same device. To achieve these goals, the conversion efficiency of the first-generation transducers must be significantly improved. Hybrid JJs can be used in novel types of qubits or quantum components. The fluxonium, gatemon, and ferrotransmon qubits offer alternative designs and notions to search novel solutions for scalability[12].

### 23.4 Concluding Remarks

Superconducting qubits have emerged as a leading platform for scalable quantum computing, with significant progress achieved in coherence times, gate fidelities, and processor integration. However, the path to realizing a universal quantum machine remains challenging and requires a multi-effort approach. Achieving fault-tolerant quantum computation requires innovations in materials, fabrication techniques, and circuit designs to enhance qubit quality and uniformity. Simultaneously, scaling processors to millions of qubits necessitates advanced cryogenic systems, compact packaging, high-density wiring, and the development of multiplexed control and readout schemes.

Beyond quantum hardware, quantum error correction protocols and their implementation remain a cornerstone for achieving reliable computation. The exploration of alternative qubit designs, such as fluxonium, gatemon, and ferrotransmon, alongside the integration of hybrid technologies like optical-to-microwave transducers or cryogenic electronics, highlights the potential for novel solutions to scalability challenges.

As the field continues to evolve, an approach balancing performance, scalability, and practical feasibility will be essential to push the boundaries of quantum computing and unlock its potential.

## Acknowledgements

A.B, D.M. and F.T. acknowledge support from Pathfinder EIC 2023 project ”FERROMON-Ferrotransmons and Ferrogatemons for Scalable Superconducting Quantum Computers”; D.M. and F.T. also from the PNRR-MUR CN00000013-ICSC and PE0000023-NQSTI projects and the COST Action SUPERQUMAP (CA21144).